\documentclass[preprint,prc,tightenlines,nofootinbib,superscriptaddress,floatfix]{revtex4}
\usepackage{hyperref}
\usepackage[T1]{fontenc}
\usepackage[utf8]{inputenc}
\usepackage{color}
\usepackage{slashed}
\usepackage{nicefrac}

\usepackage{longtable}
\usepackage{amssymb}
\usepackage{amsmath}

\usepackage{booktabs}
\usepackage{graphicx}

\renewcommand{\arraystretch}{1.35}

\newcommand{\po}{\phantom{1}}

\newcommand{\ph}{\phantom{-}}

\newcommand{\punkt}{\;\;\mbox{.}} 
\newcommand{\stack}[2]{\,\substack{\pm #1 \\ \pm #2}}

\allowdisplaybreaks

\begin{document}
\title{Nucleon polarizabilities in covariant baryon chiral perturbation theory with explicit $\Delta$ degrees of freedom}

\author{M.~Th\"urmann}
\email[]{markus.thuermann@rub.de}
\affiliation{Institut f\"ur Theoretische Physik II, Ruhr-Universit\"at Bochum, D-44780 Bochum, Germany}

\author{E.~Epelbaum} 
\email[]{evgeny.epelbaum@rub.de}
\affiliation{Institut f\"ur Theoretische Physik II, Ruhr-Universit\"at
  Bochum, D-44780 Bochum, Germany}

\author{A.~M.~Gasparyan}
\email[]{ashot.gasparyan@rub.de}
\affiliation{Institut f\"ur Theoretische Physik II, Ruhr-Universit\"at Bochum, D-44780 Bochum, Germany}
\affiliation{NRC “Kurchatov Institute” - ITEP, B. Cheremushkinskaya 25, 117218 Moscow, Russia}

\author{H.~Krebs}
\email[]{hermann.krebs@rub.de}
\affiliation{Institut f\"ur Theoretische Physik II, Ruhr-Universit\"at Bochum, D-44780 Bochum, Germany}

\begin{abstract} 
We compute various nucleon polarizabilities
in chiral perturbation theory implementing the $\Delta$-full ($\Delta$-less)
approach up to order $\epsilon^3 + q^4$ ($q^4$) in the small-scale (chiral) expansion.
The calculation is carried out using the covariant formulation of
$\chi$PT by utilizing the
extended on-mass shell renormalization scheme.
Except for the spin-independent dipole polarizabilities used  to fix the
values of certain low-energy constants, our
results for the nucleon polarizabilities are pure predictions.
We compare our calculations with available experimental data and other theoretical results.
The importance of the explicit treatment of the $\Delta$ degree of
freedom in the effective field theory description of 
the nucleon polarizabilities is analyzed. 
We also study the convergence of the $1/m$ expansion and analyze the
efficiency of the heavy-baryon approach 
for the nucleon polarizabilities.
\end{abstract}

\maketitle

\section{Introduction}
Understanding the structure of the nucleon is one of the key challenges in the physics of strong interactions,
and quantum chromodynamics (QCD) in particular.
One of the most direct ways to access the nucleon structure is to use electromagnetic probes.
In the present work we focus on the nucleon polarizabilites, which characterize the (second-order) response 
of the nucleon to an applied electromagnetic field.
In recent decades, the nucleon polarizabilites have been intensively studied both experimentally and theoretically.
At the moment, the dipole scalar (spin-independent) polarizabilites of both the proton and the neutron
are determined fairly well by various methods \cite{Tanabashi:2018oca} as well as the
forward and backward spin polarizabilites of the proton \cite{Ahrens:2001qt,Dutz:2003mm,Camen:2001st}.
Recent measurements of double-polarized Compton scattering at the Mainz Microtron
allowed one to extract also other proton spin polarizabilites
\cite{Martel:2014pba, Paudyal:2019mee}.

There are also experimental results for some of the generalized ($Q^2$-dependent) polarizabilites of the proton and 
the neutron
\cite{Prok:2008ev,Amarian:2004yf,Guler:2015hsw,Gryniuk:2016gnm,Deur:2019pew}.

From the theoretical side, a significant progress has been made
using lattice simulations \cite{Detmold:2010ts, Hall:2013dva, Lujan:2014kia, Freeman:2014kka, Chang:2015qxa, Bignell:2017lnd, Bignell:2018acn},
i.e.~by directly solving QCD in the non-perturbative regime on a
discrete Euclidean space-time grid.
However, one is not yet in the position to perform an accurate determination of the nucleon polarizabilites 
calculated on the lattice for physical pion masses.

Another systematic theoretical approach is provided by
effective field theories, in particular, by chiral perturbation theory
($\chi$PT), see \cite{Bernard:1991rq,Bernard:1991ru} for pioneering studies of the nucleon's
electromagnetic polarizabilities in this framework.
Chiral perturbation theory is an effective field theory of the standard model
consistent with its symmetries and the ways they are broken.
It allows one to expand hadronic observables in powers of the small
parameter $q$ defined as  the 
ratio of the typical soft scales such as the pion mass $M$ and external-particle 3-momenta $|\vec p|$
and the hard scale $\Lambda_{\rm b}$ of the order of the $\rho$-meson
mass.   
The effective chiral Lagrangian is expanded in powers of derivatives and the pion mass.
In the nucleon sector, an additional complication arises due to the presence of 
an extra mass scale, namely the nucleon mass, which can potentially break the power counting.
One way to circumvent this problem is to perform the $1/m$ expansion on the level of the 
effective Lagrangian. This leads to the so-called heavy-baryon approach.
The heavy-baryon scheme has been intensively used for the analysis of many 
hadronic  reactions including the nucleon Compton scattering (and,
therefore, nucleon polarizabilites), see
e.g.~\cite{Bernard:1995dp,VijayaKumar:2000pv,Gellas:2000mx,Hemmert:1996rw}
and \cite{Bernard:1995dp,Bernard:2007zu} for review articles.
The heavy-baryon expansion is, however, known to violate certain analytic properties of the $S$-matrix \cite{Becher:1999he},
which may lead to a slower convergence of the chiral expansion.
This feature has also been observed in the actual calculations of the nucleon polarizabilites.

An alternative approach to processes involving nucleons consists in keeping the covariant structure of the effective Lagrangian
and absorbing the power-counting  breaking terms by a redefinition of the lower order low-energy constants \cite{Becher:1999he, Fuchs:2003qc}.
In this work, we adopt a version of the covariant approach
known as the extended on-mass-shell renormalization scheme (EOMS) \cite{Gegelia:1999gf,Fuchs:2003qc}.
When necessary, we will slightly modify this scheme in order to enable
a direct comparison to the heavy-baryon results (see e.g.~\cite{Siemens:2016hdi}).

Another obstacle for the rapid convergence of the chiral expansion in the single-nucleon systems is the 
presence of the $\Delta$(1232)-resonance that is located close to the
pion-nucleon threshold and is known to strongly couple to the pion-nucleon channel.
This introduces another small scale $\Delta \equiv m_\Delta-m\approx
2\, M $, which leads to the appearance of terms of order 
$\mathcal{O}(M /\Delta)$ in the expansion of observables.
A natural way to improve this situation is to include the $\Delta$-isobar field explicitly into the effective Lagrangian.
We follow here the so-called small-scale-expansion (SSE) scheme by treating the scale $\Delta$ on the same footing 
as $M $ or $|\vec p|$ \cite{Hemmert:1997ye}. The universal
expansion parameter is then called $\epsilon$. For recent applications
of this theoretical approach to various processes in the
single-nucleon sector see \cite{Bernard:2012hb,Yao:2016vbz,Siemens:2016jwj,Rijneveen:2020qbc}. 
In this work, we compare the efficiency and convergence of both the $\Delta$-full and $\Delta$-less schemes
by calculating various nucleon polarizabilites up to orders $\epsilon^3 + q^4$ and $q^4$, respectively.
Our analysis is particularly instructive since we calculate a set of higher-order polarizabilites, which do not
depend on any free parameters.
We also perform the $1/m$ expansion of our results in order to analyze the efficiency of the heavy baryon approach
for the nucleon polarizabilites.

There is an alternative scheme for the chiral expansion in the
presence of explicit $\Delta$ degree of freedom
\cite{Pascalutsa:2002pi} 
called the $\delta$-counting. The main difference from the small-scale
expansion is a different power counting assignment for 
the $\Delta$-nucleon mass difference $\Delta$ by assuming the
hierarchy of scales $M\ll\Delta\ll\Lambda_{\rm b}$.
In such an approach, loop diagrams with several $\Delta$-lines are
suppressed in contrast with the calculations within the small-scale
expansion,
see \cite{Geng:2008bm,Alarcon:2011zs,Alarcon:2020wjg,Alarcon:2020icz} for recent applications. 
We compare our results with the ones obtained within the
$\delta$-counting and discuss the importance of such contributions. 

As a stringent test of our scheme, we also compare our results
with the fixed-$t$ dispersion-relations analyses of \cite{Babusci:1998ww,Holstein:1999uu,OlmosdeLeon:2001zn,Drechsel:2002ar,Hildebrandt:2003fm,Pasquini:2007hf}.
This method is based solely on the  principles of analyticity and
unitarity\footnote{For the application of a scheme that combines
  effective field theory with
dispersion-relations technique for the problem under consideration see
\cite{Gasparyan:2010xz,Gasparyan:2011yw}.}
and
therefore defines an important benchmark for theoretical approaches.

Our paper is organized as follows. The effective Lagrangian
and the power counting relevant for the construction of the Compton scattering amplitude 
within $\chi$PT as well as the renormalization of the low-energy
constants (LECs) are given in Section~\ref{sec:ChPT}.
In Section~\ref{sec:Formalism}, the formalism for the Compton scattering is described
and the nucleon polarizabilities are introduced. 
The numerical results for the nucleon polarizabilites are presented in Section~\ref{sec:numerical-results}.
We summarize our results in Section~\ref{sec:summary}. Appendices
\ref{sec:q3-loop}-\ref{sec:loop_integrals} collect the analytic expressions for the
nucleon polarizabilites.

\section{Compton scattering in chiral perturbation theory}\label{sec:ChPT}
\subsection{Effective Lagrangian}\label{sec:Lagrangian}
The description of nucleon Compton scattering in $\chi$PT relies on an effective Lagrangian. 
The effective Lagrangian relevant for the problem at hand to the order we are working consists of the following terms
\begin{align}
  \mathcal{L}_\mathrm{eff}=\mathcal{L}_{\pi\pi}^{(2)}+\mathcal{L}_{\pi\pi}^{(4)}+\mathcal{L}^{(4)}_{WZW}
    +\mathcal{L}_{\pi N}^{(1)}+\mathcal{L}_{\pi N}^{(2)}+\mathcal{L}_{\pi N}^{(3)}
    +\mathcal{L}_{\pi N}^{(4)}+\mathcal{L}^{(1)}_{\pi N \Delta}+\mathcal{L}^{(2)}_{\pi N \Delta}+\mathcal{L}^{(1)}_{\pi \Delta \Delta}\,,
\label{eq:Lagrangian}
\end{align}
where $\mathcal{L}_{WZW}$ stays for the Wess-Zumino-Witten term \cite{Wess:1971yu,Witten:1983tw}.
This Lagrangian is built in terms of the pion field through the SU(2)
matrix $U=u^2=1+\frac{i}{F}\vec\tau\cdot\vec\pi-\frac{1}{2F^2}\vec\pi^2+\dots$ 
($F$ is the pion decay constant in the chiral limit), 
the nucleon field $N$ and the Rarita-Schwinger-spinor $\Delta$-field $\psi_i^\mu$. 
The electromagnetic field $A^\mu$ enters via $v_\mu=-\tfrac{1}{2}(1+\tau_3)eA_\mu$ ($e>0$ is the proton charge).

Here, we list only the terms in the pion-nucleon 
Lagrangian \cite{Fettes:2000gb} appearing in the course of calculating the nucleon polarizabilites:
\begin{align}
\mathcal{L}^{(1)}_{\pi N} = &\ \bar{N}(i\slashed{D}-m+\frac{g_A}{2}\slashed{u}\gamma_5)N \\
\mathcal{L}^{(2)}_{\pi N} = 
&\ c_1\bar{N}\left<\chi_+\right>N - \frac{c_2}{8m^2}\left(\bar{N}\left<u_\mu u_\nu\right>D^{\mu\nu}N + \text{h.c}\right) + \frac{c_3}{2}\left<u\cdot u\right>  \nonumber \\ 
& \hspace{0.25cm} +c_4 i\bar{N}\left[u_\mu,u_\nu\right]\sigma^{\mu\nu}N + \frac{c_6}{2m} \bar{N}F^{+}_{\phantom{+}\mu\nu}N + \frac{c_7}{2m}\bar{N}\left<F^+_{\phantom{+}\mu\nu}\right>\sigma^{\mu\nu}N + \ldots \\
\mathcal{L}^{(3)}_{\pi N} = 
&\ \frac{d_6}{2m}\left(i\left[D^\mu,\tilde{F}^{+}_{\phantom{+}\mu\nu}\right]D^\nu +\text{h.c.}\right) + \frac{d_7}{2m}\left(i\left[D^\mu,\left<F^{+}_{\phantom{+}\mu\nu}\right>\right]D^\nu + \text{h.c.}\right) + \ldots \\
\mathcal{L}^{(4)}_{\pi N} =  
&\bar{N}\Big(
-\frac{e_{54}}{2}\left[D^\alpha,\left[D_\alpha,\left<F^{+}_{\phantom{+}\mu\nu}\right>\right]\right]\sigma^{\mu\nu}
 -\frac{e_{74}}{2}\left[D^\alpha\left[D_\alpha,\tilde{F}^+_{\phantom{+}\mu\nu}\right]\right]\sigma^{\mu\nu} \nonumber \\ 
&- \frac{e_{105}}{2}\left<F_{\phantom{+}\mu\nu}^+\right>\left<\chi_+\right>\sigma^{\mu\nu}
 - \frac{e_{106}}{2} \tilde{F}^+_{\phantom{+}\mu\nu}\left<\chi_+\right>\sigma^{\mu\nu}\nonumber \\ 
&+e_{89}\left<F^{+}_{\phantom{+}\mu\nu}\right>\left<F^{+\mu\nu}\right>
+  e_{91}\tilde{F}^{+}_{\phantom{+}\mu\nu}\left<F^{+\mu\nu}\right>
+ e_{93}\left<\tilde{F}^{+}_{\phantom{+}\mu\nu}\tilde{F}^{+\mu\nu}\right> \nonumber \\ 
&+\frac{e_{118}}{2}\left<F^{-\mu\nu}F^{-}_{\ph \mu\nu}+F^{+\mu\nu}F^{+}_{\phantom{+}\mu\nu}\right>
\Big)N\nonumber \\ 
&+\Big[\bar{N}\Big(
-\frac{e_{90}}{4m^2}\left<F^{+}_{\phantom{+}\alpha\mu}\right>\left<F^{+\alpha}_{\phantom{+\alpha}\nu}\right> 
 - \frac{e_{92}}{4m^2}\tilde{F}^{+}_{\phantom{+}\alpha\mu}\left<F^{+\alpha}_{\phantom{+\alpha}\nu}\right>
 - \frac{e_{94}}{4m^2} \left<\tilde{F}^{+}_{\phantom{+}\alpha\mu}\tilde{F}^{+\alpha}_{\phantom{+\alpha}\nu}\right>  \nonumber \\
& -\frac{e_{117}}{8m^2}\left<F^{-}_{\ph \alpha\mu}F^{-\alpha}_{\phantom{-\alpha}\nu}+
F^{+}_{\phantom{+}\alpha\mu}F^{+\alpha}_{\phantom{+\alpha}\nu}\right>\Big)\{D^\mu,D^\nu\}N + \text{h.c.}\Big] \,,
\end{align}
and the  terms relevant for the $\mathcal{O}(\epsilon^3)$ calculations from the $\pi N\Delta$ and $\pi \Delta\Delta$ Lagrangians: 
\begin{align}
\mathcal{L}^{(1)}_{\pi N \Delta} &= \frac{h_A}{2}\left(\bar{\Psi}^{\mu}_i\left<\tau_i u_\mu\right> N + \text{h.c.}\right)\,, \nonumber\\
\mathcal{L}^{(2)}_{\pi N \Delta} &= \frac{b_1}{4} \left(i \bar{\Psi}^{\mu}_i \left<\tau_i F^+_{\mu\alpha}\right>\gamma^\alpha\gamma_5 N + \text{h.c.}\right)\,,\nonumber\\
{\cal L}_{\pi\Delta\Delta}^{(1)} &= \bar\Psi_i^\mu
\,\Big(\frac{i}{4}\bigl\{[\gamma_\mu , \,  \gamma_\nu],
\gamma_\alpha\bigr\}\,D_{ij}^\alpha  - \frac{m_\Delta}{2} [\gamma_\mu , \,  \gamma_\nu] \delta_{ij}\Big) \, \Psi_j^\nu\,.
\end{align}
The covariant derivatives and the chiral vielbein are defined as follows:
\begin{align}
D_\mu &= \partial_\mu + \Gamma_\mu\,, \quad D_{ij}^\mu = (\partial^\mu + \Gamma^\mu) \, \delta_{ij} 
- i\epsilon_{ijk}\langle\tau^k\Gamma^\mu\rangle~,\nonumber\\
\Gamma_\mu &= \frac{1}{2}\left[u^\dagger\partial_\mu u 
+u\partial_\mu u^\dagger-i(u^\dagger v_\mu u + u v_\mu u^\dagger)\right]\,,\nonumber\\
u_\mu &= i\left[u^\dagger\partial_\mu u - u\,\partial_\mu u^\dagger - i(u^\dagger v_\mu u - u\, v_\mu u^\dagger)\right]\,,
\end{align}
while the vector field strength tensors are given by
\begin{align}
F_{\mu\nu}^\pm &= u v_{\mu\nu} u^\dagger \pm u^\dagger v_{\mu\nu} u\,, \quad \quad
\tilde
                   F_{\mu\nu}^+=F_{\mu\nu}^+-\tfrac{1}{2}\left<F^{+}_{\phantom{+}\mu\nu}\right>\,, \quad \quad
v_{\mu\nu} = \partial_\mu v_\nu - \partial_\nu v_\mu\, .
\end{align}
Notice that the definition of $b_1$ differs from the one in \cite{Hemmert:1997ye} by a factor of $m$
but is consistent with that of \cite{Bernard:2012hb}.
All redundant off-shell parameters in $\mathcal{L}_{\pi N \Delta}$ and
${\cal L}_{\pi\Delta\Delta}$ are set to zero (see the discussion in 
\cite{{Tang:1996sq,Krebs:2009bf}}). 

For the remaining terms in Eq.~(\ref{eq:Lagrangian}) and further
notations we refer the reader to
\cite{Gasser:1983yg,Fettes:2000gb,Hemmert:1997ye,Hemmert:1997wz}). 

\subsection{Power counting}\label{sub:powercounting}
To calculate the nucleon Compton-scattering amplitude one needs to select the relevant Feynman diagrams
according to their order $D$, which is determined  by the power-counting formula \cite{{Weinberg:1991um}}
\begin{align}
D=1+2L+\sum_{n}(2n-2)V_{2n}^M+\sum_d(d-1)V_d^B \label{eq:powercounting},
\end{align}
where $ L $ is the number of loops, 
$ V^M_{2n} $ is the number of vertices from  $\mathcal{L}_{\pi\pi}^{(2n)}$
and $ V^B_d $ is the total number of vertices from $\mathcal{L}_{\pi N}^{(d)}$, 
$\mathcal{L}^{(d)}_{\pi N \Delta}$ and $\mathcal{L}^{(d)}_{\pi \Delta \Delta}$.
Note that in the small-scale-expansion scheme, the nucleon and delta lines are counted on the same footing.
In this work, we label purely nucleonic contributions (containing no $\Delta$ lines) as $ q^D $ and those involving $\Delta$'s as $ \epsilon^D $.

\begin{figure}[ht]\centering 
\includegraphics[width=0.7\textwidth]{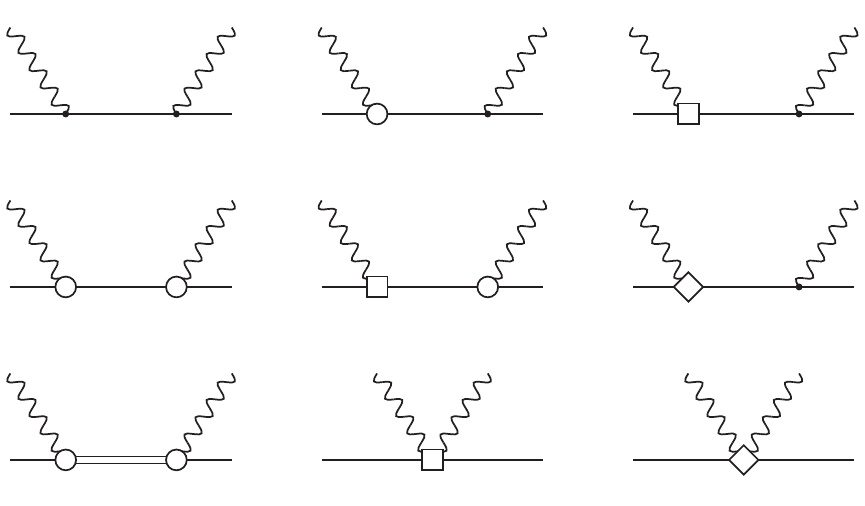}
\caption{Tree-level diagrams for nucleon Compton scattering which are
  taken into account in our analysis. Vertices of order $\mathcal{O}(q)$,
  $\mathcal{O}(q^2)$, $\mathcal{O}(q^3)$ and $\mathcal{O}(q^4)$ are denoted by dots, circles,
  squares and diamonds, respectively. Solid, wavy and double lines refer to
  nucleons, photons and $\Delta$-isobars, respectively.
  Time-reversed and crossed diagrams as well as the 
diagrams with insertions of the nucleon self-energy contact terms are not shown.}
\label{fig:e3-tree-diagrams}
\end{figure}

The tree-level diagrams are shown in Fig.~\ref{fig:e3-tree-diagrams}.
Most of the nucleon pole diagrams do not contribute to the polarizabilites (as the Born terms are subtracted by definition, see Section~\ref{sec:Formalism})
but are necessary for the renormalization of subdiagrams.
Only the nucleon pole diagrams with the $d_6$ and $d_7$ vertices generate a small residual 
non-pole contribution to the generalized polarizabilites due to the specific form 
of the corresponding effective Lagrangian.

On the other hand, the $\Delta$-pole graph provides a very important contribution to the nucleon 
polarizabilites.
The pion $t$-channel exchange diagram with the anomalous $\pi^0\gamma\gamma$ coupling is 
not included in the definition of the polarizabilites either and is, therefore, not shown.
Also not shown are the $\gamma N\to\gamma N$ contact terms from $\mathcal{L}^{(4)}_{\pi N}$.

\begin{figure}[ht]\centering 
\includegraphics[width=0.96\textwidth]{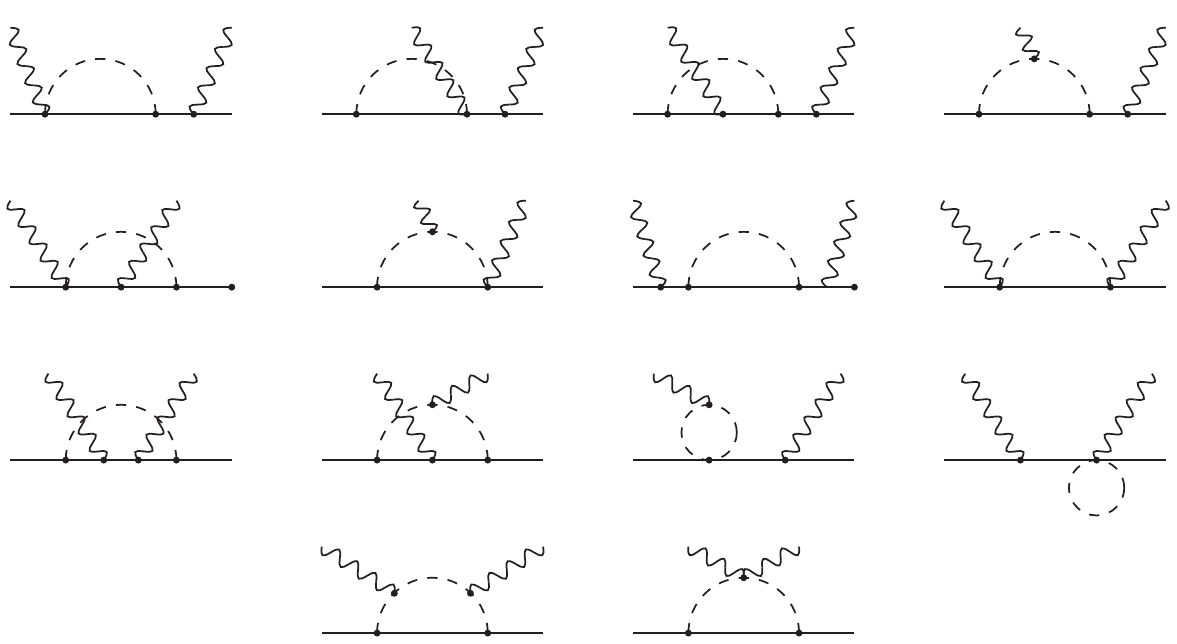}
\caption{$\mathcal{O}(q^3)$ loop diagrams for nucleon Compton scattering. Dashed
  lines refer to pions. All vertices are from the leading
order Lagrangians $\mathcal{L}_{\pi\pi}^{(2)}$ and $\mathcal{L}_{\pi N}^{(1)}$. Time-reversed and crossed diagrams are not shown.}
\label{fig:q3-diagrams}
\end{figure}

\begin{figure}[ht]\centering 
\includegraphics[width=0.96\textwidth]{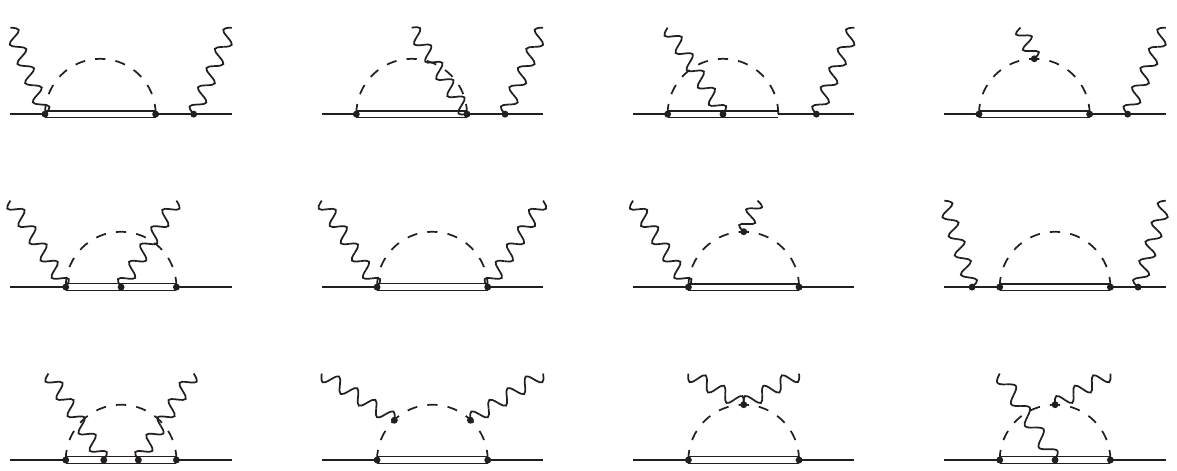}
\caption{$\mathcal{O}(\epsilon^3)$ loop diagrams for nucleon Compton scattering. All vertices are from the leading
order Lagrangians $\mathcal{L}_{\pi\pi}^{(2)}$, $\mathcal{L}_{\pi N}^{(1)}$, $\mathcal{L}^{(1)}_{\pi N \Delta}$ and $\mathcal{L}^{(1)}_{\pi \Delta \Delta}$.
Double lines denote the $\Delta$.
Time-reversed and crossed diagrams are not shown.}
\label{fig:e3-diagrams}
\end{figure}

\begin{figure}[ht]\centering 
\includegraphics[width=0.96\textwidth]{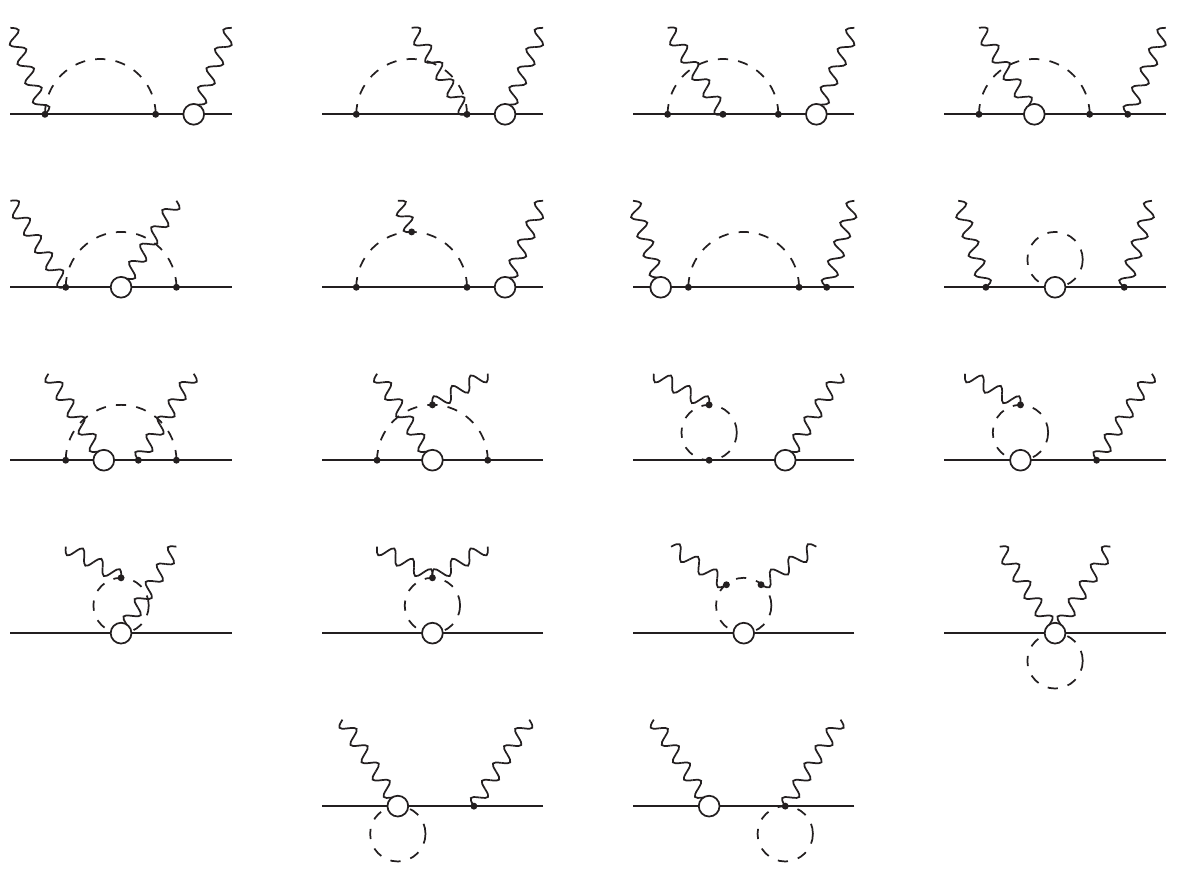}
\caption{$\mathcal{O}(q^4)$ loop diagrams for nucleon Compton scattering. Dots denote the leading order vertices and
circles denote the vertices from $\mathcal{L}_{\pi N}^{(2)}$. Time-reversed and crossed diagrams are not shown.}
\label{fig:q4-diagrams}
\end{figure}

Loop diagrams start to contribute at order $q^3$($\epsilon^3$). The corresponding sets of diagrams are
shown in Fig.~\ref{fig:q3-diagrams} for the $q^3$-loops and in Fig.~\ref{fig:e3-diagrams} for the $\epsilon^3$-loops.
The subleading $q^4$-loop diagrams are shown in Fig.~\ref{fig:q4-diagrams}.

\subsection{Renormalization}\label{sub:renormalization}
The ultraviolet divergencies appearing in loop integrals are treated
by means of dimensional regularization. 
Divergent parts of the integrals are cancelled by the corresponding counter terms 
of the Lagrangian, and the resulting amplitude is
expressed in terms of the finite quantities such as renormalized low-energy constants, physical masses and coupling constants.
Due to the presence of an additional hard scale (the nucleon or $\Delta$ mass), 
baryonic loops contain power-counting-violating terms \cite{Gasser:1987rb}.
Since such terms are local, they can be absorbed by a redefinition of the low-energy constants of the effective Lagrangian.
In this work, we adopt the extended on-mass-shell renormalization scheme (EOMS) \cite{Fuchs:2003qc} in
a combination with on-shell renormalization conditions for the nucleon mass and magnetic moments.

For the nucleon mass and wave-function renormalization, we impose the on-shell conditions
\begin{align}
\Sigma_{N}(m_N)=0 \ \text{and} \ 
\Sigma_{N}'(m_N)=0,
\end{align}
with $\Sigma_{N}(\slashed p)$ being the nucleon self-energy. 
By doing so, we fix the bare nucleon mass $ m $ and the field normalization factor $ Z_N $. 
The explicit formulae relating the physical and bare parameters can be found in \cite{Yao:2016vbz}. 
In what follows, we will denote the physical nucleon mass by $m$, which will not lead to a confusion
since the bare nucleon mass will not be discussed anymore.

In a complete analogy with the nucleon field, we renormalize the $\Delta$ field.
However, at the order we are working there are no loop corrections to the $\Delta$ self-energy.
For the calculation of the static nucleon polarizabilites, we us the real Breit-Wigner mass of the $\Delta$.
The precise value of the renormalized $\Delta$ mass is irrelevant under the kinematic conditions considered.
On the other hand, for calculation of the dynamical nucleon polarizabilites, 
in order to be able to describe the $\Delta$ region,
we implement the complex-mass scheme \cite{Denner:1999gp,Denner:2006ic} for the $\Delta$ resonance and use
the complex $\Delta$ pole mass taking the resonance width into account explicitly.

For the renormalized constants $\bar{c}_6$ and $\bar{c}_7$, we use the on-shell condition for the nucleon magnetic moments:
\begin{align}
\bar{c}_6 = \kappa_p - \kappa_n \ \text{ and }  \ \bar{c}_7 = \kappa_n  \,. \label{eq:c6c7}
\end{align}
The explicit relation between $\bar{c}_6$ and $\bar{c}_7$ and the bare constants $c_6$ and $c_7$
is given in Appendix~\ref{sec:counter_terms}.

For the remaining low-energy constants $\xi_i$  we employ the EOMS renormalization scheme.
The renormalized LECs $\bar\xi_i$ are related to the bare quantities as follows:
\begin{align} 
&\xi_i = \bar{\xi}_i - \frac{\beta_{\xi_i}}{F^2} \frac{A_0(M ^2)}{2 M ^2}
 + \frac{\Delta_{\xi_i}}{16\pi^2F^2}\,,\nonumber\\
&\xi_i \in \{d_6, d_7, e_{54}, e_{74}, e_{89}, e_{89}, e_{90}, e_{91}, e_{92}, e_{93}, e_{94}, e_{117}, e_{118}\}\,,\label{eq:LECshift}
\end{align}
with the $\beta$ functions:
\begin{align}
& \beta_{d_6} = -\frac{1-g_A^2}{6}+\frac{40h_A}{81}\,, \
\beta_{d_7} = \frac{5h_A^2}{54} \,,\nonumber\\
&\beta_{e_{54}} = 0\,, \ \beta_{e_{74}}= \frac{1-g_A^2+4c_4}{12m} \,,\nonumber\\
&\beta_{e_{91}} = \beta_{e_{92}} = 0 \,,\
 2\beta_{e_{89}} +\beta_{e_{93}} +  \beta_{e_{118}}  = \frac{c_2}{12} \,,\nonumber\\
& 2\beta_{e_{90}} +\beta_{e_{94}} +  \beta_{e_{117}}  = - \frac{c_2}{3} \,,\label{eq:beta_functions}
\end{align}
and the finite shifts
\begin{align}
 \Delta_{d_6} = -\frac{g_A^2c_6}{8} \,, \
\Delta_{d_7} = \frac{3g_A^2(c_6+2c_7)}{16}\,.
\end{align}
The constants $e_i$  do not receive finite shifts due to the power-counting violation
because we do not consider loop diagrams of order higher than $\mathcal{O}(q^4)$.
The finite shifts for $d_6$ and $d_7$ reproduce those obtained in \cite{Fuchs:2003ir} 
(note a different definition of the LECs).
The constants $e_{54}$ and $e_{74}$ do not contribute to the nucleon polarizabilites after subtracting the
Born terms. Nevertheless, we provide the corresponding $\beta$ functions for completeness.
The LECs $e_{89}$, $e_{90}$, $e_{93}$, $e_{94}$, $e_{117}$, $e_{118}$ enter  the nucleon
Compton scattering amplitude only in the linear combinations $2e_{89}+e_{93}+e_{118}$
and $2e_{90}+e_{94}+e_{117}$, for which the $\beta$ functions are given in Eq.~(\ref{eq:beta_functions}).

The pion tadpole function in $d\approx 4 $ dimensions is equal to (see Eq.~(\ref{eq:loop_integrals}))
\begin{align}
A_0(M)
=- 2 M ^2\left(\bar{\lambda}+\frac{1}{32\pi^2}\ln\left(\frac{M ^2}{\mu^2}\right)\right)\,,\nonumber\\
\bar{\lambda}=\frac{1}{16\pi^2}\left(\frac{1}{d-4}+
\frac{1}{2}(\gamma_\text{E}-\ln(4\pi)-1)\right)\,.
\end{align}
Here, $ \gamma_\text{E} $ is the Euler-Mascheroni constant and $ \mu $ is the renormalization scale. 
The divergencies remaining after the renormalization of the LECs are treated in the $\widetilde{\mathrm MS}$ \cite{Fuchs:2003qc,Gasser:1983yg} 
scheme, i.e. we set $ \bar{\lambda}=0 $.
We have checked that the residual renormalization scale dependence of
the amplitude is of a higher order than we are working.

In what follows, we will omit the bars over the renormalized LECs.

\section{Formalism}\label{sec:Formalism}
We consider nucleon Compton scattering $\gamma N\to \gamma N$ with the momenta of 
the initial (final) proton and photon denoted as $p$ ($p'$) and $q$ ($q'$), respectively.
We study the cases of real Compton scattering with $q^2=q'^2=0$ and of double virtual Compton scattering
with $Q^2=-q^2=-q'^2$.

In order to calculate the nucleon polarizabilites, we decompose the scattering amplitude $T(q^2,z,\omega)$ in the Breit frame,
where $\omega$ and $z$ are the photon energy and scattering angle,
in terms of twelve functions $A_i$:
\begin{align} 
T(q^2,z,\omega) = 2m\sum_{i = 1}^{12}A_i(q^2,z,\omega)\chi_i\,, \label{eq:amplitude}
\end{align}
with
\begin{align}
\chi_{1} &= \vec{\epsilon}\cdot \vec{\epsilon}^* \,,\nonumber\\
\chi_{2} &= (\hat{q}\times \vec{\epsilon})\cdot(\hat{q}'\times \vec{\epsilon}') \,,\nonumber\\
\chi_{3} &= \hat{q}\cdot \vec{\epsilon}\; \hat{q}\cdot \vec{\epsilon}'^* + \hat{q}'\cdot \vec{\epsilon}\;\hat{q}'\cdot \vec{\epsilon}'^* \,,\nonumber\\
\chi_{4} &= \hat{q}\cdot \vec{\epsilon}\; \hat{q}'\cdot \vec{\epsilon}'^*\,,\nonumber\\
\chi_{5} &= i\sigma\cdot\vec{\epsilon}\times\vec{\epsilon}'^* \,,\nonumber\\
\chi_{6} &= i\sigma\cdot(\hat{q}\times\vec{\epsilon})\times(\hat{q}'\times\vec{\epsilon}') \,,\nonumber\\
\chi_{7} &= i(\hat{q}\cdot\vec{\epsilon}\times\vec{\epsilon}'^*\vec{\sigma}\cdot\hat{q} + \hat{q}'\cdot\vec{\epsilon}\times\vec{\epsilon}'^*\vec{\sigma}\cdot\hat{q}')\,,\nonumber\\
\chi_{8} &= i(\hat{q}\cdot\vec{\epsilon}\times\vec{\epsilon}'^*\vec{\sigma}\cdot\hat{q}' + \hat{q}'\cdot\vec{\epsilon}\times\vec{\epsilon}'^*\vec{\sigma}\cdot\hat{q})\,,\nonumber\\
\chi_{9} &= i\hat{q}\cdot \vec{\epsilon}\; \hat{q}'\cdot \vec{\epsilon}'^*\vec{\sigma}\cdot\hat{q}\times \hat{q}' \,,\nonumber\\
\chi_{10} &=  i(\hat{q}\cdot \vec{\epsilon}\; \hat{q}\cdot \vec{\epsilon}'^*\vec{\sigma}\cdot\hat{q}\times \hat{q}' +  \hat{q}'\cdot \vec{\epsilon}\; \hat{q}'\cdot \vec{\epsilon}'^*\vec{\sigma}\cdot\hat{q}\times \hat{q}') \,,\nonumber\\
\chi_{11} &= i(\hat{q}'\cdot \vec{\epsilon}'^*\vec{\sigma}\cdot\vec{\epsilon}\times\hat{q} - \hat{q}\cdot \vec{\epsilon}\; \vec{\sigma}\cdot\vec{\epsilon}'^*\times\hat{q}')\,,\nonumber\\
\chi_{12} &= i(\hat{q}'\cdot \vec{\epsilon}'^*\vec{\sigma}\cdot\vec{\epsilon}\times\hat{q}' - \hat{q}\cdot \vec{\epsilon}\; \vec{\sigma}\cdot\vec{\epsilon}'^*\times\hat{q}) 
\,. \label{eq:final-decomposition}
\end{align}
The initial (final) photon polarization vector $ \epsilon_\mu$ ($ \epsilon'_\mu$) is defined in the Coulomb gauge ($ \epsilon_0=\epsilon'_0=0$).
The amplitude (\ref{eq:amplitude}) is supposed to be sandwiched between the Pauli spinors of the initial and final nucleon.

Given the presence of the Pauli matrices $\vec{\sigma}$ in Eq.~(\ref{eq:final-decomposition}), one can see
that there are four spin-independent structures $\chi_{1}-\chi_{4}$ and eight spin-dependent structures $\chi_{5}-\chi_{12}$.
All $\chi_i$ obey crossing-invariance. 
For real Compton scattering, only 
$\chi_1$, $\chi_2$, $\chi_5$, $\chi_6$, $\chi_7$, $\chi_8$ survive. 

The Born terms have to be  subtracted from the amplitude as explained, e.g., in \cite{Lensky:2017bwi}
in order to exclude the contributions with unexcited nucleons in the intermediate state.
This procedure essentially reduces to subtracting the tree-level $Q^2=0$ nucleon-pole diagrams  
with the nucleon charge and magnetic moments replaced by the 
full Dirac and Pauli form factors calculated consistently within our scheme applying the
same power counting. The anomalous pion $t$-channel exchange diagram is
also excluded from the definition of the polarizabilites.

The amplitudes $A_i$ can be expressed in terms of the nucleon polarizabilites by performing an expansion in $\omega$ around $\omega=0$:
\begin{align} 
A_1(\omega) &= \frac{4\pi E_N}{m}\bigg[\alpha_{\text{E1}}\omega^2 + \frac{\omega^4}{12}(2z\alpha_{\text{E2}}-\beta_{\text{M2}}+12\alpha_{\text{E1}\nu}) 
+\frac{\omega^6}{2700}\left((30z^2-2)\alpha_{\text{E3}}-20z\beta_{\text{M3}}\right. \nonumber \\ 
&\hspace{1cm} + \left. 450z\alpha_{\text{E2}\nu}-225\beta_{\text{M2}\nu}+2700\alpha_{\text{E1}\nu^2}\right) + \mathcal{O}(\omega^8) \bigg]\,, \nonumber  \\
A_2(\omega) &= \frac{4\pi E_N}{m}\bigg[\beta_{\text{M1}}\omega^2 + \frac{\omega^4}{12}(2z\beta_{\text{M2}}-\alpha_{\text{E2}}+12\beta_{\text{M1}\nu}) 
+\frac{\omega^6}{2700}\left((30z^2-2)\beta_{\text{M3}}-20z\alpha_{\text{E3}}\right. \nonumber \\ 
&\hspace{1cm} + \left. 450z\beta_{\text{M2}\nu}-225\alpha_{\text{E2}\nu}+2700\beta_{\text{M1}\nu^2}\right) + \mathcal{O}(\omega^8) \bigg]\,, \nonumber  \\
A_5(\omega) &= \frac{4\pi E_N}{m}\bigg[(\gamma_{\text{E1E1}}-\gamma_{\text{E1M2}})\omega^3+\frac{\omega^5}{5}\left(20z\gamma_{\text{E2E2}}+5\gamma_{\text{E1E1}\nu}-10z\gamma_{\text{E2M3}}\right. \nonumber \\ 
&\hspace{1cm} - \left. 5\gamma_{\text{E1M2}\nu}+2\gamma_{\text{M2E3}}+5\gamma_{\text{M2M2}}\right) + \mathcal{O}(\omega^7) \bigg]\,, \nonumber \\
A_6(\omega) &= \frac{4\pi E_N}{m}\bigg[(\gamma_{\text{M1M1}}-\gamma_{\text{M1E2}})\omega^3+\frac{\omega^5}{5}\left(20z\gamma_{\text{M2M2}}+5\gamma_{\text{M1M1}\nu}-10z\gamma_{\text{M2E3}}\right. \nonumber \\ 
&\hspace{1cm} - \left. 5\gamma_{\text{M1E2}\nu}+2\gamma_{\text{E2M3}}+5\gamma_{\text{E2E2}}\right) + \mathcal{O}(\omega^7) \bigg]\,, \nonumber \\
A_7(\omega) &= \frac{4\pi E_N}{m}\bigg[\gamma_{\text{E1M2}}\omega^3 + \frac{\omega^5}{5}\left(15z\gamma_{\text{E2M3}}+5\gamma_{\text{E1M2}\nu}-7\gamma_{\text{M2E3}}-
\!10\gamma_{\text{M2M2}}\right) + \mathcal{O}(\omega^7) \bigg]\,, \nonumber \\
A_8(\omega) &= \frac{4\pi E_N}{m}\bigg[\gamma_{\text{M1E2}}\omega^3 + \frac{\omega^5}{5}\left(15z\gamma_{\text{M2E3}}
+5\gamma_{\text{M1E2}\nu}-7\gamma_{\text{E2M3}}-10\gamma_{\text{E2E2}}\right) + \mathcal{O}(\omega^7) \bigg]\,,
\label{eq:Ai}
\end{align}
where $E_N$ is the nucleon energy.
We also introduce the linear combinations corresponding to the 
forward and backward spin polarizabilites $\gamma_0$ and $\gamma_\pi$
\begin{align}
\gamma_0 &= -\gamma_{\text{E1E1}}-\gamma_{\text{M1M1}}-\gamma_{\text{E1M2}}-\gamma_{\text{M1E2}} \,,\nonumber \\
\gamma_\pi &= -\gamma_{\text{E1E1}}+\gamma_{\text{M1M1}}-\gamma_{\text{E1M2}}+\gamma_{\text{M1E2}}\,,\label{eq:gamma0_gammapi} 
\end{align}
the higher-order forward spin polarizabilty
\begin{align}
\bar{\gamma}_0 &= -\gamma_{\text{E1E1}\nu}-\gamma_{\text{M1M1}\nu}-\gamma_{\text{M1E2}\nu}-\gamma_{\text{E1M2}\nu} \nonumber \\
&\phantom{=} -\gamma_{\text{E2E2}}-\gamma_{\text{M2M2}}-\tfrac{8}{5}(\gamma_{\text{E2M3}}+\gamma_{\text{M2E3}})  \,, \label{eq:gamma0bar} 
\end{align}
as well as the longitudinal-transverse spin polarizability
\begin{align}
\delta_{\text{LT}} &= -\frac{1}{6}\frac{d^3}{d\omega^3}
\bigg\{\frac{m}{4\pi E_N}\Big[A_5(\omega)+A_{11}(\omega)+A_{12}(\omega)\Big]\bigg\}\bigg\arrowvert_{\omega=0} \,. \label{eq:delta-LT}
\end{align}

There are similar but different amplitude decompositions used in the literature, 
which leads to different relations of those amplitudes to the nucleon polarizabilites.
For ease of comparison, we provide the transformation matrix from the 
vector of amplitudes $A_\text{this work}=(A_1, A_2, A_5, A_6, A_7, A_8 )$ defined in Eq.~(\ref{eq:final-decomposition})
to the vector of amplitudes $A_\text{LMP}=(A_1, A_2, A_3, A_4, A_5, A_6 )$ considered in \cite{Lensky:2015awa}
\begin{align}
A_\text{LMP}=L A_\text{this work}\,,\
L= \left(\begin{array}{cccccc}
1 & z & 0 & 0 & 0 & 0 \\
0 & -1 & 0 & 0 & 0 & 0 \\
0 & 0 & 1 & z & 2 & 2z \\
0 & 0 & 0 & 1 & 0 & 0 \\
0 & 0 & 0 & -1 & 0 & -1 \\
0 & 0 & 0 & 0 & -1 & 0
\end{array}\right) \,. \label{eq:transformation}
\end{align}

In this work, we also analyze the so-called dynamical polarizabilites defined in terms of the center-of-mass multipoles
as follows (see, e.g., \cite{Griesshammer:2001uw,Hildebrandt:2003fm,Lensky:2015awa,Guiasu:1978dz}):
\begin{align}
&\alpha_{\text{E}l}(\omega) = l^2(2l-1)!!\frac{(l+1)f_{EE}^{l+}+lf_{EE}^{l-}}{\omega^{2l}}  \,,\
\beta_{\text{M}l}(\omega) = l^2(2l-1)!!\frac{(l+1)f_{MM}^{l+}+lf_{MM}^{l-}}{\omega^{2l}} \,, \nonumber\\
&\gamma_{\text{E}l\text{E}l}(\omega) = \frac{f_{EE}^{l+}-f_{EE}^{l-}}{\omega^{2l+1}} \,,  \
\gamma_{\text{E}l\text{M}l\pm 1}(\omega) = (2l\pm 1)!\frac{f_{EM}^{l\pm}}{\omega^{2l\pm 1}} \,,\nonumber\\
&\gamma_{\text{M}l\text{M}l}(\omega) = \frac{f_{MM}^{l+}-f_{MM}^{l-}}{\omega^{2l+1}} \,, \
\gamma_{\text{M}l\text{E}l\pm 1}(\omega) = (2l\pm 1)!\frac{f_{ME}^{l\pm}}{\omega^{2l\pm 1}}\,,\label{eq:definition_dynamical}
\end{align}
for $l=1,2$.
Note that in contrast to the equations above, in Eq.~(\ref{eq:definition_dynamical}), $\omega$ denotes the center-of-mass photon energy.

\section{Results}\label{sec:numerical-results}
We are now in the position to present our numerical results for various proton and neutron polarizabilities
calculated up to order $\mathcal{O}(\epsilon^3 + q^4)$.
Specifically, we consider the following polarizabilities: 
spin-independent (scalar) dipole, quadrupole, octupole, dispersive dipole and quadrupole polarizabilities
as well as dipole, quadrupole and dispersive dipole spin polarizabilities.
We also discuss selected generalized (i.e.~$Q^2$-dependent) 
and dynamical (i.e.~energy-dependent) polarizabilities.

As already mentioned above, most of the results we present are pure
predictions and contain no free parameters. 
The only exceptions are the spin-independent dipole polarizabilities
$\alpha_{\text{E1}}$ and $\beta_{\text{M1}}$ at order $\mathcal{O}(q^4)$ or $\mathcal{O}(\epsilon^3 + q^4)$, which are fitted
to the experimental values. All remaining parameters are taken from other processes and are collected in 
Tables~\ref{tab:constants}, \ref{tab:c_i} and \ref{tab:e_i}.
\begin{table}[ht] \centering 
  \begin{ruledtabular} 
  \begin{tabular*}{\textwidth}{@{\extracolsep{\fill}}cccccccccc}
    $\alpha_{EM}^{-1}$ & $M $ [MeV]  & $F_\pi$ [MeV] & $m$ [MeV] & $m_\Delta$ [MeV] &
                                                                     $g_A$
    &
     $c_6$ & $c_7$ & $h_A$  & $b_1 \; [m^{-1}]$ \\
\hline 
$137.036$ &  $138.04$ & $92.21$ & $938.9$ & $1232$ & $1.27$ 
& $3.706$ & $-1.913$  & $1.43$ & $-4.98$\\
  \end{tabular*}
 \end{ruledtabular} 
\caption{Parameters used in the current work. The values of
$\alpha_{EM}$, $M $, $m$, $m_\Delta$, $g_A$, $F_\pi$ are taken from
\cite{Tanabashi:2018oca}. The LECs  
$c_6$ and $c_7$ are related to the proton and
neutron magnetic moment and $d_6$ and $d_6$ with the
proton and neutron charge radii \cite{Fuchs:2003ir}.
The values of the LECs $b_1$ and $h_A$ are extracted
from the electromagnetic and strong width of the $\Delta$-resonance,
respectively, see  \cite{Bernard:2012hb} for details and explicit
expressions. For the static polarizabilities we use the real $\Delta$ mass as 
given in the Table, whereas for the generalized polarizabilities we use the 
pole mass $m_\Delta = (1210  -  50i)\text{ MeV}$.}
\label{tab:constants}
\end{table}

\begin{table}[ht] \centering 
\renewcommand{\arraystretch}{0.75} 
  \begin{ruledtabular}
  \begin{tabular*}{\textwidth}{@{\extracolsep{\fill}}lllll}    
&$\ph q^3$ & $\ph q^4$ &$\ph \epsilon^3$& $\ph q^4+\epsilon^3$\\\hline 
$c_1 \; [m^{-1}]$ &$-0.94\pm 0.02$&$-1.05 \pm 0.03$ &$-1.05 \pm 0.03$ &$-1.05 \pm 0.03$\\
$c_2 \; [m^{-1}]$ &$\ph 2.39\pm 0.03$ &$\ph 3.15 \pm 0.03$ &$\ph 0.96 \pm 0.11$ &$\ph 0.96 \pm 0.11$ \\
$c_3 \; [m^{-1}]$ &$-4.60\pm 0.05$ &$-5.35 \pm 0.06$ &$-2.13\pm 0.19$&$-2.13\pm 0.19$\\
  \end{tabular*}
  \end{ruledtabular}  
\caption{Numerical values of the low energy constants used in the
  current work as determined by matching the solution of Roy-Steiner
  equations for $\pi N$ scattering \cite{Ditsche:2012fv} to chiral perturbation
  theory in \cite{Siemens:2016jwj}. The values for $q^4+\epsilon^3$ correspond
to the $\epsilon^3$ calculation of
\cite{Siemens:2016jwj}. }
\label{tab:c_i}
\end{table}

\begin{table}[ht] \centering 
\renewcommand{\arraystretch}{0.75} 
  \begin{ruledtabular}
 \begin{tabular*}{\textwidth}{@{\extracolsep{\fill}}rrrrr}      
 &$\ph q^3$ & $\ph q^4$ &$\ph \epsilon^3$& $\ph q^4+\epsilon^3$\\
\midrule[0.2pt]
$d_6 \; [m^{-2}]$ &$-0.61$ \cite{Fuchs:2003ir} &$-0.61$ \cite{Fuchs:2003ir} &$-0.80 \pm 0.04$ &$-1.27 \pm 0.05$\\
$d_7 \; [m^{-2}]$&$-0.43$ \cite{Fuchs:2003ir} &$-0.44$ \cite{Fuchs:2003ir} &$-0.44 \pm 0.01$ &$-0.46 \pm 0.01$ \\
$e_{91} \; [m^{-3}]$&$\ldots$  &$-0.04 \pm 0.22$ &$\ldots$ &$-0.46\pm 0.23$ \\
$e_{92} \; [m^{-3}]$&$\ldots$ &$-0.29 \pm 0.79$&$\ldots$ &$-0.22 \pm 0.80$\\
$(2e_{89}+e_{93}+e_{118}) \; [m^{-3}]$&$\ldots$ &$-0.07 \pm 0.23$&$\ldots$  &$-2.53 \pm 0.50$\\
$(2e_{90}+e_{94}+e_{117}) \; [m^{-3}]$&$\ldots$ &$-1.76 \pm 0.80$&$\ldots$  &$\ph 2.02 \pm 1.20$\\
 \end{tabular*}
  \end{ruledtabular} 
\caption{Numerical values of the low-energy constants obtained from the fit to the 
empirical values of the electric radius of the proton and neutron ($d_i$) and the proton and neutron spin-independent
polarizabilites ($e_i$). Note that the $e_i$ enter at fourth order and therefore do not have values for $q^3$ and $\epsilon^3$.}
\label{tab:e_i}
\end{table}

In the course of the calculation we have used 
our own code written in $Mathematica$ \cite{mathematica12.0} 
and FORM \cite{Kuipers:2012rf} for the analytical calculation of Feynman diagrams.
The numerical evaluation of loop integrals have been performed with help of
the \emph{Mathematica}  package \emph{Package}-{\textbf X} \cite{Patel:2016fam}. 
We have also used our own Fortran code for estimating the theoretical errors.

For our complete results at order $\mathcal{O}(\epsilon^3 + q^4)$, we also provide estimations of
the theoretical errors originating from two sources, namely the uncertainties
in the input parameters and the errors caused by the truncation of the small-scale expansion.
For the latter uncertainty, we adopt the Bayesian model used in \cite{Epelbaum:2019zqc,Epelbaum:2019kcf}
based on the ideas developed in
\cite{Furnstahl:2015rha,Melendez:2017phj,Melendez:2019izc}, see
\cite{Rijneveen:2020qbc} for a recent application to radiative pion
photoproduction. 
The observables are assumed to be expanded in parameter $Q$ given by
\begin{align}
Q = {\rm max} \bigg( \frac{M ^{\rm eff}}{\Lambda_{\rm b}}, \; \frac{\sqrt{Q^2}}{\Lambda_{\rm b}}, \; \frac{\omega}{\Lambda_{\rm b}} \bigg)\,,
\label{Eq:expansion_parameter}
\end{align}
where $Q^2$ on the right-hand side is the virtuality of the photon, and $\omega$ is the photon energy
in the case of dynamical polarizabilities.
The soft and hard scales are chosen to be $M^{\rm eff}= 200$~MeV
and $\Lambda_{\rm b}=700$~MeV in accordance with \cite{Epelbaum:2019wvf}. 
Following \cite{Epelbaum:2019zqc,Epelbaum:2019kcf}, we utilize the Gaussian prior distribution for the expansion coefficients $c_i$:
\begin{align}
 {\rm pr} (c_i | \bar c ) = \frac{1}{\sqrt{2 \pi} \bar c} \, e^{-
   c_i^2/(2 \bar c^2 )}, \quad \quad {\rm pr} ( \bar c ) = \frac{1}{\ln ( \bar c_> / \bar c_< )} \, 
\frac{1}{\bar c} \, \theta (\bar c - \bar c_< ) \, \theta (\bar c_> -
\bar c )\,,
  \label{Eq:prior}
\end{align}  
with the cut offs $\bar c_<=0.5$ and $\bar c_>=10$. Further details on
the employed Bayesian model can be found in \cite{Epelbaum:2019zqc,Epelbaum:2019kcf}. 

In the following sections, we provide a
detailed comparison of our results with the available
experimental/empirical  data as well as with other theoretical approaches 
based on chiral perturbation theory and on fixed-$t$ dispersion relations. 
We also discuss generalized polarizabilities,
investigate the convergence pattern of the $1/m$-expansion for the calculated polarizabilities
and compare the results of covariant $\chi$PT with the heavy baryon approach.
Last but not least, we emphasize that the resulting large absolute numerical
values of the octupole polarizabilites are merely due to the numerical
factors 
in their definition, which makes them consistent with the definition
of the polarizabilites for composite systems.

\subsection{Scalar dipole polarizabilites}\label{sub:scalar_dipole_polarizabilities}
We start by considering the spin-independent dipole nucleon polarizabilites $\alpha_{\text{E1}}$ and
$\beta_{\text{M1}}$. The results of the calculations at order $\mathcal{O}(q^4)$ and $\mathcal{O}(\epsilon^3 + q^4)$ 
as well as the individual contributions from orders  $\mathcal{O}(q^3)$, $\mathcal{O}(q^4)$ pion-nucleon loops,
$\mathcal{O}(\epsilon^3)$ $\pi\Delta$-loops and tree-level $\Delta$-pole graphs are presented
in Table~\ref{tab:spin-independent-dipole}.
\begin{table}[ht] \centering
\renewcommand{\arraystretch}{0.75} 
  \begin{ruledtabular}
 \begin{tabular*}{\textwidth}{@{\extracolsep{\fill}}lllll}    
& \multicolumn{2}{c}{Proton} & \multicolumn{2}{c}{Neutron} \\
& $\ph \po \alpha_{\text{E1}}$ & $\ph \po \beta_{\text{M1}}$ & $\ph \po \alpha_{\text{E1}}$& $\ph \po \beta_{\text{M1}}$\\
\midrule[0.2pt]
$q^3$ (without $\Delta$)&$\ph\po7.04$&$\po {-1.85}$ &$\ph \po 9.51$ & $\po {-1.10}$\\
$q^4$ (without $\Delta$)&$\ph\po 4.16$&$\ph\po 4.35$ &$\ph\po 2.09$ & $\ph\po 4.80$  \\
Total (without $\Delta$)&$\ph 11.20$ &$\ph\po 2.50$ &$\ph 11.60$ & $\ph \po 3.70$  \\
\midrule[0.2pt]
$q^3$ &$\ph\po7.04$&$\po {-1.85}$ &$\ph \po 9.51$ & $\po {-1.10}$\\
$\epsilon^3$ $\pi\Delta$ loop &$\po{-1.45}$ &$\ph \po 5.54$ &$\ph \po 2.78$ & $\ph \po 0.96$ \\
$\epsilon^3$ $\Delta$ tree&$\po{-3.78}$ &$\ph 11.96$ &$\po {-3.78}$& $\ph 11.96$ \\
$q^4$ &$\ph\po 9.40$&$-13.16$ &$\ph\po 3.10$ & $\po {-8.12}$  \\
\textbf{Total} &\boldmath$\po 11.20$&\boldmath$\po\po 2.50$&\boldmath$\po 11.60$ &\boldmath $\po\po 3.70$  \\
\midrule[0.2pt]
\midrule[0.2pt]
$\mathcal{O}(p^3)$ $\pi$N loops \cite{Lensky:2015awa}&$\ph \po 6.8$&$\po {-1.8}$&$\ph \po 9.4$&$\po {-1.1}$\\
$\mathcal{O}(p^{7/2})$ $\pi\Delta$ loops \cite{Lensky:2015awa}&$\ph \po 4.4$&$\po {-1.4}$&$\ph \po 4.4$&$\po {-1.4}$\\
$\Delta$ pole \cite{Lensky:2015awa}&$\po {-0.1}$&$\ph \po 7.1$ &$\po {-0.1}$&$\ph \po 7.1$ \\
Total \cite{Lensky:2015awa}&$\ph 11.2\pm 0.7$&$\ph \po 3.9 \pm 0.7$&$\ph 13.7\pm 3.1$&$\ph \po 4.6\pm 2.7$\\
\midrule[0.2pt]
Fixed-$t$ DR \cite{Babusci:1998ww, OlmosdeLeon:2001zn}&$\ph 12.1$ &$\ph\po 1.6$ &$\ph 12.5$ &$\ph\po 2.7$ \\  
HB$\chi$PT fit \cite{McGovern:2012ew}&$\ph 10.65\pm 0.50$ &$\ph\po 3.15\pm 0.50$ &$\ph 11.55\pm 1.5$ &$\ph\po  3.65\pm 1.5$ \\
B$\chi$PT fit \cite{Lensky:2014efa}&$\ph 10.6\pm 0.5$ &$\ph\po 3.2\pm 0.5$ &$\ph \ldots$ &$\ph\po  \ldots$ \\
PDG \cite{Tanabashi:2018oca}&$\ph 11.2\pm 0.4$ &$\ph\po 2.5\pm 0.4$ &$\ph 11.6\pm 1.5$ &$\ph\po  3.7\pm 2.0$ \\
 \end{tabular*}
 \end{ruledtabular} 
\caption{Numerical values for the spin-independent dipole polarizabilities of the proton and the neutron in $10^{-4}\text{fm}^3$. 
The values are compared with the results calculated 
within the $\delta$-counting scheme and obtained using fixed-$t$ dispersion relations.}
\label{tab:spin-independent-dipole}
\end{table}
At order $\mathcal{O}(q^4)$, there appear low energy constants in the effective Lagrangian that contribute to
the nucleon Compton scattering. We adjust four relevant linear
combinations of them ($e_{91}$, $e_{92}$, $2e_{89}+e_{93}+e_{118}$ and $2e_{90}+e_{94}+e_{117}$)
in such a way as to reproduce  
the empirical values of the proton and neutron spin-independent dipole polarizabilities, see Table~\ref{tab:e_i}.

In the case of the $\Delta$-less theory, the contribution at order $\mathcal{O}(q^4)$ 
for the electric polarizability $\alpha_{\text{E1}}$ of the proton (neutron) is
about two (five) times smaller than the one at order $\mathcal{O}(q^3)$,
which is an indication of a reasonable convergence of the chiral expansion.
For the magnetic polarizabilities $\beta_{\text{M1}}$,
due to some cancellations among $\mathcal{O}(q^3)$ loops,
the contributions at order $\mathcal{O}(q^4)$ are larger than the ones at order $\mathcal{O}(q^3)$
but are, nevertheless, comparable with those for the $\alpha_{\text{E1}}$.

In the $\Delta$-full scheme, the $\mathcal{O}(q^4)$ terms (that differ from the ones in the $\Delta$-less
case by the values of $c_i$'s and $e_i$'s) are significantly larger.
This feature can be traced back to the sizable $\mathcal{O}(\epsilon^3)$ contributions,
especially from the $\Delta$-pole tree-level diagrams, that need to be
compensated by adjusting
the relevant contact terms. Such contributions appear to be demoted 
to higher orders in the $\Delta$-less scheme.   
Their importance for other polarizabilites will, however, be
demonstrated below.
Thus, a seemingly better convergence of the $\Delta$-less approach for the dipole 
polarizabilites can be argued to be accidental.
Notice further that the convergence issues are not really relevant for the dipole spin-independent polarizabilites
at the order we are working due to the presence of the corresponding
compensating contact terms in the Lagrangian. 

In Table~\ref{tab:spin-independent-dipole},  we also provide for comparison the
values for the dipole spin-independent polarizabilites  
obtained by analyzing experimental data using fixed-$t$ dispersion
relations \cite{Babusci:1998ww, OlmosdeLeon:2001zn}, 
and by fitting experimental data employing various versions of the 
$\delta$-counting schemes (with the loop diagrams calculated utilizing the covariant \cite{Lensky:2014efa}
or heavy-baryon approach \cite{McGovern:2012ew}).
It is particularly instructive to compare our results with \cite{Lensky:2015awa},
where the individual contributions calculated within the $\delta$-counting scheme are presented.
Such a comparison allows one to analyze the importance of
the explicit $\Delta$ degrees of freedom and the sensitivity
of the results to employed counting schemes for the $\Delta$-nucleon
mass difference. 
There are two main sources of differences between our approach and the
one used in \cite{Lensky:2015awa}
(apart from slightly different numerical values of the coupling constants).
First, different terms in the effective Lagrangian corresponding to the $\gamma N \Delta$ vertex 
are used. The $\gamma N \Delta$ Lagrangian of \cite{Lensky:2015awa} contains 
two terms with the so-called magnetic and electric $\gamma N \Delta$-couplings $g_M$ and $g_E$:
\begin{align} 
\mathcal{L}_{\gamma N\Delta} = 
\frac{3e}{2m(m_\Delta+m)}\bar{N}T_3^\dagger(i g_M \tilde{F}^{\mu\nu}-g_E\gamma_5F^{\mu\nu})\partial_\mu\Delta_\nu + h.c.\,,
\end{align}
which in our scheme correspond to the $b_1$- and $h_1$-terms (the contribution from the
$h_1$-term is of a higher order in our power counting and does not appear in the current calculations).
The two prescriptions are identical when both the nucleon and the Delta are on the mass shell.
Otherwise, the difference is compensated by local contact terms of a
higher order in the $1/m$-expansion,
see \cite{Tang:1996sq, Pascalutsa:2000kd, Krebs:2009bf,
  Krebs:2008zb} for a related discussion.
Such off-shell effects manifest themselves, e.g., in the tree-level $\Delta$-contribution to the magnetic
polarizability $\beta_{\text{M1}}$. Although the residue of the $\Delta$ pole in the magnetic channel is the
same in both schemes (the constants $b_1$ and $g_M$ are roughly in agreement with each other when calculating
the magnetic $\gamma N \Delta$ transition form factor), the full result differs almost by a factor of two due
to the presence of the non-pole (background) terms. 
The non-vanishing (and sizable) contribution of the $\Delta$-tree-level
diagrams to the electric polarizabilites  $\alpha_{\text{E1}}$  
is in our scheme a pure $1/m$-effect caused by the induced electric
$\gamma N \Delta$ coupling stemming from the particular form 
of the effective Lagrangian. On the other hand, the $\Delta$ tree-level contribution to $\alpha_{\text{E1}}$ 
is negligible in the $\delta$-counting scheme because of the smallness
of the electric $\gamma N \Delta$ coupling $g_E$.
Note that terms proportional to $g_E^2$ ($h_1^2$) start to contribute
only at order $\epsilon^5$ in the small-scale-expansion scheme. 
The observed dependence of the considered polarizabilites on the off-shell effects
might be an indication of the importance of such higher order contributions.
Fortunately, such $1/m$ effects are strongly suppressed for
higher-order polarizabilites as will be shown below.

The second difference between the two schemes is related to 
power-counting of various diagrams with
internal $\Delta$-lines.
While the $\pi N$ loops in \cite{Lensky:2015awa} at order
$\mathcal{O}(p^3)$ are identical with the ones included in our $\mathcal{O}(q^3)$ results, 
the diagrams with two and three $\Delta$-lines inside the loop
are suppressed in the $\delta$-counting and are not included in their leading-order $\pi \Delta$-loop amplitude.
On the other hand, such diagrams are required by gauge invariance
(notice, however, that in the Coulomb gauge, their contribution 
is suppressed by a factor $1/m$). 
In any case, we observe a significant difference between the size  
of the $\epsilon^3$ $\pi\Delta$-loop contributions in our scheme and
the $\mathcal{O}(p^{7/2})$ ones
of \cite{Lensky:2015awa} involving only the $\pi\Delta$-loops with a single $\Delta$-line.

\subsection{Dipole spin polarizabilites}\label{sub:dipole_spin_polarizabilities}
Next, we consider the dipole spin polarizabilites $\ph \gamma_{\text{E1E1}}$, 
$\ph \gamma_{\text{M1M1}}$, $\ph \gamma_{\text{E1M2}}$ and $\ph \gamma_{\text{M1E2}}$.
These quantities are less sensitive to the short range dynamics as the relevant contact 
terms appear at order $\mathcal{O}(q^5)$.
Therefore, one expects a better convergence pattern for them.
At the order we are working, the spin polarizabilites are predictions and do not
depend on any free parameters.
The numerical values of the spin polarizabilites for the proton and neutron are collected in Table~\ref{tab:spin-proton}.
\begin{table}[ht] \centering
\renewcommand{\arraystretch}{0.75} 
 \begin{ruledtabular}
 \begin{tabular*}{\textwidth}{@{\extracolsep{\fill}}lllll}      
& $\ph \gamma_{\text{E1E1}}^{(p)}$ & $\ph \gamma_{\text{M1M1}}^{(p)}$ & $\ph \gamma_{\text{E1M2}}^{(p)}$& $\ph \gamma_{\text{M1E2}}^{(p)}$ \\
\midrule[0.2pt]
$q^3$  (without $\Delta$)& $-3.46$& $-0.13$& $\ph 0.57$ & $\ph 0.95$\\
$q^4$  (without $\Delta$)& $-0.01$& $\ph 0.49$ & $-0.25$ & $\ph 0.56$ \\
Total  (without $\Delta$)& $-3.47$& $\ph 0.36$& $\ph 0.32$& $\ph 1.51$ \\
\midrule[0.2pt]
$q^3$ & $-3.46$& $-0.13$& $\ph 0.57$ & $\ph 0.95$\\
$\epsilon^3$ $\pi\Delta$ loops & $-0.11$& $\ph 0.58$& $\ph 0.48$& $-0.79$\\
$\epsilon^3$ $\Delta$ tree & $-1.07$& $\ph 3.85$& $-0.88$ & $\ph 1.74$ \\
$q^4$ & $- 0.01$& $\ph 0.49$ & $-0.25$ & $\ph 0.56$ \\
\textbf{Total} &\!\!\! \boldmath $-4.65 \stack{0.12}{0.44}$&\; \boldmath $ 4.80\stack{0.43}{0.44}$&\!\!\! \boldmath $-0.08 \stack{0.11}{0.08}$&\; \boldmath $2.47\stack{0.21}{0.26}$ \\
\midrule[0.2pt]
\midrule[0.2pt]
$\mathcal{O}(p^3)$ $\pi$N loops \cite{Lensky:2015awa}& $-3.4$& $-0.1$& $\ph 0.5$& $\ph 0.9$\\
$\mathcal{O}(p^{7/2})$ $\pi\Delta$ loops \cite{Lensky:2015awa}& $\ph 0.4$& $-0.2$& $\ph 0.1$& $-0.2$\\
$\Delta$ pole \cite{Lensky:2015awa}& $-0.4$& $\ph 3.3$ & $-0.4$& $\ph 0.4$ \\
Total \cite{Lensky:2015awa}& $-3.3\pm 0.8$& $\ph 2.9\pm 1.5$& $\ph 0.2\pm 0.2$& $\ph 1.1\pm 0.3$\\
\midrule[0.2pt]
Fixed-$t$ DR \cite{Babusci:1998ww}& $-3.4$ & $\ph 2.7$ & $\ph 0.3$ & $\ph 1.9$\\  
Fixed-$t$ DR \cite{Holstein:1999uu, Hildebrandt:2003fm, Pasquini:2007hf}& $-4.3$ & $\ph 2.9$ & $-0.02$ & $\ph 2.2$\\
HB$\chi$PT fit \cite{McGovern:2012ew, Griesshammer:2015ahu}& $-1.1\pm 1.9$ & $\ph 2.2 \pm 0.8$ & $-0.4 \pm 0.6$ & $\ph 1.9 \pm 0.5$ \\
MAMI 2015 \cite{Martel:2014pba}& $-3.5\pm 1.2$ & $\ph 3.16 \pm 0.85$ & $-0.7\pm 1.2$ & $\ph 1.99\pm 0.29$ \\
MAMI 2018 \cite{Paudyal:2019mee} & $-3.18 \pm 0.52$ & $\ph 2.98 \pm 0.43$ & $-0.44\pm 0.67$ & $\ph 1.58\pm 0.43$  \\
\\
& $\ph \gamma_{\text{E1E1}}^{(n)}$ & $\ph \gamma_{\text{M1M1}}^{(n)}$ & $\ph \gamma_{\text{E1M2}}^{(n)}$& $\ph \gamma_{\text{M1E2}}^{(n)}$ \\
\midrule[0.2pt]
$q^3$  (without $\Delta$)& $-4.86$& $-0.17$& $\ph 0.61$ & $\ph 1.36$\\
$q^4$  (without $\Delta$)& $-0.46$& $\ph 1.42$ & $-0.59$ & $\ph 0.76$ \\
Total  (without $\Delta$)& $-5.32$& $\ph 1.25$& $-0.02$& $\ph 2.12$ \\
\midrule[0.2pt]
$q^3$ & $-4.86$& $-0.17$& $\ph 0.61$ & $\ph 1.36$\\
$\epsilon^3$ $\pi\Delta$ loops & $\ph 0.22$& $\ph 0.12$& $\ph 0.11$& $-0.27$\\
$\epsilon^3$ $\Delta$ tree & $-1.07$& $\ph 3.85$& $-0.88$ & $\ph 1.74$ \\
$q^4$ & $- 0.46$& $\ph 1.42$ & $-0.59$ & $\ph 0.76$ \\
\textbf{Total} &\!\!\! \boldmath $-6.17\stack{0.12}{0.56}$&\; \boldmath $ 5.22\stack{0.42}{0.59}$&\!\!\! \boldmath $-0.75 \stack{0.10}{0.20}$&\;\, \boldmath $3.59 \stack{0.20}{0.37}$ \\
\midrule[0.2pt]
\midrule[0.2pt]
$\mathcal{O}(p^3)$ $\pi$N loops \cite{Lensky:2015awa}& $-4.7$& $-0.2$& $\ph 0.6$& $\ph 1.3$\\
$\mathcal{O}(p^{7/2})$ $\pi\Delta$ loops \cite{Lensky:2015awa}& $\ph 0.4$& $-0.2$& $\ph 0.1$& $-0.2$\\
$\Delta$ pole \cite{Lensky:2015awa}& $-0.4$& $\ph 3.3$ & $-0.4$& $\ph 0.4$ \\
Total \cite{Lensky:2015awa}& $-4.7\pm 1.1$& $\ph 2.9\pm 1.5$& $\ph 0.2\pm 0.2$& $\ph 1.6\pm 0.4$\\
\midrule[0.2pt]
Fixed-$t$ DR \cite{Babusci:1998ww}& $-5.6$ & $\ph 3.8$ & $-0.7$ & $\ph 2.9$\\  
Fixed-$t$ DR \cite{Drechsel:2002ar, Hildebrandt:2003fm, Lensky:2015awa}& $-5.9$ & $\ph 3.8$ & $-0.9$ & $\ph 3.1$\\
HB$\chi$PT fit \cite{McGovern:2012ew, Griesshammer:2015ahu}& $-4.0\pm 1.9$ & $\ph 1.3 \pm 0.8$ & $-0.1 \pm 0.6$ & $\ph 2.4 \pm 0.5$ \\

 \end{tabular*}
 \end{ruledtabular}                                                                                                                
\caption{Numerical values for the dipole spin polarizabilities of the proton (upper table) and the neutron (lower table) in $10^{-4}\text{fm}^4$. 
The upper errors originate from the uncertainty 
in the input parameters, the lower errors come from the truncation of the 
small-scale expansion. 
The values are compared with the results calculated 
within the $\delta$-counting scheme and obtained using fixed-$t$ dispersion relations. }
\label{tab:spin-proton}
\end{table}

We also provide theoretical errors for our complete scheme at order
$\mathcal{O}(q^4+\epsilon^3)$. The upper error reflects the uncertainty 
in the input parameters, whereas the lower value is the Bayesian
estimate of the error coming from the truncation of the 
small-scale expansion. 

The experimental values in Table~\ref{tab:spin-proton}
are obtained from the dispersion-relation analysis of the 
double-polarized Compton scattering asymmetries $\Sigma_3$ and $\Sigma_{2x}$ \cite{Martel:2014pba},
and, in a newer experiment, also $\Sigma_{2z}$ \cite{Paudyal:2019mee}.

Our predictions for the proton spin polarizabilites at order
$\mathcal{O}(q^4+\epsilon^3)$ agree with the experimental values of  \cite{Martel:2014pba}
within the errors
with only a slight deviation for $\gamma_{\text{M1M1}}$. 
The deviation from the values extracted in the recent MAMI experiment \cite{Paudyal:2019mee} are somewhat larger.
Note that the $\Delta$-less approach fails to reproduce $\gamma_{\text{M1M1}}$ for the proton because of the missing
$\Delta$-pole contribution, which would appear as a contact term at order $\mathcal{O}(q^5)$.

The contributions of order $\mathcal{O}(q^4)$ are in all cases significantly smaller than the leading terms of order 
$\mathcal{O}(q^3+\epsilon^3)$ in the $\Delta$-full scheme (except for $\gamma_{\text{E1M2}}$ where the leading-order result is 
small due to cancellations between individual contributions), which is
an indication of a reasonable convergence of the small-scale expansion. 
The smallness of the $\mathcal{O}(q^4)$-terms can probably also be traced back
to the fact that the diagrams containing 
$c_1$, $c_2$ and $c_3$ vertices do not contribute to spin polarizabilites.
Our $\Delta$-full  results
also agree well with
the values obtained from the fixed-$t$ dispersion relations 
for the proton and the neutron, except for $\gamma_{\text{M1M1}}$,
where our prediction appears to be somewhat larger.

In Table~\ref{tab:spin-combined-proton}, we present the results for the 
forward and backward spin polarizabilites $\gamma_{0}$ and $\gamma_{\pi}$
which are the linear combinations of 
the four spin polarizabilites and can be more easily accessed experimentally.
For these quantities, the agreement with the experimental values is slightly worse, as can be seen
from Table ~\ref{tab:spin-combined-proton}.
\begin{table}[ht] \centering
\renewcommand{\arraystretch}{0.75} 
\begin{ruledtabular}
 \begin{tabular*}{\textwidth}{@{\extracolsep{\fill}}lllll}        
& $\ph\gamma_{0}^{(p)}$ & $\ph\gamma_{\pi}^{(p)}$ & $\ph\bar{\gamma}_{0}^{(p)}$& $\ph\delta_{\text{LT}}^{(p)}$ 
\\
\midrule[0.2pt]
$q^3$  (without $\Delta$)& $\ph 2.08$& $\ph \po 3.72$& $\ph 2.20$ & $\ph 1.54$\\
$q^4$  (without $\Delta$) & $-0.80$& $\ph\po 1.31$& $-0.37$& $\ph 0.58$\\
Total  (without $\Delta$) & $\ph 1.28$& $\ph \po 5.03$& $\ph 1.83$& $\ph 2.12$\\
\midrule[0.2pt]
$q^3$ & $\ph 2.08$& $\ph \po 3.72$& $\ph 2.20$ & $\ph 1.54$\\
$\epsilon^3$ $\pi\Delta$ loops & $-0.16$& $\po {-0.58}$& $- 0.01$& $\ph 1.21$\\
$\epsilon^3$ $\Delta$ tree & $-3.64$& $\ph\po 7.55$& $- 1.24$& $-0.36$\\
$q^4$ & $-0.80$& $\ph\po 1.31$& $-0.37$& $\ph 0.58$\\
\textbf{Total} & \!\boldmath $-2.53\stack{0.40}{0.31}$ & \!\boldmath $\ph 12.00\stack{0.83}{1.10}$& \!\boldmath $\ph 0.58\stack{0.13}{0.15}$& \!\boldmath $\ph 2.98\stack{0.08}{0.30}$\\
\midrule[0.2pt]
\midrule[0.2pt]
$\mathcal{O}(p^3)$ $\pi$N loops \cite{Lensky:2015awa}& $\ph 2.0$& $\po\ph 3.6$& $\ph 2.1$& $\ph -$\\
$\mathcal{O}(p^{7/2})$ $\pi\Delta$ loops \cite{Lensky:2015awa}& $-0.1$& $\po{-0.9}$& $-0.01$& $\ph -$\\
$\Delta$ pole \cite{Lensky:2015awa}& $-2.8$& $\po\ph 4.4$ & $-1.0$& $\ph -$ \\
Total \cite{Lensky:2015awa}& $-0.9\pm 1.4$& $\po\ph 7.2\pm 1.7$& $\ph 1.1\pm 0.5$&  $\ph -$\\
\midrule[0.2pt]
Fixed-$t$ DR \cite{Babusci:1998ww}& $-1.5$ & $\po\ph 7.8$ & $\ph -$ & $\ph -$\\  
Fixed-$t$ DR \cite{Holstein:1999uu, Hildebrandt:2003fm,
   Pasquini:2007hf}& $-0.8$ & $\po\ph 9.4$ & $\ph 0.6$ & $\ph -$\\
HB$\chi$PT fit \cite{McGovern:2012ew, Griesshammer:2015ahu}& $-2.6\pm  1.9$ & $\po\ph 5.6 \pm 1.9$
                                      & $\ph -$ & $\ph -$ \\
Experiment \cite{Ahrens:2001qt,Dutz:2003mm,Camen:2001st}& $-1.01\pm 0.13$& $\po\ph 8.0 \pm 1.8$& $\ph -$& $\ph -$\\
\midrule[0.2pt]
B$\chi$PT \cite{Bernard:2012hb}& $-1.74\pm 0.40$ & $\ph -$& $\ph -$& $\ph 2.40 \pm 0.01$ \\
\\
& $\ph\gamma_{0}^{(n)}$ & $\ph\gamma_{\pi}^{(n)}$ & $\ph\bar{\gamma}_{0}^{(n)}$& $\ph\delta_{\text{LT}}^{(n)}$ 
\\
\midrule[0.2pt]
$q^3$  (without $\Delta$) & $\ph 3.06$& $\ph \po 5.45$& $\ph 3.06$ & $\ph 2.41$\\
$q^4$  (without $\Delta$) & $-1.13$& $\ph\po 3.23$& $-0.46$& $\ph 0.50$\\
Total  (without $\Delta$) & $\ph 1.93$& $\ph \po 8.68$&$\ph 2.60$& $\ph 2.91$\\
\midrule[0.2pt]
$q^3$ & $\ph 3.06$& $\ph \po 5.45$& $\ph 3.06$ & $\ph 2.41$\\
$\epsilon^3$ $\pi\Delta$ loops & $-0.18$& $\po {-0.49}$& $- 0.01$& $\ph 0.33$\\
$\epsilon^3$ $\Delta$ tree & $-3.64$& $\ph\po 7.55$& $- 1.24$& $-0.36$\\
$q^4$ & $-1.13$& $\ph\po 3.23$& $-0.46$& $\ph 0.50$\\
\textbf{Total} & \!\boldmath $-1.89\stack{0.40}{0.38}$ & \!\boldmath $\ph 15.73\stack{0.83}{1.62}$& \!\boldmath $\ph 1.35\stack{0.13}{0.17}$& \!\boldmath $\ph 2.88\stack{0.06}{0.28}$\\
\midrule[0.2pt]
$\mathcal{O}(p^3)$ $\pi$N loops \cite{Lensky:2015awa}& $\ph 3.0$&
                                                                  $\ph\po
                                                                  5.3$&
                                                                        $\ph 2.9$& $\ph -$\\
$\mathcal{O}(p^{7/2})$ $\pi\Delta$ loops \cite{Lensky:2015awa}&
                                                                $-0.1$& $\po {-0.9}$& $-0.01$& $\ph -$\\
$\Delta$ pole \cite{Lensky:2015awa}& $-2.8$& $\po\ph 4.5$ & $-1.0$&
                                                                    $\ph
                                                                    - $ \\
Total \cite{Lensky:2015awa}& $\ph 0.03\pm 1.4$& $\po\ph 9.0\pm 2.0$&
                                                                     $\ph
                                                                     1.9\pm
                                                                     0.7$&
                                                                           $\ph - $\\
\midrule[0.2pt]
Fixed-$t$ DR \cite{Babusci:1998ww}& $-0.4$ & $\ph 13.0$ & $\ph -$ &
                                                                    $\ph -$\\  
Fixed-$t$ DR \cite{Lensky:2015awa, Hildebrandt:2003fm,
  Drechsel:2002ar}& $-0.1$ & $\ph 13.7$ & $\ph -$ & $\ph -$\\
HB$\chi$PT fit \cite{McGovern:2012ew, Griesshammer:2015ahu}& $\ph
                                                             0.5\pm
                                                             1.9$&
                                                                   $\po\ph
                                                                   7.6
                                                                   \pm
                                                                   1.9$
                                      & $\ph -$ & $\ph -$ \\
\midrule[0.2pt]
B$\chi$PT \cite{Bernard:2012hb}& $-0.77\pm 0.40$ & $\po\ph - $& $\ph -$& $\ph 2.38 \pm 0.03$ \\
 \end{tabular*}
 \end{ruledtabular}     
\caption{Numerical values for the combined polarizabilities $\gamma_{0}$, $\gamma_{\pi}$, 
$\bar{\gamma}_{0}$ and $\delta_{\text{LT}}$ of the proton (upper table) and the neutron (lower table). All values except for $\bar{\gamma}_0$ 
are given in $10^{-4}\text{fm}^4$ while $\bar{\gamma}_0$ is given in $10^{-4}\text{fm}^6$. 
The values are compared with various results either calculated within the $\delta$-counting scheme and obtained using fixed-$t$ dispersion relations.
The results of \cite{Bernard:2012hb} are equivalent with our
calculations without the $q^4$-contribution. For remaining notation see Table \ref{tab:spin-proton}.}
\label{tab:spin-combined-proton}
\end{table}

As in the case of scalar dipole polarizabilites, we compare our $\Delta$-tree-level and $\Delta$-loop
contributions with \cite{Lensky:2015awa} in order to analyze the differences of the two $\Delta$-full approaches
and the size of the unphysical off-shell terms.
For the spin polarizabilites, the off-shell effects (which we identify
with the difference of the  $\Delta$-tree-level terms in two 
schemes considered) are smaller but, nevertheless, comparable to
theoretical errors or even larger.
This might indicate that our theoretical errors are somewhat underestimated.
This should not come as a surprise because the Bayesian model for the error
estimation that we implement is not fully trustworthy as long as 
only two orders in the expansion in terms of the small parameter $Q$
are used as an input.
Notice further that we treat the order $q^4 + \epsilon^3$ results as being
the full fourth-order predictions when estimating truncation errors.   
The off-shell contributions add up constructively for the forward and backward spin polarizabilites 
(as can be seen in Table~\ref{tab:spin-combined-proton}),
which explains the worse agreement with experiment for these linear combinations.

The $\Delta$-loop terms are also different in the $\epsilon$- and $\delta$-counting schemes, which points
to the non-negligible contribution of the diagrams with multiple $\Delta$-lines. 
Note, however, that the overall absolute values of the $\epsilon^3$ $\Delta$-loops are, on average,
smaller than in the case of the scalar dipole polarizabilites and than the typical values of the
dipole spin polarizabilites.
Therefore, spin polarizabilites appear to be less sensitive to such details.
On the other hand, the suppression of the $\epsilon^3$ $\Delta$-loops
does not exclude the possibility that 
the $\epsilon^4$ $\Delta$-loops
(with order $\mathcal{O}(q^2)$ $\gamma N\Delta$ and $\gamma \Delta\Delta$ vertices),
which are not included in the current study,
yield important contributions, see also the discussion in subsection~\ref{sub:Q-polarizabilities}.

\subsection{Higher-order polarizabilities}\label{sub:Higher-order_polarizabilities}
In this subsection, we focus on higher-order nucleon polarizabilites including scalar quadrupole, dipole dispersive,
octupole and quadrupole dispersive, as well as spin quadrupole and dipole dispersive polarizabilites.
All relevant numerical values are collected in Tables~\ref{tab:spin-independent-2-proton}-\ref{tab:spin-quadrupole}
(we also provide the values for the higher-order forward spin polarizabilities 
$\bar \gamma_0$ in Table~\ref{tab:spin-combined-proton}).
Note that unnaturally large values of the scalar quadrupole and, especially, octupole polarizabilites are related
to the traditional $l$-dependent normalization factor in the
definition of these polarizabilites and have no physical meaning. 
\begin{table}[ht] \centering
\renewcommand{\arraystretch}{0.75} 
\begin{ruledtabular}
 \begin{tabular*}{\textwidth}{@{\extracolsep{\fill}}lllll}        
& $\ph \alpha_{\text{E2}}^{(p)}$& $\ph \beta_{\text{M2}}^{(p)}$ & $\ph\po \alpha_{\text{E1}\nu}^{(p)}$& $\ph\po \beta_{\text{M1}\nu}^{(p)}$ \\
\midrule[0.2pt]
$q^3$ (without $\Delta$)&$\ph 14.1$&$\po {-8.7}$&$\ph\po 0.8$&$\ph\po 1.9$\\
$q^4$ (without $\Delta$)&$\ph 16.1$&$- 15.9$&$\po {-4.1}$ &$\ph\po 4.2$\\
Total (without $\Delta$)&$\ph 30.2$ &$-24.6$ &$\po {-3.3}$ &$\ph\po 6.1 $\\
\midrule[0.2pt]
$q^3$ &$\ph 14.1$&$\po {-8.7}$&$\ph\po 0.8$&$\ph\po 1.9$\\
$\epsilon^3$ $\pi\Delta$ loops &$\ph\po 5.8$&$\po {-6.1}$&$\po {-0.9}$ &$\ph\po 1.1$\\
$\epsilon^3$ $\Delta$ tree &$\ph\po 1.3$&$\po {-4.7}$&$\po {-1.6}$ &$\ph\po 5.0$\\
$q^4$ &$\ph\po 8.3$&$\po {-7.6}$&$\po {-2.2}$ &$\ph\po 2.3$\\
\textbf{Total} &\;\;\boldmath $29.5\stack{0.9}{3.4}$ &\!\boldmath $-27.1\stack{1.3}{3.1} $ &\;\boldmath $ -3.9\stack{0.4}{0.8} $ &
\;\;\,\boldmath $10.2\stack{0.7}{1.0}$\\
\midrule[0.2pt]
\midrule[0.2pt]
$\mathcal{O}(p^3)$ $\pi$N loops \cite{Lensky:2015awa}&$\ph 13.5$&$\po {-8.4}$&$\ph\po 0.7$&$\ph\po 1.8$\\
$\mathcal{O}(p^{7/2})$ $\pi\Delta$ loops \cite{Lensky:2015awa}&$\ph\po 3.2$&$\po {-2.7}$&$\po {-0.6}$&$\ph\po 0.6$\\
$\Delta$ pole \cite{Lensky:2015awa}&$\ph\po 0.6$&$\po {-4.5}$ &$\po {-1.5}$ &$\ph\po 4.7$ \\
Total \cite{Lensky:2015awa}&$\ph 17.3\pm 3.9$&$-15.5\pm 3.5$&$\po {-1.3}\pm 1.0$&$\ph\po 7.1\pm 2.5$\\
\midrule[0.2pt]
Fixed-$t$ DR \cite{Babusci:1998ww, OlmosdeLeon:2001zn},&$\ph 27.5$ &$-22.4$&$\po {-3.8}$ &$\ph\po 9.1$\\  
Fixed-$t$ DR \cite{Hildebrandt:2003fm,Holstein:1999uu}&$\ph 27.7$ &$-24.4$&$\po {-3.9}$ &$\ph\po 9.3$\\
\\
& $\ph \alpha_{\text{E2}}^{(n)}$& $\ph \beta_{\text{M2}}^{(n)}$ & $\ph \po \alpha_{\text{E1}\nu}^{(n)}$& $\ph \po \beta_{\text{M1}\nu}^{(n)}$ \\
\midrule[0.2pt]
$q^3$ (without $\Delta$)&$\ph 12.9$&$\po {-9.0}$ &$\ph\po 2.2$&$\ph\po 1.9$\\
$q^4$ (without $\Delta$)&$\ph 16.0$ &$-15.6$ &$\po {-3.9}$&$\ph\po 3.9$\\
Total (without $\Delta$)&$\ph 29.0$ &$-24.6$ &$\po {-1.7}$ &$\ph\po 5.8 $ \\
\midrule[0.2pt]
$q^3$ &$\ph 12.9$&$\po {-9.0}$ &$\ph\po 2.2$&$\ph\po 1.9$\\
$\epsilon^3$ $\pi\Delta$ loops &$\ph\po 6.2$&$\po {-6.0}$ &$\po {-1.2}$&$\ph\po 1.3$\\
$\epsilon^3$ $\Delta$ tree &$\ph\po 1.3$&$\po {-4.7}$ &$\po {-1.6}$ &$\ph\po 5.0$\\
$q^4$ &$\ph\po 8.2$ &$\po {-7.3}$ &$\po {-2.0}$&$\ph\po 1.9$\\
\textbf{Total} &\;\;\boldmath $28.7\stack{0.9}{3.3}$ &\!\boldmath $-27.0\stack{1.3}{3.1}$ &\;\boldmath $-2.6\stack{0.4}{0.7}$ &\;\;\boldmath $10.2\stack{0.7}{1.0}$ \\
\midrule[0.2pt]
\midrule[0.2pt]
$\mathcal{O}(p^3)$ $\pi$N loops \cite{Lensky:2015awa}&$\ph 12.4$&$\po {-8.7}$&$\ph\po 2.1$&$\ph\po 1.8$\\
$\mathcal{O}(p^{7/2})$ $\pi\Delta$ loops \cite{Lensky:2015awa}&$\ph\po 3.2$&$\po {-2.7}$&$\po {-0.6}$&$\ph\po 0.6$\\
$\Delta$ pole \cite{Lensky:2015awa}&$\ph\po 0.6$&$\po {-4.5}$ &$\po {-1.5}$ &$\ph\po 4.7$ \\
Total \cite{Lensky:2015awa}&$\ph 16.2\pm 3.7$&$-15.8\pm 3.6$&$\ph\po 0.1\pm 1.0$&$\ph\po 7.2\pm 2.5$\\
\midrule[0.2pt]
Fixed-$t$ DR \cite{Babusci:1998ww}&$\ph 27.2$ &$-23.5$&$\po {-2.4}$ &$\ph\po 9.2$\\  
Fixed-$t$ DR \cite{Drechsel:2002ar,Hildebrandt:2003fm,Lensky:2015awa}&$\ph 27.9$ &$-24.3$&$\po {-2.8}$ &$\ph\po 9.3$\\
 \end{tabular*}
 \end{ruledtabular}       
\caption{Numerical values for the dispersive and the quadropole polarizabilities for the proton (upper table) and the neutron (lower table) in $10^{-4}\text{fm}^5$. 
The values are compared with the results calculated in 
$\delta$-counting $\chi$PT and obtained using fixed-$t$ dispersion
relation. For remaining notation see Table \ref{tab:spin-proton}.}
\label{tab:spin-independent-2-proton}
\end{table}

\begin{table}[ht] \centering
\renewcommand{\arraystretch}{0.75} 
\begin{ruledtabular}
 \begin{tabular*}{\textwidth}{@{\extracolsep{\fill}}lllllll}         
& $\ph\po\alpha_{\text{E3}}^{(p)}$ & $\ph\po\beta_{\text{M3}}^{(p)}$ & $\ph\alpha_{\text{E2}\nu}^{(p)}$&$\ph\beta_{\text{M2}\nu}^{(p)}$ &$\ph\alpha_{\text{E1}\nu^2}^{(p)}$& $\ph\beta_{\text{M1}\nu^2}^{(p)}$ \\
\midrule[0.2pt]
$q^3$ (without $\Delta$)&$\ph 134.7$ &$\po {-95.6}$ &$-22.5$ &$\ph 17.4$ & $\ph\po 6.7$ & $-3.4$\\
$q^4$ (without $\Delta$)&$\ph 123.4$ &$-118.6$ &$-21.2$ &$\ph 20.6$ & $\ph\po 3.0$ & $-2.9$\\
Total (without $\Delta$)&$\ph 258.1$ &$-214.1 $ &$-43.8$ &$\ph 38.0$ & $\ph\po 9.7$ & $-6.3 $ \\
\midrule[0.2pt]
$q^3$ &$\ph 134.7$ &$\po {-95.6}$ &$-22.5$ &$\ph 17.4$ & $\ph\po 6.7$ & $-3.4$\\
$\epsilon^3$ $\pi\Delta$ loops &$\ph\po 52.3$ &$\po {-48.7}$ &$\po {-9.1}$ &$\ph\po 8.7$ & $\ph\po 1.3$ & $-1.3$\\
$\epsilon^3$ $\Delta$ tree &$\po\po {-1.2}$&$\ph\po\po 4.3$ &$\ph\po 1.7$ &$\po {-5.8}$ & $\po{-1.0}$ & $\ph 2.5$\\
$q^4$ &$\ph\po 61.5$ &$\po {-59.}8$ &$-10.5$ &$\ph 10.2$ & $\ph\po 1.4$ & $- 1.4$\\
\textbf{Total} &\;\;\boldmath $247.2\stack{6.1\po}{27.2}$ &\!\!\boldmath $-199.8\stack{6.5\po}{23.8}$ &\!\!\,\boldmath $-40.4\stack{1.3}{4.5}$ &\;\;\boldmath $30.4\stack{1.7}{3.9}$ &\;\;\;\;\boldmath $8.4\stack{0.3}{0.8}$ &\!\!\boldmath $-3.7\stack{0.4}{0.5}$ \\
\midrule[0.2pt]
\midrule[0.2pt]
& $\ph\po\alpha_{\text{E3}}^{(n)}$ & $\ph\po\beta_{\text{M3}}^{(n)}$ &$\ph\alpha_{\text{E2}\nu}^{(n)}$& $\ph\beta_{\text{M2}\nu}^{(n)}$ & $\ph\alpha_{\text{E1}\nu^2}^{(n)}$& $\ph\beta_{\text{M1}\nu^2}^{(n)}$ \\
\midrule[0.2pt]
$q^3$  (without $\Delta$)&$\ph 136.2$ &$\po {-95.3}$ &$-25.1$ &$\ph 17.3$ &$\ph\po 8.3$ &$-3.6$\\
$q^4$  (without $\Delta$)&$\ph 123.4$ &$-118.8$ &$-21.4$ &$\ph 21.0$ &$\ph\po 3.1$ &$-3.1$\\
Total  (without $\Delta$)&$\ph 259.6$ &$-214.1$ &$-46.5$ &$\ph 38.2$ &$\ph 11.4$ &$-6.7$ \\
\midrule[0.2pt]
$q^3$ &$\ph 136.2$ &$\po {-95.3}$ &$-25.1$ &$\ph 17.3$ &$\ph\po 8.3$ &$-3.6$\\
$\epsilon^3$ $\pi\Delta$ loops &$\ph\po 52.2$ &$\po {-48.7}$&$\po {-9.0}$ &$\ph\po 8.7$ &$\ph\po 1.3$ &$-1.3$\\
$\epsilon^3$ $\Delta$ tree &$\po\po {-1.2}$ &$\ph\po\po 4.3$ &$\ph\po 1.7$ &$\po {-5.8}$ &$\po{-1.0}$ &$\ph 2.5$\\
$q^4$ &$\ph\po 61.6$ &$\po {-60.0}$ &$-10.7$ &$\ph 10.5$ &$\ph\po 1.5$ &$-3.6$\\
\textbf{Total} &\;\;\boldmath $248.8\stack{6.1\po}{27.3}$ &\!\!\boldmath $-199.7\stack{6.5\po}{23.8}$ &\!\!\boldmath $-43.2\stack{1.3}{4.7}$ &\;\;\boldmath $ 30.6\stack{1.7}{3.9}$ &\;\;\boldmath $ 10.1\stack{0.3}{1.0}$ &\!\boldmath $-4.0\stack{0.4}{1.2}$ \\
\end{tabular*}
 \end{ruledtabular}        
\caption{Numerical values for spin-independent octupole polarizabilities $\alpha_{\text{E3}}$ and $\beta_{\text{M3}}$, 
quadrupole dispersive polarizabilities $\alpha_{\text{E2}\nu}$ and $\beta_{\text{M2}\nu}$ 
as well as higher dipole dispersive polarizabilities $\alpha_{\text{E2}\nu^2}$ and $\beta_{\text{M2}\nu^2}$ 
the proton (denoted with $(p)$) and the neutron (denoted with
$(n)$). All values are given in $10^{-4}\text{fm}^7$. For remaining notation see Table \ref{tab:spin-proton}.}
\label{tab:spin-independent-3}
\end{table}

\begin{table}[ht] \centering
\renewcommand{\arraystretch}{0.75} 
 \begin{ruledtabular}
 \begin{tabular*}{\textwidth}{@{\extracolsep{\fill}}lllll}         
& $\gamma_{\text{E2E2}}^{(p)}$ & $\gamma_{\text{M2M2}}^{(p)}$ & $\gamma_{\text{E2M3}}^{(p)}$& $\gamma_{\text{M2E3}}^{(p)}$ \\
\midrule[0.2pt]
$q^3$  (without $\Delta$)& $-7.56$& $\ph\po 1.16$& $\ph 5.78$& $\ph 4.85$ \\
$q^4$  (without $\Delta$)& $-0.46$& $\po {-0.63}$& $\ph 0.32$& $-1.84$\\
Total  (without $\Delta$)& $-8.02$ & $\ph\po 0.53$& $\ph 6.10$ & $\ph 3.02$\\
\midrule[0.2pt]
$q^3$ & $-7.56$& $\ph\po 1.16$& $\ph 5.78$& $\ph 4.85$ \\
$\epsilon^3$ $\pi\Delta$ loops & $\ph 0.30$& $\ph\po 0.22$& $-0.40$& $- 0.49$\\
$\epsilon^3$ $\Delta$ tree & $-1.01$& $-10.16$& $\ph 1.13$& $-3.11$\\
$q^4$ & $-0.46$& $\po {-0.63}$& $\ph 0.32$& $-1.84$\\
\textbf{Total} &\!\boldmath $-8.74 \,\substack{\pm 0.11 \\ \pm 0.84}$ &\;\boldmath $-9.42 \,\substack{\pm 1.11 \\ \pm 0.86}$&\;\;\,\boldmath $6.84 \, \substack{\pm 0.13\\ \pm 0.63}$ &\!\boldmath $-0.59 \, \substack{\pm 0.35\\ \pm 0.62}$\\
\\
& $\gamma_{\text{E2E2}}^{(n)}$ & $\gamma_{\text{M2M2}}^{(n)}$ & $\gamma_{\text{E2M3}}^{(n)}$& $\gamma_{\text{M2E3}}^{(n)}$ \\
\midrule[0.2pt]
$q^3$  (without $\Delta$)& $-1.18$& $\ph\po 1.97$& $\ph 5.59$& $\ph 3.56$ \\
$q^4$  (without $\Delta$)& $-0.12$& $\po {-2.27}$& $\ph 0.40$& $-1.66$\\
Total  (without $\Delta$)& $-1.30$ & $\po {-0.30}$& $\ph 5.98$ & $\ph 1.89$\\
\midrule[0.2pt]
$q^3$ & $-1.18$& $\ph\po 1.97$& $\ph 5.59$& $\ph 3.56$ \\
$\epsilon^3$ $\pi\Delta$ loops & $\ph 0.33$& $\ph\po 0.13$& $-0.44$& $- 0.33$\\
$\epsilon^3$ $\Delta$ tree & $-1.01$& $-10.16$& $\ph 1.13$& $-3.11$\\
$q^4$ & $-0.12$& $\po {-2.27}$& $\ph 0.40$& $-1.66$\\
\textbf{Total} & \!\boldmath $-1.99 \,\substack{\pm 0.12\\ \pm 0.18}$ & \!\!\boldmath $-10.33 \,\substack{\pm 1.11\\ \pm 1.10}$& \;\;\,\boldmath $6.67 \,\substack{\pm 0.14 \\ \pm 0.61}$ &\!\!\! \boldmath $-1.55 \,\substack{\pm 0.35 \\ \pm 0.56}$\\
\\
& $\gamma_{\text{E1E1}\nu}^{(p)}$ & $\gamma_{\text{M1M1}\nu}^{(p)}$ & $\gamma_{\text{E1M2}\nu}^{(p)}$& $\gamma_{\text{M1E2}\nu}^{(p)}$ 
\\
\midrule[0.2pt]
$q^3$  (without $\Delta$)& $-3.26$& $\ph 0.40$& $- 0.29$ & $\ph 0.85$\\
$q^4$  (without $\Delta$)& $\ph 0.0003$& $\ph 0.24$& $-0.12$& $\ph 0.28$\\
Total  (without $\Delta$)& $-3.26$ & $\ph 0.64$& $-0.41$& $\ph 1.13$\\
\midrule[0.2pt]
$q^3$ & $-3.26$& $\ph 0.40$& $- 0.29$ & $\ph 0.85$\\
$\epsilon^3$ $\pi\Delta$ loops & $\ph 0.02$& $-0.02 $& $\ph 0.02$& $\ph 0.001$\\
$\epsilon^3$ $\Delta$ tree & $-0.49$& $\ph 1.73$& $-0.56$& $\ph 0.70$\\
$q^4$ & $\ph 0.0003$& $\ph 0.24$& $-0.12$& $\ph 0.28$\\
\textbf{Total} &\!\!\! \boldmath $-3.72 \stack{0.05}{0.35}$ &\!\!\! \boldmath $\ph 2.35\stack{0.19}{0.21}$&\!\!\! \boldmath $-0.95\stack{0.06}{0.09}$&\!\!\! \boldmath $\ph 1.83\stack{0.08}{0.17}$ \\
\\
& $\gamma_{\text{E1E1}\nu}^{(n)}$ & $\gamma_{\text{M1M1}\nu}^{(n)}$ & $\gamma_{\text{E1M2}\nu}^{(n)}$& $\gamma_{\text{M1E2}\nu}^{(n)}$ 
\\
\midrule[0.2pt]
$q^3$  (without $\Delta$)& $-4.62$& $\ph 0.46$& $-0.29$ & $\ph 1.23$\\
$q^4$  (without $\Delta$)& $-0.10$& $\ph 0.50$& $-0.18$& $\ph 0.28$\\
Total  (without $\Delta$)& $-4.72$ & $\ph 0.97$& $-0.47$& $\ph 1.51$ \\
\midrule[0.2pt]
$q^3$ & $-4.62$& $\ph 0.46$& $-0.29$ & $\ph 1.23$\\
$\epsilon^3$ $\pi\Delta$ loops & $\ph 0.03$& $-0.03$& $\ph 0.03$& $- 0.01$\\
$\epsilon^3$ $\Delta$ tree & $-0.49$& $\ph 1.73$& $-0.56$& $\ph 0.70$\\
$q^4$ & $-0.10$& $\ph 0.50$& $-0.18$& $\ph 0.28$\\
\textbf{Total} &\!\!\! \boldmath $-5.17\stack{0.05}{0.49}$ &\!\!\! \boldmath $\ph 2.67\stack{0.19}{0.27}$&\!\!\! \boldmath $-1.00\stack{0.06}{0.10}$&\!\!\! \boldmath $\ph 2.19\stack{0.08}{0.20}$ \\
 \end{tabular*}
 \end{ruledtabular}      
\caption{Numerical values for the quadrupole spin polarizabilities $\gamma_{\text{E2E2}}$, 
$\gamma_{\text{M2M2}}$, $\gamma_{\text{E2M3}}$ and $\gamma_{\text{M2E3}}$ and for the dispersive spin polarizabilities $\gamma_{\text{E1E1}\nu}$, 
$\gamma_{\text{M1M1}\nu}$, $\gamma_{\text{E1M2}\nu}$ and $\gamma_{\text{M1E2}\nu}$ of the proton (indicated with $(p)$) 
and the neutron (indicated with $(n)$). All values are given in
$10^{-6}\text{fm}^5$. For remaining notation see Table \ref{tab:spin-proton}.}
\label{tab:spin-quadrupole}
\end{table}

We summarize the general features of the higher-order polarizabilites.
Both $\Delta$-less and $\Delta$-full schemes give roughly the same results, except for the channels where the
$\Delta$-tree-level contribution is significant, i.e.~for magnetic multipoles.
Note that in the $\Delta$-less approach, such contributions would appear only at extremely high 
orders, which makes the $\Delta$-less framework rather inefficient.

The second observation concerns the loop contributions.
While for all spin polarizabilites, the $\epsilon^3$-$\Delta$-loops and 
the $q^4$-loops are strongly suppressed,
for scalar polarizabilites the situation is different.
In the $\Delta$-less scheme, the  $q^4$-loops are comparable with the $q^3$-loops
or larger, which spoils convergence.
On the other hand, in the $\Delta$-full scheme, a significant part of the
$q^4$-loop contributions is shifted to the $\epsilon^3$-$\Delta$-loops.
This happens due to the $\Delta$-resonance saturation of the low-energy constants $c_i$,
in particular $c_2$ and $c_3$ \cite{Bernard:1996gq, Krebs:2018jkc}, which do not contribute to 
the spin polarizabilites.
As a result, the convergence pattern of the $\Delta$-full scheme looks
very convincing for both scalar and spin polarizabilites. 
The only exceptions are the $\gamma_{\text{M2E3}}$ polarizabilites,
where the $\epsilon^3$ result is unnaturally small due to accidental
cancellations 
between the $q^3$-loops and the $\Delta$-tree-level  contributions.

Our predictions at order $\mathcal{O}(q^4+\epsilon^3)$ for all scalar quadrupole
and dipole dispersive polarizabilites of the proton and the neutron 
agree within errors with the results based on fixed-$t$ dispersion
relations, see Table~\ref{tab:spin-independent-2-proton}. 
Note that the predictions of the $\delta$-counting scheme of \cite{Lensky:2015awa} do not reproduce  
the fixed-$t$ dispersion relations values for $\alpha_{\text{E2}}$ and
$\beta_{\text{M2}}$. The main difference to our result in this
channel comes from the $q^4$-loops and $\epsilon^3$-$\Delta$-loops. 
On the other hand, the difference in  the tree-level-$\Delta$ contributions appears very small,
indicating the insignificance of the off-shell effects, as one would expect for such high-order polarizabilites.

\subsection{Generalized polarizabilities}\label{sub:Q-polarizabilities}
\begin{figure}[!ht]\centering
\includegraphics[width=0.96\textwidth]{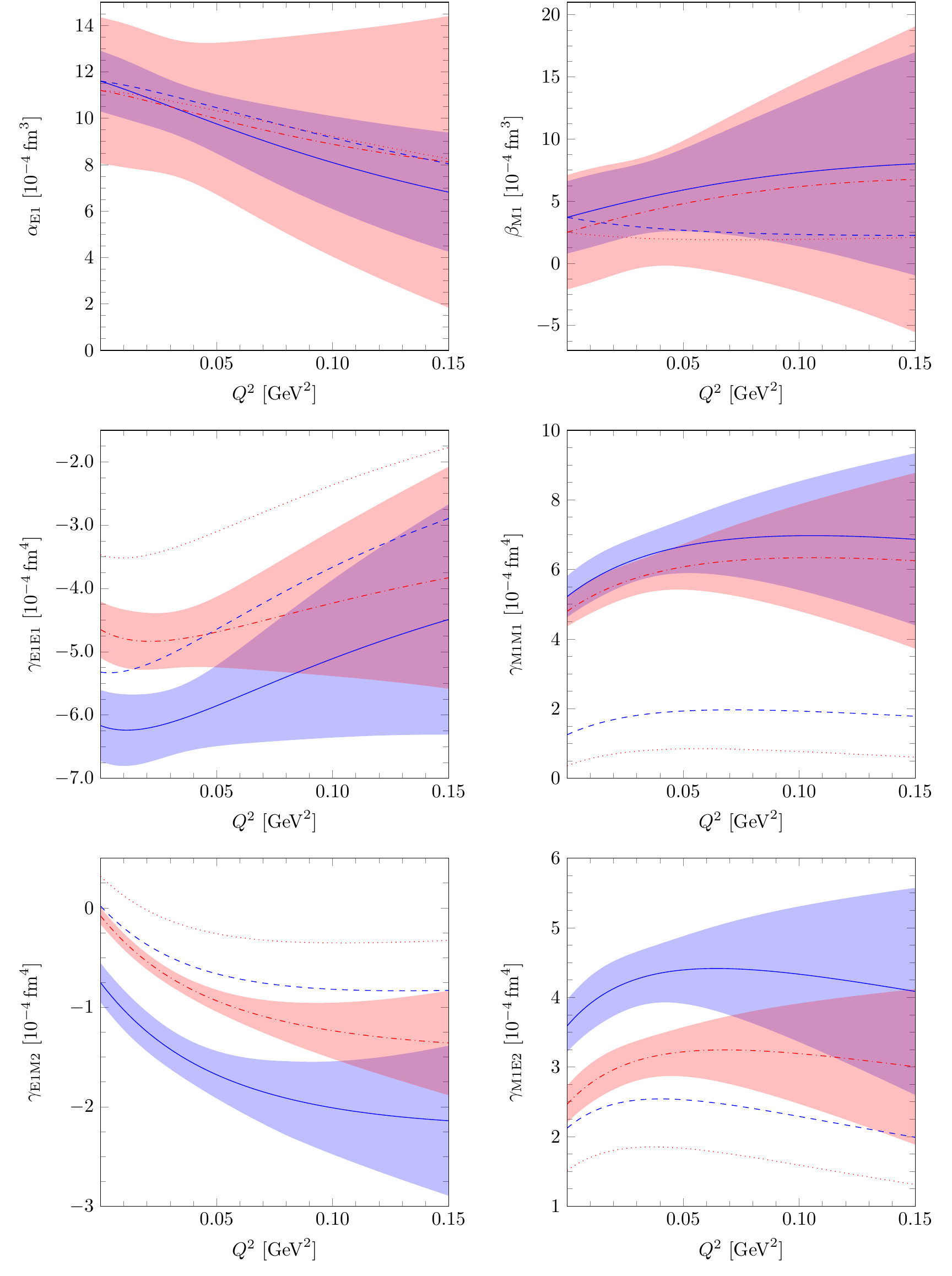}
\caption{$Q^2$-dependence of the scalar and spin polarizabilites for the proton (dotted and dash-dotted lines) and the neutron (dashed and solid lines).
The dotted and dashed lines correspond to the $\Delta$-less $\mathcal{O}(q^4)$ results, whereas the dash-dotted and solid lines correspond to the $\Delta$-full $\mathcal{O}(\epsilon^3+q^4)$ results.
The bands indicate the theoretical truncation errors.}
\label{Fig:Q2_combined}
\end{figure}

Now are now in the position to discuss the generalized ($Q^2$-dependent) nucleon polarizabilites.
We consider the doubly virtual Compton scattering with the initial and final virtuality of the photon equal to $Q^2$.
In Fig.~\ref{Fig:Q2_combined}, the scalar and spin dipole polarizabilites for the proton and the neutron are plotted
as a function of $Q^2$, and the $\Delta$-full and $\Delta$-less schemes are compared.
The scalar polarizabilites at $Q^2=0$ are adjusted to the empirical values, see subsection~\ref{sub:scalar_dipole_polarizabilities}.
The difference of the $\Delta$-full and $\Delta$-less spin polarizabilites at $Q^2=0$ was discussed in subsection~\ref{sub:dipole_spin_polarizabilities}
and can be considered as a higher-order contact-term contribution.
Therefore, we focus here on the $Q^2$-dependence of the polarizabilites relative to their $Q^2=0$ values.
For the spin polarizabilites and for the electric scalar polarizability, the $\Delta$-full and $\Delta$-less curves
go almost parallel to each other, whereas for the magnetic scalar polarizabilites the slope and the curvature of the
curves are opposite in sign. This is due to a significant contribution of the $\Delta$-tree-level contribution in this channel.
It should be emphasized that the scalar generalized polarizabilites contribute to the Lamb shift of muonic hydrogen, see e.g.~\cite{Hagelstein:2015egb}.

\begin{figure}[!ht]\centering
\includegraphics[width=0.92\textwidth]{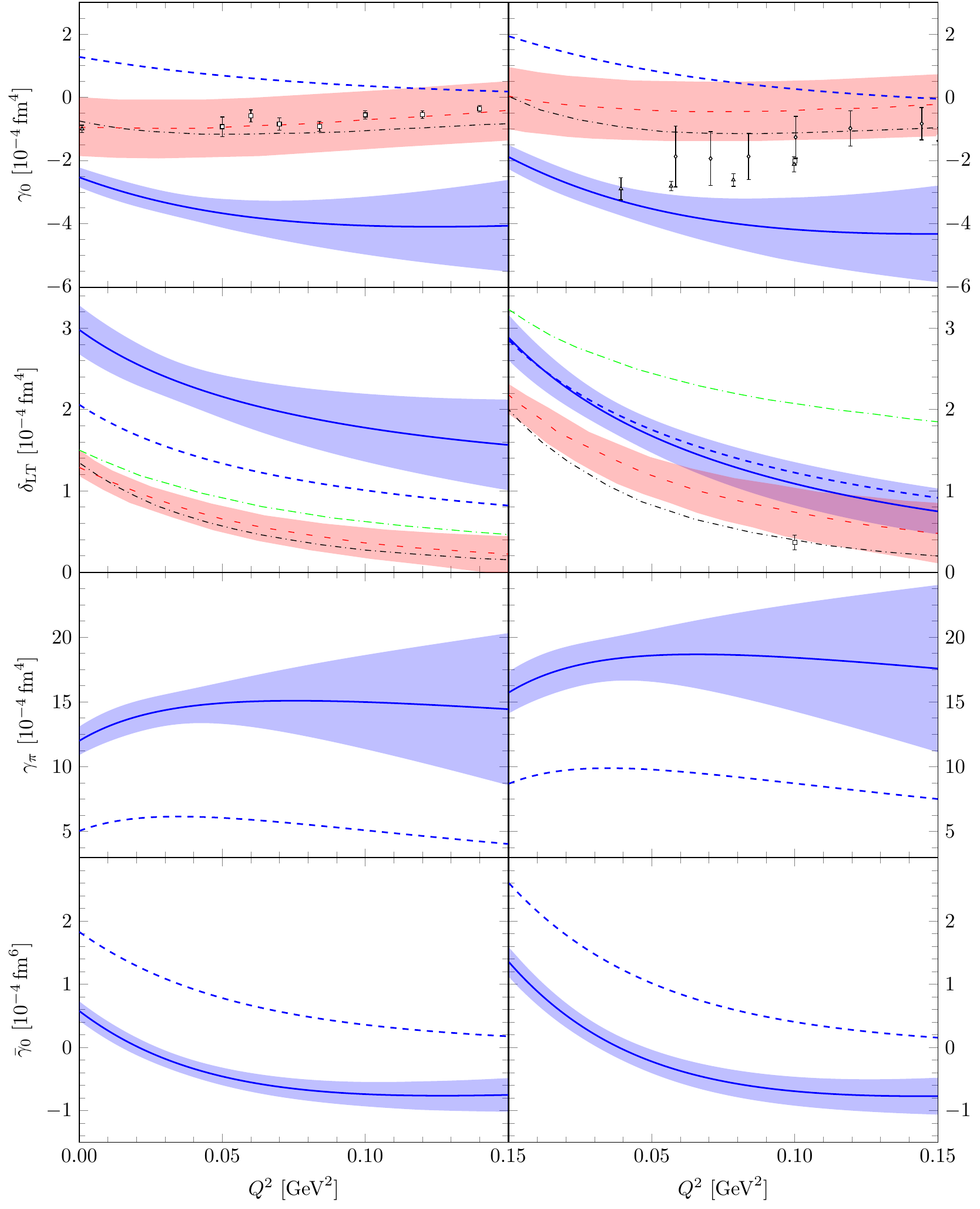}
\caption{$Q^2$-dependencies of the forward polarizabilities $\gamma_{0}$
and $\delta_{\text{LT}}$, the backward polarizability $\gamma_{\pi}$ and
the combined higher-order polarizability $\bar{\gamma}_{0}$ for the
proton (left) and the neutron (right). The thick solid blue lines
indicate our $\Delta$-full $\mathcal{O}(q^4+\epsilon^3)$ calculations
with a $1\sigma$ truncation error band corresponding to $68\%$
degree-of-belief intervals and the thick dashed blue lines
show our $\Delta$-less $\mathcal{O}(q^4)$ calculations. The red loosely
dashed lines represent the NLO B$\chi$PT 
calculation from \cite{Alarcon:2020icz} with the red error bands. The black dash-dotted line
presents the MAID model predictions from \cite{Drechsel:2002ar} (proton) and
\cite{Amarian:2004yf} (neutron). The green double-dash-dotted line is the
$\mathcal{O}(p^4)$ calculation from \cite{Kao:2002cp}. Empirical data are: for
$\gamma_0^{(p)}$ from \cite{Gryniuk:2016gnm} (triangle) and
\cite{Prok:2008ev} (squares); for $\gamma_0^{(n)}$ from \cite{Deur:2019pew}
(preliminary, triangles), \cite{Amarian:2004yf} (square) and
\cite{Guler:2015hsw} (diamonds); for $\delta_{\text{LT}}^{(n)}$ from
\cite{Amarian:2004yf}.}
\label{Fig:spin_3}
\end{figure}

We also present the $Q^2$-dependence of several combined spin
polarizabilites, for some of which the experimental data are
available, see Fig.~\ref{Fig:spin_3} (their limiting values for
$Q^2=0$ are collected in Table~\ref{tab:spin-combined-proton}). 
We observe no improvement as compared to \cite{Bernard:2012hb} (pure $\mathcal{O}(\epsilon^3)$ calculation)
due to the inclusion of the $\mathcal{O}(q^4)$ contributions. In fact, the description of $\gamma_0$ for the proton
is even worse. A possible source of such a discrepancy could be a missing
contribution of the $\Delta$-loop diagrams at order  $\mathcal{O}(\epsilon^4)$, 
as was suggested in \cite{Bernard:2012hb}. On the other hand,
taking into account a much better description of the data in
\cite{Lensky:2014dda} 
(within the $\delta$-counting scheme)
and the fact that the disagreement of our result with experiment for
the value of $\gamma_0$ for the proton at $Q^2=0$ was caused by the
large contribution  
from the induced electric $\gamma N\Delta$-coupling (as a $1/m$
effect), one may expect the improvement to be achieved after 
including the relevant higher-order $\gamma N\Delta$-vertices from the
effective Lagrangian analogously to \cite{Lensky:2014dda}. 

\subsection{Dynamical polarizabilities}\label{sub:energy-polarizabilities}
\begin{figure}[ht]\centering 
\includegraphics[width=0.82\textwidth]{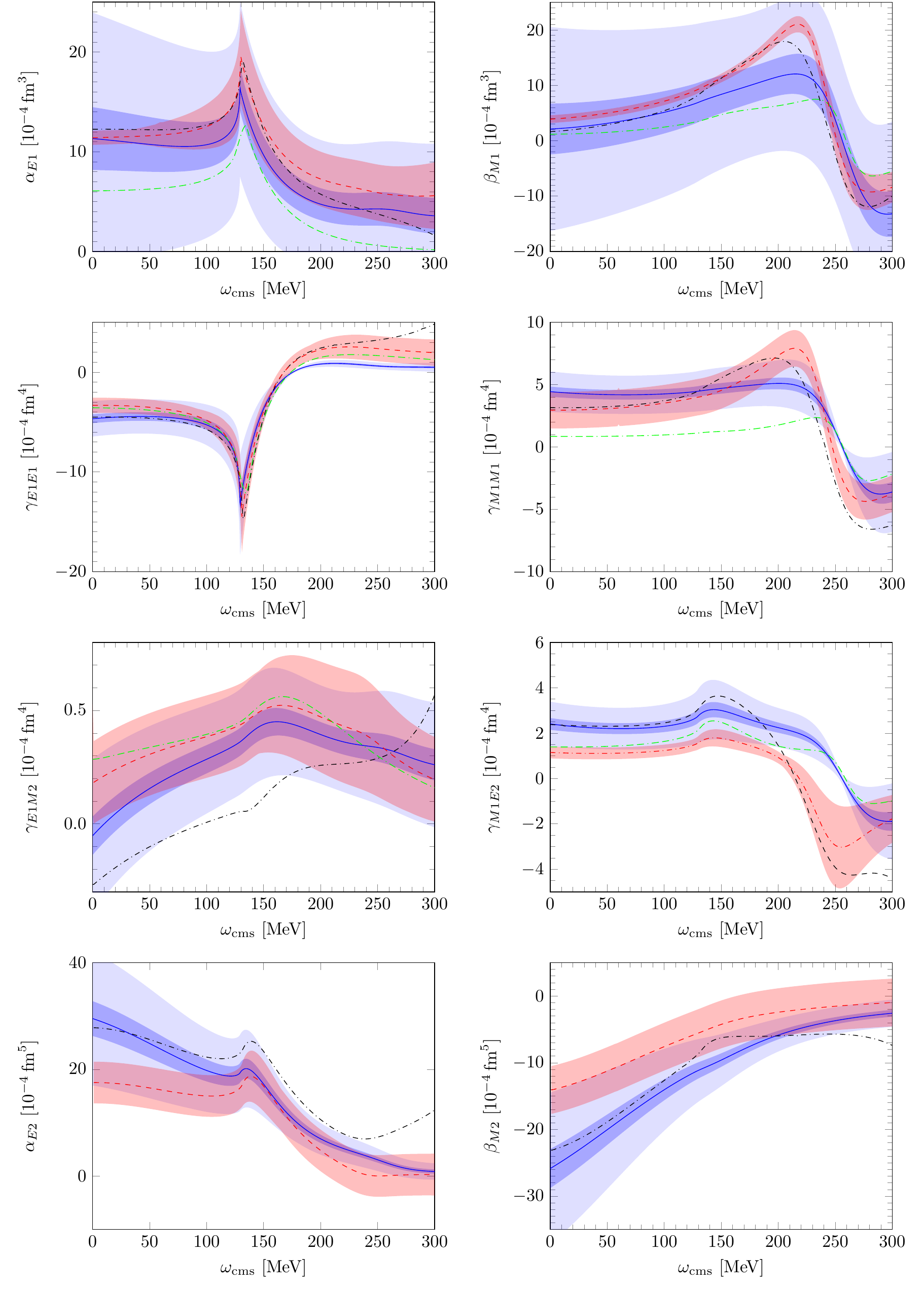}
\caption{The $\omega$-dependence of the real parts of the dipole polarizabilities and the spinless quadrupole polarizabilites 
$\alpha_{\text{E2}}$ and $\beta_{\text{M2}}$ for the proton. 
The solid blue lines represent our $\Delta$-full $\mathcal{O}(q^4+\epsilon^3)$ result.
The inner (outer) blue bands stand for the $1\sigma$ ($2\sigma$) truncation error.
The red dashed lines are the B$\chi$PT calculation \cite{Lensky:2015awa}
with the red error bands. 
The black dash-dotted lines correspond to the fixed-$t$
dispersion-relations calculation \cite{Hildebrandt:2003fm} and the green
double-dash-dotted lines correspond to the results of
\cite{Aleksejevs:2013cda}. }
\label{Fig:dynamic_proton_1}
\end{figure}

\begin{figure}[ht]\centering 
\includegraphics[width=0.82\textwidth]{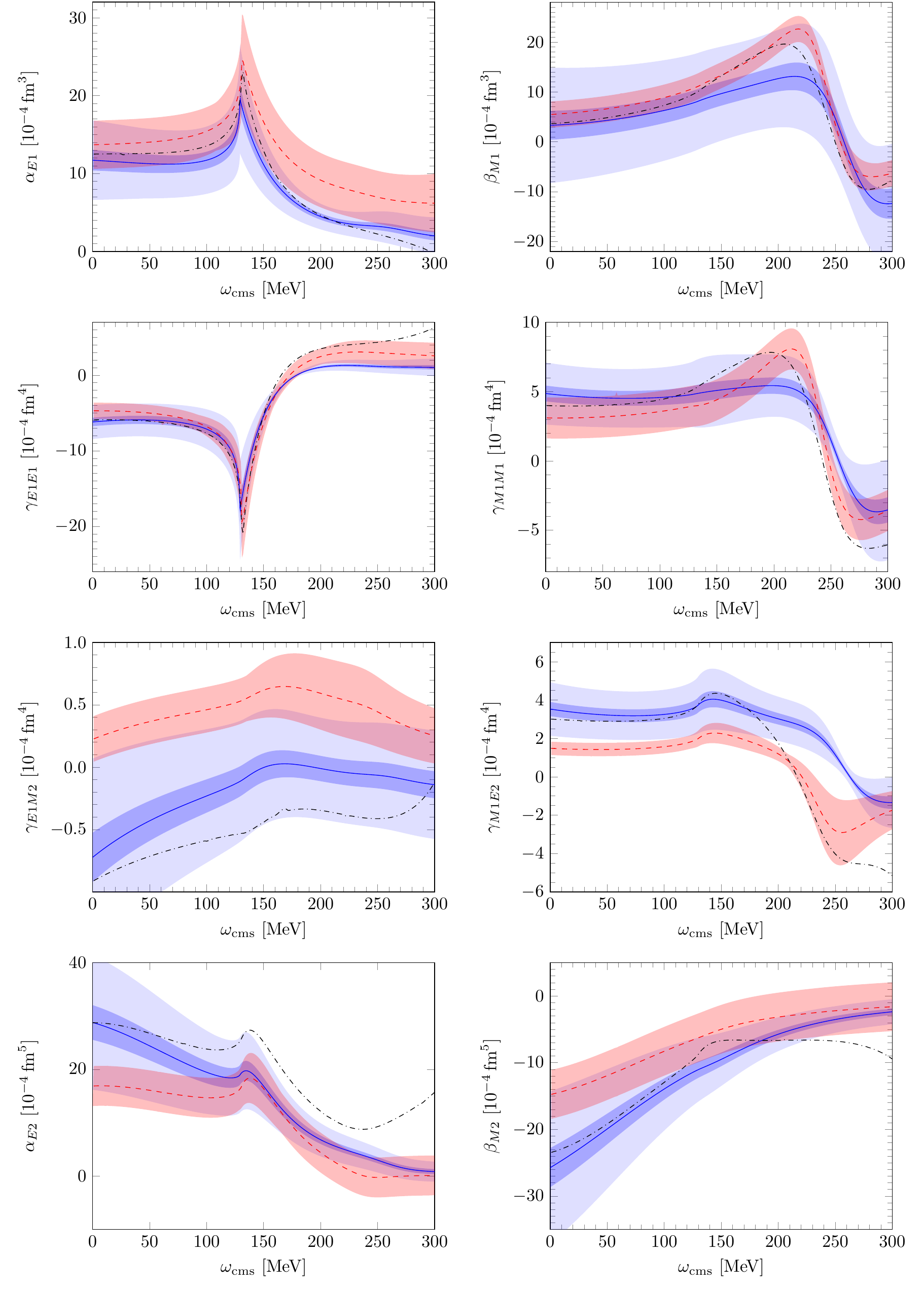}
\caption{The $\omega$-dependence of the real parts of the dipole polarizabilities and the spinless quadrupole polarizabilites 
$\alpha_{\text{E2}}$ and $\beta_{\text{M2}}$ for the neutron. 
The solid blue lines represent our $\Delta$-full $\mathcal{O}(q^4+\epsilon^3)$ result.
The inner (outer) blue bands stand for the $1\sigma$ ($2\sigma$) truncation error.
The red dashed lines are the B$\chi$PT calculation \cite{Lensky:2015awa}
with the red error bands. 
The black dash-dotted lines correspond to the fixed-$t$
dispersion-relations calculation \cite{Hildebrandt:2003fm}.}
\label{Fig:dynamic_neutron_1}
\end{figure}

One can also probe
the electromagnetic structure of the nucleon 
by looking at dynamical (energy-dependent) polarizabilites that describe the response
to the nucleon electromagnetic excitations at arbitrary energy.
In Figs.~\ref{Fig:dynamic_proton_1}, \ref{Fig:dynamic_neutron_1}, we present the energy dependence
of the dipole and spinless quadrupole polarizabilites up to the center-of-mass energy 
$\omega_\text{CM}=300$~MeV. For comparison, also shown are the 
results obtained using the $\delta$-counting scheme \cite{Lensky:2015awa},
the fixed-$t$ dispersion relations \cite{Hildebrandt:2003fm},
and the Computational Hadronic Model \cite{Aleksejevs:2013cda}\footnote{We have extracted those data points from Ref.~\cite{Lensky:2015awa}.}.
The $1\sigma$ and $2\sigma$ truncation errors corresponding to 
$68\%$ and $95\%$
degree-of-belief intervals
are shown as bands in the figures. 
Our results agree rather well with the ones of the fixed-$t$
dispersion relations at $\omega_\text{CM}=0$ (except for
$\gamma_{E1M2}$). 
Therefore, it is natural to compare the two approaches at non-zero energies.
As can be seen from the figures, the deviation of our results from those
of the fixed-$t$ dispersion relations increases with energy, 
which may provide yet another indication that our theoretical errors
are underestimated (as discussed in subsection~\ref{sub:Higher-order_polarizabilities}), 
and the convergence of the small-scale expansion becomes slower $\omega_\text{CM}\gtrsim150-200$~MeV. 
However, for $\alpha_{E2}$ a large discrepancy (beyond $2\sigma$) between the two theoretical frameworks
is observed already for $\omega_\text{CM}\gtrsim 100-150$~MeV.
This could be due to 
the aforementioned large induced electric 
$\gamma N \Delta$-coupling in our scheme, whose effect 
increases with energy.

\newpage
\subsection{Heavy-Baryon Expansion}\label{sub:hb-expansion}
In this subsection, we study the convergence of the $1/m$-expansion
of our results (the nucleon-$\Delta$ mass difference $\Delta$ is kept
finite and constant) 
obtained within the covariant framework.
analyzing such an expansion we can test the efficiency of the heavy-baryon approach by reproducing some of its contributions
appearing at higher orders. 
We present the $1/m$-expansion for the dipole scalar and spin polarizabilities in 
Tables~\ref{tab:HB-alpha-E1}-\ref{tab:HB-gamma-M1E2} starting from the leading order (LO) static ($m^0$) results up to 
the order $1/m^5$ (N$^5$LO).
Obviously, the static results as well as the $1/m$-corrections to the leading-order terms
coincide with the corresponding heavy-baryon calculations, see \cite{Bernard:1995dp,VijayaKumar:2000pv,Gellas:2000mx,Hemmert:1996rw}.
\begin{table}[ht] \centering 
\renewcommand{\arraystretch}{0.75} 
\begin{tabular}{@{}lllllllllllll@{}}
\toprule
$\alpha_{\text{E1}}^{(p)}$& $\ph q^3$ & $\ph q^4$ & $\ph \epsilon^3_{\text{Loop}}$& $\ph \epsilon^3_{\text{Tree}}$ & $\ph \text{Full}$ & \hspace{1cm} &
$\alpha_{\text{E1}}^{(n)}$& $\ph q^3$ & $\ph q^4$ & $\ph \epsilon^3_{\text{Loop}}$& $\ph \epsilon^3_{\text{Tree}}$ & $\ph \text{Full}$ \\
\cmidrule[0.2pt]{1-6}
\cmidrule[0.2pt]{8-13}
$\operatorname{LO}$ & $12.78$ & $\po 8.86$ & $\ph 7.86$ & $\ph \ldots$ & $\ph 29.50$ & $$&
$\operatorname{LO}$ & $12.78$ & $\po 2.67$ & $\ph 7.86$ & $\ph \ldots$ & $\ph 23.31$\\
$\operatorname{NLO}$& $\po 8.47$ & $\po 8.86$ & $\ph 2.89$ & $-5.60$ & $\ph 14.61$ & $$&
$\operatorname{NLO}$&$\po 9.67$ & $\po 2.67$ & $\ph 4.94$ & $-5.60$ & $\ph 11.68$\\
$\operatorname{N}^2\!\operatorname{LO}$ & $\po 6.60$ & $\po 9.53$ & $\ph 0.97$ & $-2.98$ & $\ph 14.13$& $$ &
$\operatorname{N}^2\!\operatorname{LO}$ & $\po 9.50$ & $\po 3.22$ & $\ph 3.03$ & $-2.98$ & $\ph 12.77$\\
$\operatorname{N}^3\!\operatorname{LO}$ & $\po 7.01$ & $\po 9.40$ & $-1.03$ & $-4.11$ & $\ph 11.26$& $$&
$\operatorname{N}^3\!\operatorname{LO}$ & $\po 9.51$ & $\po 3.10$ & $\ph 2.70$ & $-4.11$ & $\ph 11.19$\\
$\operatorname{N}^4\!\operatorname{LO}$ & $\po 7.04$ & $\po 9.40$ & $-1.33$ & $-3.65$ & $\ph 11.46$& $$&
$\operatorname{N}^4\!\operatorname{LO}$ & $\po 9.51$ & $\po 3.09$ & $\ph 2.83$ & $-3.65$ & $\ph 11.78$\\
$\operatorname{N}^5\!\operatorname{LO}$ & $\po 7.04$ & $\po 9.40$ & $-1.49$ & $-3.83$ & $\ph 11.12$ & $$&
$\operatorname{N}^5\!\operatorname{LO}$ & $\po 9.51$ & $\po 3.09$ & $\ph 2.75$ & $-3.83$ & $\ph 11.52$\\
\cmidrule[0.2pt]{1-6}
\cmidrule[0.2pt]{8-13}
Full & $\po 7.04$ & $\po 9.40$ & $-1.45$ & $-3.78$ & $\ph 11.20$& $$ &
Full & $\po 9.51$ & $\po 3.09$ & $\ph 2.78$ & $-3.78$ & $\ph 11.60$\\
\bottomrule
\end{tabular}
\caption{Numerical values for the $1/m$-expansion of
  $\alpha_{\text{E1}}$. Note that the $\epsilon^3_{\text{Tree}}$ only starts at NLO.}
\label{tab:HB-alpha-E1}
\end{table}
\begin{table}[ht] \centering
\renewcommand{\arraystretch}{0.75} 
\begin{tabular}{@{}lllllllllllll@{}}
\toprule
$\beta_{\text{M1}}^{(p)}$& $\ph q^3$ & $\ph q^4$ & $\ph \epsilon^3_{\text{Loop}}$& $\ph \epsilon^3_{\text{Tree}}$ & $\ph \text{Full}$ & \hspace{1cm} &
$\beta_{\text{M1}}^{(n)}$& $\ph q^3$ & $\ph q^4$ & $\ph \epsilon^3_{\text{Loop}}$& $\ph \epsilon^3_{\text{Tree}}$ & $\ph \text{Full}$\\

\cmidrule[0.2pt]{1-6}
\cmidrule[0.2pt]{8-13}
$\operatorname{LO}$ & $\ph 1.28$ & $-12.33$ & $\po 1.36$ & $11.96$ & $2.26$ & $$ &
$\operatorname{LO}$ & $\ph 1.28$ & $-7.35$ & $\po 1.36$ & $11.96$ & $7.24$\\

$\operatorname{NLO}$& $\ph 0.56$ & $-12.09$ & $\po 1.61$& $11.96$ & $2.04$ & $$ &
$\operatorname{NLO}$& $-0.88$ & $-7.60$ & $\po 0.34$ & $11.96$ & $3.83$\\

$\operatorname{N}^2\!\operatorname{LO}$ & $-2.83$ & $-13.75$ & $\po 9.05$& $11.96$ & $4.43$ & $$ &
$\operatorname{N}^2\!\operatorname{LO}$ & $-1.11$ & $-8.27$ & $\po 1.28$ & $11.96$ & $3.85$\\

$\operatorname{N}^3\!\operatorname{LO}$ & $-1.94$ & $-13.23$ & $\po 5.67$ & $11.96$ & $2.45$ & $$ &
$\operatorname{N}^3\!\operatorname{LO}$ & $-1.10$ & $-8.12$ & $\po 0.69$ & $11.96$ & $3.43$\\

$\operatorname{N}^4\!\operatorname{LO}$ & $-1.83$ & $-13.14$ & $\po 6.39$ & $11.96$ & $3.38$ & $$ &
$\operatorname{N}^4\!\operatorname{LO}$ & $-1.10$ & $-8.12$ & $\po 1.09$ & $11.96$ & $3.83$\\

$\operatorname{N}^5\!\operatorname{LO}$ & $-1.85$ & $-13.16$ & $\po 5.32$  & $11.96$ & $2.28$ & $$ &
$\operatorname{N}^5\!\operatorname{LO}$ & $-1.10$ & $-8.12$ & $\po 0.90$ & $11.96$ & $3.64$\\
\cmidrule[0.2pt]{1-6}
\cmidrule[0.2pt]{8-13}
Full & $-1.85$ & $-13.16$ & $\po 5.54$  & $11.96$ & $2.50$ & $$ &
Full & $-1.10$ & $-8.12$ & $\po 0.96$ & $11.96$ & $3.70$\\
\bottomrule
\end{tabular}
\caption{Numerical values for the $1/m$-expansion of $\beta_{\text{M1}}$.}
\label{tab:HB-beta-M1}
\end{table}

\begin{table}[ht] \centering 
\renewcommand{\arraystretch}{0.75} 
\begin{tabular}{@{}lllllllllllll@{}}
\toprule
$\gamma_{\text{E1E1}}^{(p)}$& $\ph q^3$ & $\ph q^4$ & $\ph \epsilon^3_{\text{Loop}}$& $\ph \epsilon^3_{\text{Tree}}$ & $\ph \text{Full}$ & \hspace{1cm} &
$\gamma_{E1E1}^{(n)}$& $\ph q^3$ & $\ph q^4$ & $\ph \epsilon^3_{\text{Loop}}$& $\ph \epsilon^3_{\text{Tree}}$ & $\ph \text{Full}$\\
\cmidrule[0.2pt]{1-6}
\cmidrule[0.2pt]{8-13}
$\operatorname{LO}$ & $-5.81$ & $\ph \ldots$ & $\ph 0.60$ & $\ph \ldots$ & $-5.22$ & $$ &
$\operatorname{LO}$ & $-5.81$ & $\ph \ldots$ & $\ph 0.60$ & $\ph \ldots$ & $-5.22$ \\
$\operatorname{NLO}$& $-1.38$ & $\ph \ldots$ & $\ph 0.96$& $-0.63$ & $-1.05$ & $$ &
$\operatorname{NLO}$& $-4.34$ & $\ph \ldots$ & $\ph 0.66$ & $-0.63$ & $-4.31$ \\
$\operatorname{N}^2\!\operatorname{LO}$ & $-3.03$ & $\ph 0.20$ & $-0.63$& $-1.22$ & $-4.68$ & $$ &
$\operatorname{N}^2\!\operatorname{LO}$ & $-4.83$ & $-0.37$ & $\ph 0.84$ & $-1.22$ & $-5.58$ \\
$\operatorname{N}^3\!\operatorname{LO}$ & $-3.57$ & $-0.11$ & $-1.41$ & $-1.02$ & $-6.11$ & $$ &
$\operatorname{N}^3\!\operatorname{LO}$ & $-4.86$ & $-0.49$ & $\ph 0.07$ & $-1.02$ & $-6.30$ \\
$\operatorname{N}^4\!\operatorname{LO}$ & $-3.47$ & $-0.02$ & $\ph 0.07$ & $-1.08$ & $-4.50$ & $$ &
$\operatorname{N}^4\!\operatorname{LO}$ & $-4.86$ & $-0.46$ & $\ph 0.26$ & $-1.08$ & $-6.15$ \\
$\operatorname{N}^5\!\operatorname{LO}$ & $-3.46$ & $-0.01$ & $-0.13$  & $-1.06$ & $-4.66$ & $$ &
$\operatorname{N}^5\!\operatorname{LO}$ & $-4.86$ & $-0.46$ & $\ph 0.23$ & $-1.06$ & $-6.16$ \\
\cmidrule[0.2pt]{1-6}
\cmidrule[0.2pt]{8-13}
Full & $-3.46$ & $-0.01$ & $-0.11$  & $-1.07$ & $-4.65$ & $$ &
Full & $-4.86$ & $-0.46$ & $\ph 0.22$ & $-1.07$ & $-6.17$ \\
\bottomrule
\end{tabular}
\caption{Numerical values for the $1/m$-expansion of $\gamma_{\text{E1E1}}$. The dots mark entries that do not exist e.g. the $\epsilon^3_{\text{Tree}}$ starts at NLO and therefore does not have a LO contribution.}
\label{tab:HB-gamma-E1E1}
\end{table}

\begin{table}[ht] \centering
\renewcommand{\arraystretch}{0.75} 
\begin{tabular}{@{}lllllllllllll@{}}
\toprule
$\gamma_{\text{M1M1}}^{(p)}$& $\ph q^3$ & $\ph q^4$ & $\ph \epsilon^3_{\text{Loop}}$& $\ph \epsilon^3_{\text{Tree}}$ & $\ph \text{Full}$ & \hspace{1cm} &
$\gamma_{M1M1}^{(n)}$& $\ph q^3$ & $\ph q^4$ & $\ph \epsilon^3_{\text{Loop}}$& $\ph \epsilon^3_{\text{Tree}}$ & $\ph \text{Full}$\\

\cmidrule[0.2pt]{1-6}
\cmidrule[0.2pt]{8-13}$\operatorname{LO}$ & $-1.16$ & $\ph \ldots$ & $\ph 0.21$ & $\ph 4.03$ & $\ph 3.08$ & $$ &
$\operatorname{LO}$ & $-1.16$ & $\ph \ldots$ & $\ph 0.21$ & $\ph 4.03$ & $\ph 3.08$ \\
$\operatorname{NLO}$& $\ph 1.39$ & $\ph 1.93$ & $-0.12$& $\ph 3.40$ & $\ph 6.59$ & $$ &
$\operatorname{NLO}$& $\ph 0.31$ & $\ph 2.05$ & $-0.27$ & $\ph 3.40$ & $\ph 5.49$ \\
$\operatorname{N}^2\!\operatorname{LO}$ & $\ph 0.52$ & $\ph 1.24$ & $\ph 2.21$& $\ph 3.89$ & $\ph 7.85$ & $$ &
$\operatorname{N}^2\!\operatorname{LO}$ & $-0.13$ & $\ph 1.56$ & $\ph 0.78$ & $\ph 3.89$ & $\ph 6.11$ \\
$\operatorname{N}^3\!\operatorname{LO}$ & $-0.37$ & $\ph 0.19$ & $-0.44$ & $\ph 3.87$ & $\ph 3.26$ & $$ &
$\operatorname{N}^3\!\operatorname{LO}$ & $-0.17$ & $\ph 1.39$ & $\ph 0.04$ & $\ph 3.87$ & $\ph 5.13$ \\
$\operatorname{N}^4\!\operatorname{LO}$ & $-0.15$ & $\ph 0.46$ & $\ph 0.81$ & $\ph 3.84$ & $\ph 4.96$ & $$ &
$\operatorname{N}^4\!\operatorname{LO}$ & $-0.17$ & $\ph 1.42$ & $\ph 0.23$ & $\ph 3.84$ & $\ph 5.33$ \\
$\operatorname{N}^5\!\operatorname{LO}$ & $-0.12$ & $\ph 0.50$ & $\ph 0.39$  & $\ph 3.86$ & $\ph 4.62$ & $$ &
$\operatorname{N}^5\!\operatorname{LO}$ & $-0.17$ & $\ph 1.42$ & $\ph 0.05$ & $\ph 3.86$ & $\ph 5.16$ \\

\cmidrule[0.2pt]{1-6}
\cmidrule[0.2pt]{8-13}Full & $-0.13$ & $\ph 0.49$ & $\ph 0.58$  & $\ph 3.85$ & $\ph 4.80$ & $$ &
Full & $-0.17$ & $\ph 1.42$ & $\ph 0.12$ & $\ph 3.85$ & $\ph 5.22$ \\
\bottomrule
\end{tabular}
\caption{Numerical values for the $1/m$-expansion of $\gamma_{\text{M1M1}}$. The dots mark entries that do not exist e.g. the $q^4$ starts at NLO and therefore does not have a LO contribution.}
\label{tab:HB-gamma-M1M1}
\end{table}

\begin{table}[ht] \centering 
\renewcommand{\arraystretch}{0.75} 
\begin{tabular}{@{}lllllllllllll@{}}
\toprule
$\gamma_{\text{E1M2}}^{(p)}$& $\ph q^3$ & $\ph q^4$ & $\ph \epsilon^3_{\text{Loop}}$& $\ph \epsilon^3_{\text{Tree}}$ & $\ph \text{Full}$ & \hspace{1cm} &
$\gamma_{E1M2}^{(n)}$& $\ph q^3$ & $\ph q^4$ & $\ph \epsilon^3_{\text{Loop}}$& $\ph \epsilon^3_{\text{Tree}}$ & $\ph \text{Full}$\\

\cmidrule[0.2pt]{1-6}
\cmidrule[0.2pt]{8-13}$\operatorname{LO}$ & $\ph 1.16$ & $\ph \ldots$ & $-0.21$ & $\ph \ldots$ & $\ph 0.95$ & $$ &
$\operatorname{LO}$ & $\ph 1.16$ & $\ph \ldots$ & $-0.21$ & $\ph \ldots$ & $\ph 0.95$ \\
$\operatorname{NLO}$& $\ph 0.22$ & $\ph \ldots$ & $\ph 0.01$& $-0.63$ & $-0.40$ & $$ &
$\operatorname{NLO}$& $\ph 0.49$ & $\ph \ldots$ & $\ph 0.04$ & $-0.63$ & $-0.09$ \\
$\operatorname{N}^2\!\operatorname{LO}$ & $\ph 0.57$ & $\ph 0.03$ & $\ph 1.43$& $-0.92$ & $\ph 1.11$ & $$ &
$\operatorname{N}^2\!\operatorname{LO}$ & $\ph 0.60$& $-0.43$ & $\ph 0.21$ & $-0.92$ & $-0.54$ \\
$\operatorname{N}^3\!\operatorname{LO}$ & $\ph 0.54$ & $-0.36$ & $\ph 0.84$ & $-0.88$ & $\ph 0.15$  & $$ &
$\operatorname{N}^3\!\operatorname{LO}$ & $\ph 0.61$ & $-0.63$ & $\ph 0.27$ & $-0.88$ & $-0.64$ \\
$\operatorname{N}^4\!\operatorname{LO}$ & $\ph 0.56$ & $-0.26$ & $\ph 0.05$ & $-0.88$ & $-0.52$ & $$ &
$\operatorname{N}^4\!\operatorname{LO}$ & $\ph 0.61$ & $-0.59$ & $\ph 0.03$ & $-0.88$ & $-0.84$ \\
$\operatorname{N}^5\!\operatorname{LO}$ & $\ph 0.57$ & $-0.25$ & $\ph 0.70$  & $-0.88$ & $\ph 0.14$  & $$ &
$\operatorname{N}^5\!\operatorname{LO}$ & $\ph 0.61$ & $-0.59$ & $\ph 0.16$ & $-0.88$ & $-0.70$ \\

\cmidrule[0.2pt]{1-6}
\cmidrule[0.2pt]{8-13}Full & $\ph 0.57$ & $-0.25$ & $\ph 0.48$  & $-0.88$ & $-0.08$ & $$ &
Full & $\ph 0.61$ & $-0.59$ & $\ph 0.11$ & $-0.88$ & $-0.75$ \\
\bottomrule
\end{tabular}
\caption{Numerical values for the $1/m$-expansion of $\gamma_{\text{E1M2}}$. The dots mark entries that do not exist e.g. the $\epsilon^3_{\text{Tree}}$ starts at NLO and therefore does not have a LO contribution.}
\label{tab:HB-gamma-E1M2}
\end{table}

\begin{table}[ht] \centering
\renewcommand{\arraystretch}{0.75} 
\begin{tabular}{@{}lllllllllllll@{}}
\toprule
$\gamma_{\text{M1E2}}^{(p)}$& $\ph q^3$ & $\ph q^4$ & $\ph \epsilon^3_{\text{Loop}}$& $\ph \epsilon^3_{\text{Tree}}$ & $\ph \text{Full}$ & \hspace{1cm} &
$\gamma_{M1E2}^{(n)}$& $\ph q^3$ & $\ph q^4$ & $\ph \epsilon^3_{\text{Loop}}$& $\ph \epsilon^3_{\text{Tree}}$ & $\ph \text{Full}$\\

\cmidrule[0.2pt]{1-6}
\cmidrule[0.2pt]{8-13}$\operatorname{LO}$ & $\ph 1.16$ & $\ph \ldots$ & $-0.21$ & $\ph \ldots$ & $\ph 0.95$ & $$ &
$\operatorname{LO}$ & $\ph 1.16$ & $\ph \ldots$ & $-0.21$ & $\ph \ldots$ & $\ph 0.95$ \\
$\operatorname{NLO}$& $\ph 0.76$ & $\ph 1.03$ & $-0.11$& $\ph 1.89$ & $\ph 3.56$ & $$ &
$\operatorname{NLO}$& $\ph 1.30$ & $\ph 0.96$ & $-0.08$ & $\ph 1.89$ & $\ph 4.06$ \\
$\operatorname{N}^2\!\operatorname{LO}$ & $\ph 0.89$ & $\ph 0.66$ & $-1.32$& $\ph 1.59$ & $\ph 1.82$ & $$ &
$\operatorname{N}^2\!\operatorname{LO}$ & $\ph 1.36$ & $\ph 0.87$ & $-0.33$ & $\ph 1.59$ & $\ph 3.48$ \\
$\operatorname{N}^3\!\operatorname{LO}$ & $\ph 0.98$ & $\ph 0.52$ & $-1.02$ & $\ph 1.82$ & $\ph 2.31$ & $$ &
$\operatorname{N}^3\!\operatorname{LO}$ & $\ph 1.37$ & $\ph 0.73$ & $-0.25$ & $\ph 1.82$ & $\ph 3.67$ \\
$\operatorname{N}^4\!\operatorname{LO}$ & $\ph 0.95$ & $\ph 0.56$ & $-0.80$ & $\ph 1.71$ & $\ph 2.42$ & $$ &
$\operatorname{N}^4\!\operatorname{LO}$ & $\ph 1.36$ & $\ph 0.76$ & $-0.26$ & $\ph 1.71$ & $\ph 3.57$ \\
$\operatorname{N}^5\!\operatorname{LO}$ & $\ph 0.95$ & $\ph 0.56$ & $-0.86$  & $\ph 1.76$ & $\ph 2.40$ & $$ &
$\operatorname{N}^5\!\operatorname{LO}$ & $\ph 1.36$ & $\ph 0.76$ & $-0.29$ & $\ph 1.76$ & $\ph 3.59$ \\

\cmidrule[0.2pt]{1-6}
\cmidrule[0.2pt]{8-13}Full & $\ph 0.95$ & $\ph 0.56$ & $-0.79$  & $\ph 1.74$ & $\ph 2.47$ & $$ &
Full & $\ph 1.36$ & $\ph 0.76$ & $-0.27$ & $\ph 1.74$ & $\ph 3.59$ \\
\bottomrule
\end{tabular}
\caption{Numerical values for the $1/m$-expansion of $\gamma_{\text{M1E2}}$. The dots mark entries that do not exist e.g. the $\epsilon^3_{\text{Tree}}$ starts at NLO and therefore does not have a LO contribution.}
\label{tab:HB-gamma-M1E2}
\end{table}

We first consider the convergence of the $1/m$-expansion of the individual contributions 
from the $q^3$-, $q^4$- and $\epsilon^3$-loop diagrams, and from the $\Delta$-tree-level terms.
In general, the convergence is rather slow.
The most rapid convergence is observed for the $q^3$- and $q^4$-loops.
Sometimes (e.g.~for $\alpha_{E1}$, $\gamma_{M1E2}$),
the expanded value approaches the ``exact'' one already at NLO-N$^2$LO.
In other cases, the expanded values oscillate at lower $1/m$-orders, especially
when the resulting value is small due to cancellations among various diagrams.
It is natural to expect a slower convergence for the diagrams with $\Delta$-lines
as the formal expansion parameter $\Delta/m$ is roughly twice as
large as $M /m$.
Nevertheless, the expansion for the tree-$\Delta$-graphs converges, in general, only slightly worse
than the $\pi N$-loops ($\beta_{M1}$ is accidentally $m$-independent).
On the other hand, for the $\Delta\pi$-loops, the convergence is very poor.
This set of diagrams comprises loops with one, two and three $\Delta$-lines,
and cancellations among them occur quite often.
Some of the values strongly oscillate and one hardly sees
a sign of convergence even at N$^5$LO, e.g. for $\gamma_{M1M1}$, $\gamma_{E1M2}$.

\begin{figure}[ht]\centering
\includegraphics[width=0.96\textwidth]{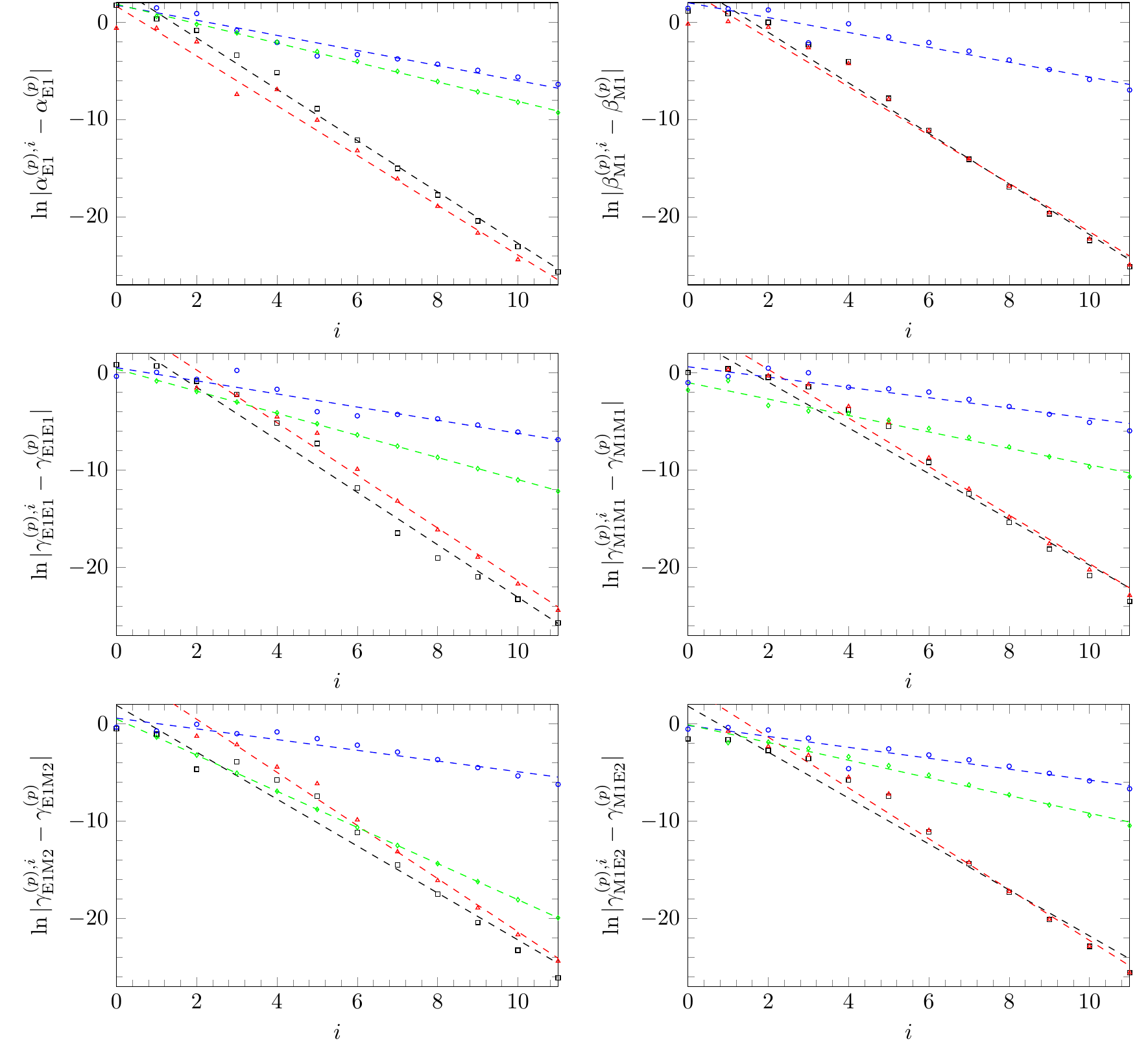}
\caption{Logarithmic difference of the absolute value between the HB-expanded contributions and the final, 
non-expanded value for the spin-independent and spin-dependent dipole polarizabilities of the proton. 
$i$ stands for the $i$-th order in the HB expansion. The black squares represent the $q^3$ contribution, 
the red triangles represent the $q^4$ contribution, the blue circles represent the $\epsilon^3$-loop contribution 
and the green diamonds represent the $\epsilon^3$-tree contributions. The dashed lines stand for the corresponding linear regression.}
\label{Fig:1_over_m_convergence_proton}
\end{figure}

\begin{figure}[ht]\centering
\includegraphics[width=0.96\textwidth]{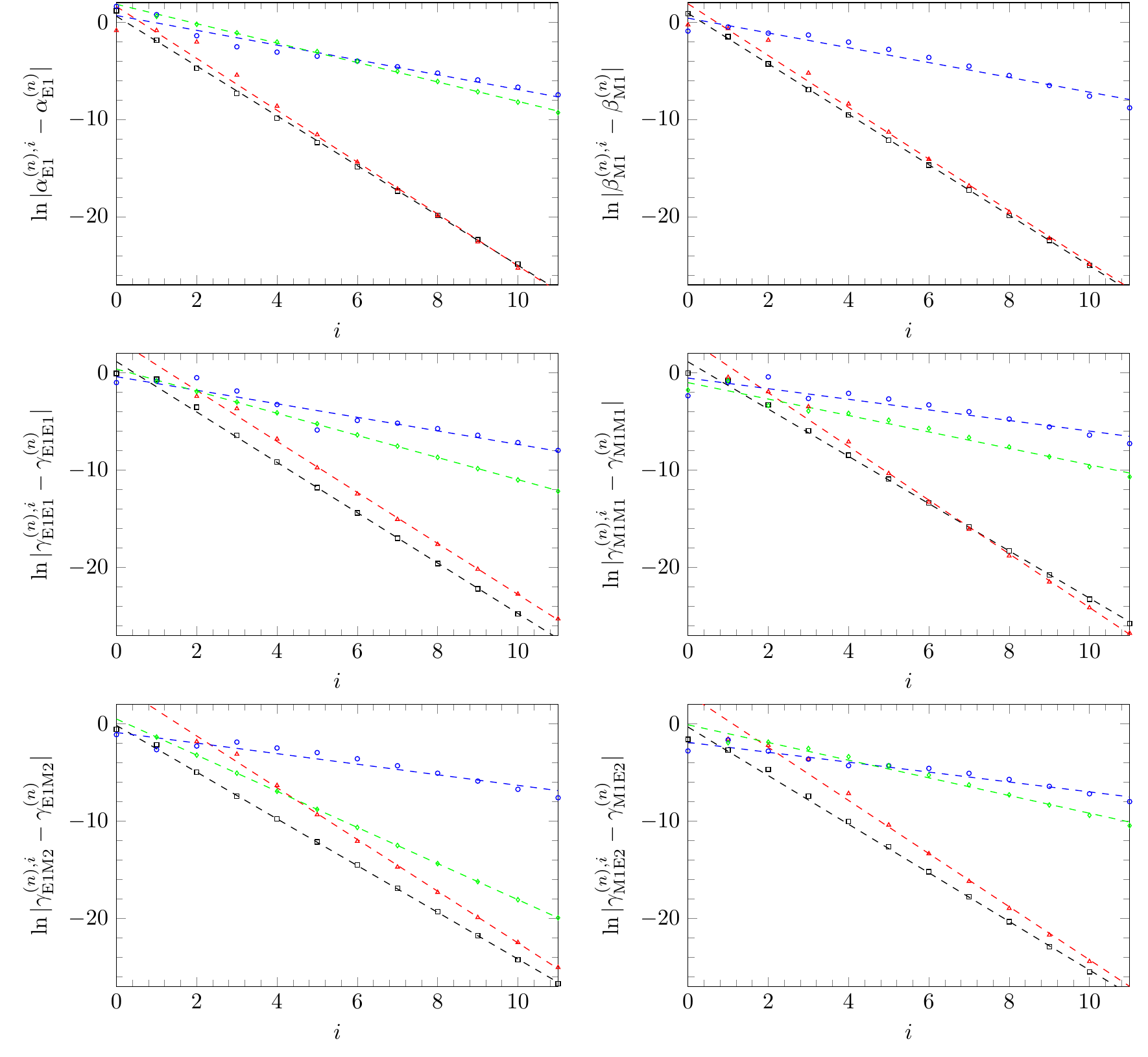}
\caption{Logarithmic difference of the absolute value between the HB-expanded contributions and the final, 
non-expanded value for the spin-independent and spin-dependent dipole polarizabilities of the neutron. 
$i$ stands for the $i$-th order in the HB expansion. The black squares represent the $q^3$ contribution, 
the red triangles represent the $q^4$ contribution, the blue circles represent the $\epsilon^3$-loop 
contribution and the green diamonds represent the $\epsilon^3$-tree contributions. 
The dashed lines stand for the corresponding linear regression.}
\label{Fig:1_over_m_convergence_neutron}
\end{figure}

Nevertheless, we have checked that the $1/m$-expansion converges in principle (formally) for all diagrams.
This is illustrated in Figs.~\ref{Fig:1_over_m_convergence_proton},~\ref{Fig:1_over_m_convergence_neutron},
where the logarithm of the remainder in the $1/m$-series is plotted against the order of expansion.
As one can see from the plots, the expanded $\Delta$-loops approach their unexpanded values very slowly,
making
such an expansion impractical.
Note that contributions of the $\Delta$-loops are smaller for the spin-dependent polarizabilities.

We now consider the $1/m$-expansion of the sum of all contributions to the nucleon polarizabilities.
As one can see in Tables~\ref{tab:HB-alpha-E1},~\ref{tab:HB-beta-M1}, the electric and magnetic scalar polarizabilities
at NLO agree rather well with the unexpanded values (for the absolute difference is large but
the relative difference is small), while the individual contributions in some cases strongly oscillate.
Such an agreement is accidental. Moreover, e.g.~the $\beta^(p)_{M1}$ at N$^2$LO deviates significantly from the 
full result and approaches it again after several
oscillations. Nevertheless, these effects can be compensated by a
redefinition of the $q^4$ contact terms.

The situation is different for the spin-dependent polarizabilities, where the NLO values 
in most cases deviate rather strongly from the unexpanded result, see Tables~\ref{tab:HB-gamma-E1E1}-\ref{tab:HB-gamma-M1E2}.
It should be emphasized that these differences can be absorbed into contact terms only at order $q^5$.

Summarizing, we conclude that the $1/m$-expansion (and, hence, the heavy-baryon scheme)
is rather inefficient for calculating nucleon polarizabilites
in the $\Delta$-full approach, which is in line with the results of the heavy-baryon calculations mentioned above.
On the other hand, the small-scale expansion seems to converge reasonably well.

\section{Summary and outlook}
\label{sec:summary}
In this work, we have presented various nucleon polarizabilities obtained within
covariant chiral perturbation theory with explicit $\Delta$(1232) degrees of freedom,
calculated up to order $\mathcal{O}(\epsilon^3+q^4)$ in the small-scale expansion.
The theoretical errors were estimated by combining the uncertainties of the input parameters
and the errors due to the truncation of the small-scale expansion calculated using 
a Bayesian model.
The results were compared with the $\Delta$-less approach at order $\mathcal{O}(q^3)$ and $\mathcal{O}(q^4)$,
as well as with the empirical values and other theoretical approaches (in particular, with 
the $\delta$-counting $\Delta$-full scheme and the fixed-$t$ dispersion-relations method).

The general conclusion of this study is that the $\Delta$-full scheme that we adopt 
is quite efficient for analyzing the nucleon polarizabilites.
It shows reasonable convergence, and the obtained results agree well
with experiment  and
the fixed-$t$ dispersion-relations values. 
The results obtained in the $\Delta$-less approach are considerably
worse both from the point of
view of convergence and agreement with experiment.

The scalar dipole polarizabilites were used as an input to adjust four low energy constants 
appearing at order $\mathcal{O}(q^4)$ in the effective Lagrangian. Therefore, we were not concerned with 
the issue of convergence for these quantities (although the convergence is far from being satisfactory).

Our predictions for the dipole spin polarizabilites $\gamma_{E1E1}$, $\gamma_{E1M2}$, $\gamma_{M1E2}$
obtained in the $\Delta$-full scheme
agree with experimental values of  \cite{Martel:2014pba} and are slightly larger for $\gamma_{M1M1}$.
The agreement is somewhat worse with the analysis of the recent MAMI experiment  \cite{Martel:2014pba}.
The same pattern is observed in the comparison with the fixed-$t$ dispersion-relations results.
On the other hand, the predictions for the forward and backward spin polarizabilites $\gamma_0$ and $\gamma+\pi$
differ noticeably from the empirical values.
Such a deviation can be explained by a sizable contributions of the ``induced'' electric 
$\gamma N \Delta$-coupling observed for these linear combinations.                                
This effect if formally suppressed by a factor of $1/m$, but
numerically it turns out to
be sizable. In the $\delta$-counting scheme of \cite{Lensky:2015awa}, the electric 
$\gamma N \Delta$-coupling constant is adjusted to data and appears to be rather small.
This is an indication that including higher-order $\Delta$-pole graphs in our scheme
might improve the results.

Due to small contributions of the $q^4$-loops to spin polarizabilites, a rather rapid convergence is achieved for all of them.

We have also analyzed several higher-order polarizabilites.
A nice convergence rate is observed for all polarizabilites calculated within the $\Delta$-full scheme.
This is, however, not the case for the higher-order scalar polarizabilites calculated in the $\Delta$-less approach.
This pattern can be understood in terms of the $\Delta$-resonance
saturation of the low-energy constants $c_2$ and $c_3$. The main
effect of the explicit treatment of the $\Delta$ is that some parts of
the $q^4$-loops are shifted to the $\epsilon^3$-loops. 

For the scalar quadrupole and dipole-dispersive polarizabilites, our predictions 
agree with the results of the fixed-$t$ dispersion-relations approach.
However, our results differ noticeably from the ones obtained in the
$\delta$-counting scheme.
This difference is caused not only by the $q^4$-loop contributions, but also
by the $\epsilon^3$-loops with multiple $\Delta$-lines, which
contribute at higher orders in the $\delta$-counting scheme.
This points to the importance of such terms also for higher-order polarizabilites.

We also studied the $Q^2$-dependence of the nucleon polarizabilites by considering 
generalized scalar and spin polarizabilities. We found that the $Q^2$-dependence
of the magnetic scalar polarizabilites is significantly different in the $\Delta$-full
and the $\Delta$-less approach.
We also observed a substantial deviation of the
$\mathcal{O}(\epsilon^3+q^4)$ results for the $Q^2$-dependent polarizabilities $\gamma_0$ for the proton and for the neutron as well as
$\delta_{\text{LT}}$ for the neutron from the available experimental
data and no improvement
compared to the $\mathcal{O}(\epsilon^3)$-results. We expect that taking into account
the $\mathcal{O}(\epsilon^4)$ terms (in particular the tree-level graphs) might
improve the description of the data.

An alternative way to study the electromagnetic structure of the nucleon is to 
consider dynamical (energy-dependent) polarizabilites.
We have analyzed the energy dependence of the dipole and spinless quadrupole polarizabilites
and compared them with other theoretical investigations.
In particular, we observed a rather large deviation 
from the fixed-$t$ dispersion-relations approach at energies $\omega_\text{CM}\gtrsim150-200$~MeV 
(in some cases for $\omega_\text{CM}\gtrsim100$~MeV), which indicates the slow convergence
of the small-scale expansion in that energy region.

Finally, we have analyzed the convergence of the $1/m$-expansion of the results obtained
in the covariant calculation for various polarizabilites.
Such an scheme allows one to see how reliable the
heavy-baryon expansion is for the evaluation of the nucleon polarizabilites.
We considered the expansion up to N$^5$LO.
Our conclusion is that 
the heavy-baryon expansion 
is not efficient for calculating nucleon polarizabilites
in the $\Delta$-full approach.
Nevertheless, the small-scale expansion seems to converge reasonably well.

A natural extension of the current work towards increasing accuracy of the results
follows from the discussion above. We expect a better accuracy and a better agreement with
the experimental data after including the $\Delta$-tree-level graphs of order $\mathcal{O}(\epsilon^4)$
with the electric $\gamma N \Delta$-coupling
as well as the $\mathcal{O}(\epsilon^4)$-loop diagrams, that is performing a complete $\mathcal{O}(\epsilon^4)$ calculation.

\acknowledgments
We are grateful to Jambul Gegelia for helpful discussions and to
Veronique Bernard and Ulf-G.~Mei{\ss}ner for sharing their insights
into the considered topics.
This work was supported in part by BMBF (contract No. 05P18PCFP1), by
DFG (Grant No. 426661267) and by DFG
   through funds provided to the Sino-German CRC 110 “Symmetries and
   the Emergence of Structure in QCD” (Grant No. TRR110). 

\newpage
\appendix

\section{\texorpdfstring{$q^3$}{q3}-Values}\label{sec:q3-loop} 

The analytic expressions for $\alpha_{\text{E1}}$, $\beta_{\text{M1}}$, $\alpha_{\text{E2}}$, $\beta_{\text{M2}}$ 
and a linear combination of the spin-dependent first order polarizabilities for both proton and neutron were already 
calculated in \cite{Lensky:2015awa} but are also given here for completeness. For convenience we define 
$\Xi_2 = \Xi \left(m^2,m,M \right) /(\mu^2-4)$ where $\mu = M /m$, $\Xi_1 = \ln\left(\mu\right)$
and
\begin{align}
 \Xi(p^2,m_1,m_2) &= \frac{1}{p^2}\sqrt{\lambda(m_1^2,m_2^2,p^2)}\ln\left(\frac{m_1^2+m_2^2+\sqrt{\lambda(m_1^2,m_2^2,p^2)}-p^2}{2m_1m_2}\right)\,,\nonumber\\
\lambda(a,b,c) &= a^2+b^2+c^2-2ab-2bc-2ac\,.
\end{align}

\subsection{Proton values}\label{sub:proton-q3} 

\subsubsection{Spin-independent first order polarizabilities}\label{subsub:proton-spin-independent-1-q3} 

\begin{align*} 
&\alpha_{\text{E1}}^{(p)} 
  =  \; 
 \frac{e^2 g_A^2}{192 \pi^3 F^2 \mu^2 \left(\mu^2-4\right)^2 m} \left[\vphantom{\frac{e}{m}} 2\left(-9 \mu^{10}+110 \mu^8-479 \mu^6+870 \mu^4-590 \mu^2+80\right) \Xi_2\right. \\
&\;\;\;\; \left.\left. + \left(-18 \mu^8+157 \mu^6-407 \mu^4+304\mu^2\right) \vphantom{\frac{e}{m}}\right]
+\frac{e^2 g_A^2\left(9 \mu^4-20 \mu^2+9\right) \Xi_1}{96 \pi^3 F^2m} \right.\,,
\\
&\beta_{\text{M1}}^{(p)}
  =   
-\frac{e^2 g_A^2}{192 \pi^3 F^2 \mu^2 \left(\mu^2-4\right) m} \left[\vphantom{\frac{e}{m}} 2\left(27 \mu^8-212 \mu^6+471 \mu^4-246 \mu^2+2\right) \Xi_2\right. \\
&\;\;\;\; \left. \left. - \left(54 \mu^6-235 \mu^4+127\mu^2 \right) \vphantom{\frac{e}{m}}\right]
+\frac{e^2 g_A^2 \left(27 \mu^4-50 \mu^2+9\right) \Xi_1}{96 \pi^3 F^2 m} \right.\,.
\end{align*}

\subsubsection{Spin-independent second order polarizabilities}\label{subsub:proton-spin-independent-2-q3} 

\begin{align*} 
&\alpha_{\text{E2}}^{(p)} 
  =   
-\frac{e^2 g_A^2}{480 \pi^3 F^2 \mu^4 \left(\mu^2-4\right)^3 m^3} \left[ \vphantom{\frac{e}{m}} \left(450 \mu^{14}-7341 \mu^{12}+16584 \mu^{10}-143010 \mu^8  \right.\right. \vphantom{\frac{e}{m}} \\
&\;\;\;\;\left.\left.+212940 \mu^6-132300 \mu^4+18624 \mu^2+1344\right) \Xi_2
- 450 \mu^{12}-5766 \mu^{10}+26437 \mu^8\right. \vphantom{\frac{e}{m}} \\
&\;\;\;\;\left. -50449 \mu^6+34592 \mu^4-3536\mu^2  \vphantom{\frac{e}{m}} \right] +\frac{e^2 g_A^2\Xi_1}{160 \pi^3 F^2 m^3} \left(150 \mu^4-347 \mu^2+170\right) \,,\\
\\
&\beta_{\text{M2}}^{(p)} 
  =   
-\frac{e^2 g_A^2}{480 \pi^3 F^2 \mu^4 \left(\mu^2-4\right)^2 m^3} \left[ \vphantom{\frac{e}{m}} \left(990 \mu^{12}-11781 \mu^{10}+49020 \mu^8-81330 \mu^6  \right.\right. \vphantom{\frac{e}{m}} \\
&\;\;\;\;\left.\left. +43020 \mu^4-2700 \mu^2+144\right) \Xi_2
+ 990 \mu^{10}-8316 \mu^8+20000 \mu^6-11137 \mu^4+92\mu^2  \vphantom{\frac{e}{m}} \right]\\
&\;\;\;\;
+\frac{e^2 g_A^2\Xi_1}{160 \pi^3F^2 m^3} \left(330 \mu^4-627 \mu^2+170\right) \,,\\
\\
&\alpha_{\text{E1}\nu}^{(p)} 
  =  \; 
 \frac{e^2 g_A^2}{5760 \pi^3 F^2 \mu^4 \left(\mu^2-4\right)^4 m^3} \left[ \vphantom{\frac{e}{m}} \left(-5130 \mu^{16}+102987 \mu^{14}-841656 \mu^{12}\right.\right. \vphantom{\frac{e}{m}} \\
 &\;\;\;\;\left.\left. +3561462 \mu^{10} -8161020 \mu^8+9522420 \mu^6-4534128 \mu^4+397824 \mu^2+6912\right) \Xi_2\right. \vphantom{\frac{e}{m}} \\
 &\;\;\;\;\left.
-5130 \mu^{14}+85032 \mu^{12}-544482 \mu^{10}+1658251 \mu^8-2335148 \mu^6+1156768 \mu^4\right. \vphantom{\frac{e}{m}} \\
&\;\;\;\;\left.-45504\mu^2  \vphantom{\frac{e}{m}} \right] +\frac{e^2 g_A^2\Xi_1}{1920 \pi^3 F^2 m^3} \left(1710 \mu^4-3549\mu^2+1210\right) \,,\\
\\
&\beta_{\text{M1}\nu}^{(p)} 
  =   
-\frac{e^2 g_A^2}{1920 \pi^3 F^2 \mu^4 \left(\mu^2-4\right)^3 m^3} \left[ \vphantom{\frac{e}{m}} \left(1890 \mu^{14}-30129 \mu^{12}+184876 \mu^{10}-538370 \mu^8  \right.\right. \vphantom{\frac{e}{m}} \\
&\;\;\;\;\left.\left. +730660 \mu^6-374140 \mu^4+33056 \mu^2+896\right) \Xi_2
+1890 \mu^{12}-23514 \mu^{10}+102731 \mu^8\right. \vphantom{\frac{e}{m}} \\
&\;\;\;\;\left. -178873 \mu^6+96560 \mu^4-4032\mu^2  \vphantom{\frac{e}{m}} \right] 
+\frac{e^2 g_A^2\Xi_1}{1920 \pi^3 F^2 m^3} \left(1890 \mu^4-3669 \mu^2+1210\right) \,.
\end{align*}

\subsubsection{Spin-dependent first order polarizabilities}\label{subsub:proton-spin-dependent-1-q3} 

\begin{align*}
&\gamma_{\text{E1E1}}^{(p)}
 =  \;  
 \frac{e^2 g_A^2}{384 \pi^3 F^2 \mu^2 \left(\mu^2-4\right)^2 m^2} \left[\vphantom{\frac{e}{m}} 2\left(-9 \mu^{10}+134 \mu^8-739 \mu^6+1790 \mu^4  -1650 \mu^2\right.\right. \vphantom{\frac{e}{m}} \\
&\;\;\;\;\left.\left.+264\right) \Xi_2- \left(18\mu^8-205 \mu^6+764 \mu^4-900 \mu^2+80\right) \vphantom{\frac{e}{m}}\right]  +\frac{e^2 g_A^2 \Xi_1}{192 \pi^3 F^2 m^2}\left(9 \mu^4\right. \vphantom{\frac{e}{m}} \\
&\;\;\;\;\left.\left.-44 \mu^2+29\right)\right. \,,\\ 
\\
&\gamma_{\text{M1M1}}^{(p)}
  =  \;  
 \frac{e^2 g_A^2}{384 \pi^3 F^2 \mu^2 \left(\mu^2-4\right)^2 m^2} \left[ \vphantom{\frac{e}{m}}2\left(-63 \mu^{10}+744 \mu^8-3059 \mu^6+4970 \mu^4 -2534 \mu^2\right.\right. \vphantom{\frac{e}{m}} \\
&\;\;\;\;\left.\left.+152\right) \Xi_2 - \left(126 \mu^8-1047 \mu^6+2462 \mu^4-1308 \mu^2+16\right) \vphantom{\frac{e}{m}}\right] +\frac{e^2 g_A^2 \Xi_1}{192 \pi^3 F^2 m^2}\left(63 \mu^4\right. \vphantom{\frac{e}{m}} \\
&\;\;\;\;\left.\left.-114 \mu^2+29\right) \right. \,,\\
\\
&\gamma_{\text{E1M2}}^{(p)}
  =   
-\frac{e^2 g_A^2}{384 \pi^3 F^2 \mu^2 \left(\mu^2-4\right)^2 m^2} \left[\vphantom{\frac{e}{m}} 2\left(\left(3 \left(\mu^2-5\right) \left(3 \mu^4-19 \mu^2+34\right) \mu^2+46\right) \mu^2 \right.\right. \vphantom{\frac{e}{m}} \\
&\;\;\;\;\left.\left.+56\right) \Xi_2
+ \left(-18 \mu^8+141 \mu^6-284 \mu^4+4 \mu^2+16\right) \vphantom{\frac{e}{m}}\right]
+\frac{e^2 g_A^2 \Xi_1}{64 \pi^3 F^2m^2} \left(3 \mu^4-4 \mu^2-1\right) \,,\\
\\
&\gamma_{\text{M1E2}}^{(p)}
  =   \; 
 \frac{e^2 g_A^2}{384 \pi^3 F^2 \mu^2 \left(\mu^2-4\right) m^2} \left[ \vphantom{\frac{e}{m}}2\left(9 \mu^8-68 \mu^6+141 \mu^4-66 \mu^2+6\right) \Xi_2 + \left(18 \mu^6\right.\right. \vphantom{\frac{e}{m}} \\
&\;\;\;\;\left.\left.-73\mu^4+30 \mu^2-4\right) \vphantom{\frac{e}{m}}\right]-\frac{e^2 g_A^2 \Xi_1}{192 \pi^3 F^2 m^2} \left(9 \mu^4-14 \mu^2+3\right)\,.
\end{align*}

\subsubsection{Spin-dependent second order polarizabilities}\label{subsub:proton-spin-dependent-2-q3} 

\begin{align*} 
&\gamma_{\text{E2E2}}^{(p)} 
  =  
-\frac{e^2 g_A^2}{138240 \pi^3 F^2 \mu^4 \left(\mu^2-4\right)^3 m^4} \left[\vphantom{\frac{e}{m}}\left(6030 \mu^{14}-107814 \mu^{12}+764796 \mu^{10} \right.\right. \vphantom{\frac{e}{m}} \\
&\;\;\;\;\left.\left. -2694300 \mu^8+4759500 \mu^6-3591840 \mu^4+577056 \mu^2+36096\right)\Xi_2  +6030 \mu^{12}\right. \vphantom{\frac{e}{m}} \\
&\;\;\;\;\left.-86709 \mu^{10}+461861 \mu^8-1079504 \mu^6+954200\mu^4-100864 \mu^2 -3328 \vphantom{\frac{e}{m}}\right] \\
&\;\; +\frac{e^2 g_A^2\Xi_1}{23040 \pi^3 F^2 m^4} \left(1005 \mu^4-3899 \mu^2+2530\right) \,,\\
\\
&\gamma_{\text{M2M2}}^{(p)} 
  =  
-\frac{e^2 g_A^2}{138240 \pi^3 F^2 \mu^4 \left(\mu^2-4\right)^3 m^4} \left[\vphantom{\frac{e}{m}}\left(29250 \mu^{14}-462354 \mu^{12}+2802636 \mu^{10} \right.\right. \vphantom{\frac{e}{m}} \\
&\;\;\;\;\left.\left.-8007300 \mu^8+10510020 \mu^6-5011200 \mu^4+379296 \mu^2+7296\right)\Xi_2 +29250 \mu^{12}\right. \vphantom{\frac{e}{m}} \\
&\;\;\;\;\left.-359979 \mu^{10}+1545103 \mu^8-2601080 \mu^6+1278024 \mu^4 -40480 \mu^2-256 \vphantom{\frac{e}{m}}\right] \\
&\;\;+\frac{e^2 g_A^2\Xi_1}{23040 \pi^3 F^2 m^4} \left(4875 \mu^4-8809 \mu^2+2530\right) \,,\\
\\
&\gamma_{\text{E2M3}}^{(p)} 
  =  \; 
 \frac{e^2 g_A^2}{69120 \pi^3 F^2 \mu^4 \left(\mu^2-4\right)^3 m^4} \left[\vphantom{\frac{e}{m}}\left(-1710 \mu^{14}+26382 \mu^{12}-153696 \mu^{10}\right.\right. \vphantom{\frac{e}{m}} \\
 &\;\;\;\;\left.\left.+407652 \mu^8 -447780 \mu^6+91920 \mu^4+28032 \mu^2+3072\right) \Xi_2
-1710 \mu^{12}+20397 \mu^{10} \right. \vphantom{\frac{e}{m}} \\
&\;\;\;\;\left.-82493 \mu^8+119300 \mu^6-19808 \mu^4-7040 \mu^2-512 \vphantom{\frac{e}{m}}\right] \\
&\;\;+\frac{e^2 g_A^2\Xi_1}{11520 \pi^3 F^2 m^4} \left(285 \mu^4-407 \mu^2-32\right) \,,\\
\\
&\gamma_{\text{M2E3}}^{(p)} 
 =  \; 
\frac{e^2 g_A^2}{69120 \pi^3 F^2 \mu^4 \left(\mu^2-4\right)^2 m^4} \left[\vphantom{\frac{e}{m}}\left(912 \mu^{12}-11118 \mu^{10}+42072 \mu^8-58260 \mu^6 \right.\right. \vphantom{\frac{e}{m}} \\
&\;\;\;\;\left.\left.+19860 \mu^4-1920 \mu^2-768\right) \Xi_2 +990 \mu^{10}-7653 \mu^8+15413 \mu^6-4112 \mu^4 +832 \mu^2\right. \vphantom{\frac{e}{m}} \\
&\;\;\;\; \left.+128\vphantom{\frac{e}{m}} \right] -\frac{e^2 g_A^2\Xi_1}{11520 \pi^3 F^2 m^4} \left(165 \mu^4-203 \mu^2+32\right) \,,\\
\\
&\gamma_{\text{E1E1}\nu}^{(p)} 
  =  \; 
 \frac{e^2 g_A^2}{46080 \pi^3 F^2 \mu^4 \left(\mu^2-4\right)^4 m^4} \left[\vphantom{\frac{e}{m}}\left(-33750 \mu^{16}+688446 \mu^{14}-5745756 \mu^{12} \right.\right. \vphantom{\frac{e}{m}} \\
 &\;\;\;\;\left.\left.+25026300 \mu^{10}-59825628 \mu^8+74731920 \mu^6-40404960 \mu^4+4788480\mu^2 \right.\right. \vphantom{\frac{e}{m}} \\
 &\;\;\;\;\left.\left.+299520\right) \Xi_2
-33750 \mu^{14}+570321 \mu^{12}-3752313 \mu^{10}+11906948 \mu^8 -18015768 \mu^6\right. \vphantom{\frac{e}{m}} \\
&\;\;\;\;\left.+10422016 \mu^4-748416 \mu^2-64512 \vphantom{\frac{e}{m}}\right] +\frac{e^2 g_A^2\Xi_1}{7680 \pi^3 F^2 m^4} \left(5625 \mu^4-13491 \mu^2+6038\right) \,,\\
\\
&\gamma_{\text{M1M1}\nu}^{(p)} 
  =  \; 
 \frac{e^2 g_A^2}{15360 \pi^3 F^2 \mu^4 \left(\mu^2-4\right)^4 m^4} \left[\vphantom{\frac{e}{m}}\left(-18270 \mu^{16}+364014 \mu^{14}-2946868 \mu^{12}  \right.\right. \vphantom{\frac{e}{m}} \\
 &\;\;\;\;\left.\left.+12320172 \mu^{10}-27795996 \mu^8+31816880 \mu^6-14929824 \mu^4+1383552\mu^2\right.\right. \vphantom{\frac{e}{m}} \\
 &\;\;\;\;\left.\left.+45056\right) \Xi_2
-18270 \mu^{14}+300069 \mu^{12}-1898193 \mu^{10}+5684492 \mu^8-7821048 \mu^6\right. \vphantom{\frac{e}{m}} \\
&\;\;\;\;\left.+3793952 \mu^4-184320 \mu^2-1024 \vphantom{\frac{e}{m}}\right] +\frac{e^2 g_A^2\Xi_1}{7680 \pi^3 F^2 m^4} \left(9135 \mu^4-17577 \mu^2+6038\right) \,,\\
\\
&\gamma_{\text{E1M2}\nu}^{(p)} 
  =   
-\frac{e^2 g_A^2}{115200 \pi^3 F^2 \mu^4 \left(\mu^2-4\right)^4 m^4} \left[\vphantom{\frac{e}{m}}\left(72090 \mu^{16}-1426818 \mu^{14}+11444436 \mu^{12}\right.\right. \vphantom{\frac{e}{m}} \\
&\;\;\;\;\left.\left.-47196324 \mu^{10}+104160132 \mu^8-114538560 \mu^6+48945312 \mu^4-3267456 \mu^2\right.\right. \vphantom{\frac{e}{m}} \\
&\;\;\;\;\left.\left.+3072\right) \Xi_2
+72090 \mu^{14}-1174503 \mu^{12}+7339287 \mu^{10}-21533144 \mu^8+28502616 \mu^6\right. \vphantom{\frac{e}{m}} \\
&\;\;\;\;\left.-12369760 \mu^4+243968 \mu^2+7168 \vphantom{\frac{e}{m}}\right] +\frac{e^2 g_A^2\Xi_1}{19200 \pi^3 F^2 m^4} \left(12015 \mu^4-21533 \mu^2+5922\right) \,,\\
\\
&\gamma_{\text{M1E2}\nu}^{(p)} 
  =   
-\frac{e^2 g_A^2}{115200 \pi^3 F^2 \mu^4 \left(\mu^2-4\right)^3 m^4} \left[\vphantom{\frac{e}{m}}\left(20790 \mu^{14}-358638 \mu^{12}+2436924 \mu^{10} \right.\right. \vphantom{\frac{e}{m}} \\
&\;\;\;\;\left.\left. -8138508\mu^8 + 13400220\mu^6 -9090480\mu^4+1224672\mu^2 +54912\right) \Xi_2+20790 \mu^{12}\right. \vphantom{\frac{e}{m}} \\
&\;\;\;\;\left.-285873 \mu^{10}+1438497 \mu^8-3110380 \mu^6+2403672 \mu^4-197280 \mu^2+3328\vphantom{\frac{e}{m}}\right] \\
&\;\;+\frac{e^2 g_A^2 \Xi_1}{19200 \pi^3 F^2 m^4} 7\left(495 \mu^4-1609 \mu^2+846\right) \,.
\end{align*}

\subsection{Neutron values}\label{sub:neutron-q3} 

\subsubsection{Spin-independent first order polarizabilities}\label{subsub:neutron-spin-independent-1-q3} 

\begin{align*} 
&\alpha_{\text{E1}}^{(n)} 
  =  
 \frac{e^2 g_A^2}{192 \pi^3 F^2 \mu^2 \left(\mu^2-4\right)^2 m} \left[\vphantom{\frac{e}{m}} 2\left(-3 \mu^6+30 \mu^4-98 \mu^2+80\right)\Xi_2
- \left(\mu^4-16\mu^2\right) \vphantom{\frac{e}{m}}\right] \\
&\;\;+\frac{e^2 g_A^2 \Xi_1}{32 \pi^3 F^2 m}\,,
\\
&\beta_{\text{M1}}^{(n)}
  =   
-\frac{e^2 g_A^2}{192 \pi^3 F^2 \mu^2 \left(\mu^2-4\right) m} \left[\vphantom{\frac{e}{m}} 2\left(3 \left(\mu^2-6\right) \mu^2+2\right)\Xi_2
- 11 \mu^2 \vphantom{\frac{e}{m}}\right]
+\frac{e^2 g_A^2 \Xi_1}{32 \pi^3 F^2 m}\,.
\end{align*}

\subsubsection{Spin-independent second order polarizabilities}\label{subsub:neutron-spin-independent-2-q3} 

\begin{align*} 
&\alpha_{\text{E2}}^{(n)} 
  =   
-\frac{e^2 g_A^2}{480 \pi^3 F^2 \mu^4 \left(\mu^2-4\right)^3 m^3} \left[\vphantom{\frac{e}{m}}\left(120\mu^{10}-1680 \mu^8+8400\mu^6-17700 \mu^4 +9984\mu^2  \right.\right. \vphantom{\frac{e}{m}}\\
&\;\;\;\;\left.\left.+1344\right) \Xi_2
+113 \mu^8-1119 \mu^6+4272 \mu^4-3536\mu^2\vphantom{\frac{e}{m}}\right]
+\frac{e^2 g_A^2 \Xi_1}{4 \pi^3 F^2 m^3} \,,\\
\\
&\beta_{\text{M2}}^{(n)} 
  =  \; 
 \frac{e^2 g_A^2}{480 \pi^3 F^2 \mu^4\left(\mu^2-4\right)^2 m^3} \left[\vphantom{\frac{e}{m}}\left(-120\mu^8 + 1200\mu^6 - 3600\mu^4 + 1500\mu^2 - 144\right) \Xi_2  \right. \vphantom{\frac{e}{m}}\\
&\;\;\;\;\left.-127 \mu^6+885 \mu^4-92\mu^2 \vphantom{\frac{e}{m}}\right]
+\frac{e^2 g_A^2 \Xi_1}{4 \pi^3 F^2 m^3} \,,\\
\\
&\alpha_{\text{E1}\nu}^{(n)} 
  =  \; 
 \frac{e^2 g_A^2}{5760 \pi^3 F^2 \mu^4 \left(\mu^2-4\right)^4 m^3} \left[\vphantom{\frac{e}{m}}\left(-240\mu^{12} + 4320\mu^{10} - 30240\mu^8 + 97380\mu^6   \right.\right. \vphantom{\frac{e}{m}}\\
&\;\;\;\;\left.\left.- 132528\mu^4 + 38784\mu^2 + 6912\right)\Xi_2
-261 \mu^{10}+4031 \mu^8-18540 \mu^6+28928 \mu^4 \right. \vphantom{\frac{e}{m}}\\
& \;\;\;\;\left.-3520\mu^2 \vphantom{\frac{e}{m}}\right]
+\frac{e^2 g_A^2 \Xi_1}{24 \pi^3 F^2 m^3} \,,\\
\\
&\beta_{\text{M1}\nu}^{(n)} 
  =   
-\frac{e^2 g_A^2}{1920 \pi^3 F^2 \mu^4 \left(\mu^2-4\right)^3 m^3} \left[\vphantom{\frac{e}{m}}\left(80\mu^{10} - 1120\mu^8 + 5600\mu^6 - 12340\mu^4 \right.\right. \vphantom{\frac{e}{m}}\\
&\;\;\;\;\left.\left. + 7136\mu^2  +896\right) \Xi_2
+73 \mu^8-687 \mu^6+3168 \mu^4-2752\mu^2 \vphantom{\frac{e}{m}}\right]
+\frac{e^2 g_A^2 \Xi_1}{24 \pi^3 F^2 m^3}\,.
\end{align*}

\subsubsection{Spin-dependent first order polarizabilities}\label{subsub:neutron-spin-dependent-1-q3} 

\begin{align*}
&\gamma_{\text{E1E1}}^{(p)}
  =   
-\frac{e^2 g_A^2}{192 \pi^3 F^2 \mu^2 \left(\mu^2-4\right) m^2} \left[\vphantom{\frac{e}{m}} \left(5 \left(\mu^2-6\right) \mu^2+22\right) \Xi_2
+ \left(10-7 \mu^2\right) \vphantom{\frac{e}{m}}\right] \\
&\;\;+\frac{5 e^2 g_A^2 \Xi_1}{192 \pi^3F^2 m^2} \,,\\ 
\\
&\gamma_{\text{M1M1}}^{(p)}
  =  \;  
 \frac{e^2 g_A^2}{192 \pi^3 F^2 \mu^2 \left(\mu^2-4\right)^2 m^2} \left[\vphantom{\frac{e}{m}} \left(-5 \mu^6+50 \mu^4-166 \mu^2+88\right) \Xi_2
- \left(3 \mu^4 \right. \right. \vphantom{\frac{e}{m}} \\
&\;\;\;\; \left.\left. -38 \mu^2+8\right) \vphantom{\frac{e}{m}}\right]
+\frac{5 e^2 g_A^2 \Xi_1}{192 \pi^3 F^2 m^2}\,,\\
\\
&\gamma_{\text{E1M2}}^{(p)}
  =  \; 
 \frac{e^2 g_A^2}{192 \pi^3 F^2 \mu^2 \left(\mu^2-4\right)^2 m^2} \left[\vphantom{\frac{e}{m}} \left(\mu^6-10 \mu^4+46 \mu^2-40\right) \Xi_2
- \left(\mu^4+10 \mu^2-8\right) \vphantom{\frac{e}{m}}\right] \\
&\;\;-\frac{e^2 g_A^2 \Xi_1}{192\pi^3 F^2 m^2} \,,\\
&\gamma_{\text{M1E2}}^{(p)}
  =  \; 
 \frac{e^2 g_A^2}{192 \pi^3 F^2 \mu^2 \left(\mu^2-4\right) m^2} \left[\vphantom{\frac{e}{m}} \left(\mu^4-6 \mu^2-2\right) \Xi_2
+ \left(3 \mu^2-2\right) \vphantom{\frac{e}{m}}\right] -\frac{e^2 g_A^2 \Xi_1}{192 \pi^3 F^2 m^2}\,.
\end{align*}

\subsubsection{Spin-dependent second order polarizabilities}\label{subsub:neutron-spin-dependent-2-q3} 

\begin{align*} 
&\gamma_{\text{E2E2}}^{(n)} 
  =  
-\frac{e^2 g_A^2}{34560 \pi^3 F^2 \mu^4 \left(\mu^2-4\right)^3 m^4} \left[\vphantom{\frac{e}{m}}\left(450\mu^{10} - 6300\mu^8 + 31410\mu^6 - 62340\mu^4   \right.\right.\vphantom{\frac{e}{m}} \\
&\;\;\;\;\left.\left.+ 29304\mu^2 + 9024\right) \Xi_2
+472 \mu^8-4927 \mu^6+15802 \mu^4-11136 \mu^2-832 \vphantom{\frac{e}{m}}\right] \\
&\;\;+\frac{e^2 g_A^2 5 \Xi_1}{384 \pi^3 F^2 m^4}\,,\\
\\
&\gamma_{\text{M2M2}}^{(n)} 
  =   
-\frac{e^2 g_A^2}{34560 \pi^3 F^2 \mu^4 \left(\mu^2-4\right)^3 m^4} \left[\vphantom{\frac{e}{m}}\left(450\mu^{10} - 6300\mu^8 + 31590\mu^6 - 74220\mu^4    \right.\right.\vphantom{\frac{e}{m}} \\
&\;\;\;\;\left.\left.+ 27864\mu^2 + 1824\right) \Xi_2
 +428 \mu^8-4495 \mu^6+16486 \mu^4-6280 \mu^2-64 \vphantom{\frac{e}{m}}\right] +\frac{e^2 g_A^2 5 \Xi_1}{384 \pi^3 F^2 m^4} \,,\\
 \\
&\gamma_{\text{E2M3}}^{(n)} 
  =  \; 
 \frac{e^2 g_A^2}{17280 \pi^3 F^2 \mu^4 \left(\mu^2-4\right)^3 m^4} \left[\vphantom{\frac{e}{m}}\left(36\mu^{10} - 504\mu^8 + 2610\mu^6 - 6780\mu^4 + 3888\mu^2  \right.\right.\vphantom{\frac{e}{m}} \\
 &\;\;\;\;\left.\left. + 768\right) \Xi_2
+14 \mu^8-155 \mu^6+1844 \mu^4-1440 \mu^2-128 \vphantom{\frac{e}{m}}\right]
-\frac{e^2 g_A^2 \Xi_1}{480 \pi^3 F^2 m^4}\,,\\
\\
&\gamma_{\text{M2E3}}^{(n)} 
  =  \; 
 \frac{e^2 g_A^2}{17280 \pi^3 F^2 \mu^4 \left(\mu^2-4\right)^2 m^4} \left[\vphantom{\frac{e}{m}}\left(36\mu^8 - 360\mu^6 + 990\mu^4 + 60\mu^2 - 192\right) \Xi_2 \right. \vphantom{\frac{e}{m}} \\
 &\;\;\;\;\left.+58 \mu^6 -355 \mu^4-32 \mu^2+32\vphantom{\frac{e}{m}}\right]
-\frac{e^2 g_A^2 \Xi_1}{480 \pi^3 F^2 m^4}\,,\\
\\
&\gamma_{\text{E1E1}\nu}^{(n)} 
  =  
-\frac{e^2 g_A^2}{11520 \pi^3 F^2 \mu^4 \left(\mu^2-4\right)^3 m^4} \left[\vphantom{\frac{e}{m}}\left(342\mu^{10} - 4788\mu^8 + 24210\mu^6 - 51780\mu^4  \right.\right.\vphantom{\frac{e}{m}} \\
&\;\;\;\;\left.\left. + 26664\mu^2 + 3744\right) \Xi_2
+276 \mu^8-2935 \mu^6+11506 \mu^4-4680 \mu^2-4032 \vphantom{\frac{e}{m}}\right] \\
&\;\;+\frac{19 e^2 g_A^2 \Xi_1}{640 \pi^3 F^2 m^4}\,,\\
\\
&\gamma_{\text{M1M1}\nu}^{(n)} 
  =  \;  
 \frac{e^2 g_A^2}{3840 \pi^3 F^2 \mu^4 \left(\mu^2-4\right)^4 m^4} \left[\vphantom{\frac{e}{m}} \left(-114\mu^{12}+2052\mu^{10}-14274\mu^8+45700\mu^6  \right.\right.\vphantom{\frac{e}{m}} \\
 &\;\;\;\;\left.\left.-65928\mu^4+21984\mu^2+8192\right) \Xi_1 -136\mu^{10}+1969\mu^8-8702\mu^6+17624\mu^4\right.\vphantom{\frac{e}{m}} \\
 &\;\;\;\;\left.-8960\mu^2-256\vphantom{\frac{e}{m}} \right]
+\frac{19 e^2 g_A^2\Xi_1}{640 \pi^3 F^2 m^4}\,,\\
\\
&\gamma_{\text{E1M2}\nu}^{(n)} 
 =  \; 
\frac{e^2 g_A^2}{14400 \pi^3 F^2 \mu^4 \left(\mu^2-4\right)^4 m^4} \left[\vphantom{\frac{e}{m}}\left(-393\mu^{12} + 7074\mu^{10} - 49923\mu^8 + 174660\mu^6  \right.\right.\vphantom{\frac{e}{m}} \\
&\;\;\;\;\left.\left.- 279684\mu^4 + 127152\mu^2 + 1536\right) \Xi_2 -294\mu^{10}+4282\mu^8-28623\mu^6+68156\mu^4\right.\vphantom{\frac{e}{m}} \\
&\;\;\;\;\left.-30016\mu^2-896\vphantom{\frac{e}{m}}\right] +\frac{e^2 g_A^2 \Xi_1}{4800 \pi^3 F^2 m^4}\,,\\
\\
&\gamma_{\text{M1E2}\nu}^{(n)} 
  =   
-\frac{e^2 g_A^2}{14400 \pi^3 F^2 \mu^4 \left(\mu^2-4\right)^3 m^4} \left[\vphantom{\frac{e}{m}}\left(393\mu^{10} - 5502\mu^8 + 27105\mu^6 - 49140\mu^4 \right.\right.\vphantom{\frac{e}{m}} \\
&\;\;\;\;\left.\left. + 19044\mu^2 + 8784\right) \Xi_2
+(\mu^2 -1)\left(492\mu^6-4613\mu^4+8404\mu^2-416\right) \vphantom{\frac{e}{m}}\right] \\
&\;\;+\frac{131 e^2 g_A^2 \Xi_1}{4800 \pi^3 F^2 m^4}\,.
\end{align*}

\section{\texorpdfstring{$\epsilon^3$}{eps3}-Tree values}\label{sec:delta3-tree} 
We give here the explicit analytic expressions for the $\Delta$-tree contribution to the nucleon polarizabilites.
\begin{align*}
\alpha_{E1} &= -\frac{b_1^2 e^2 \left(\mu_\Delta^2+\mu_\Delta+1\right)}{18 \pi  m^3 \mu_\Delta^2 \left(\mu_\Delta+1\right)} \,,\\
\beta_{M1} &= \frac{b_1^2 e^2}{18 \pi  m^3 \left(\mu_\Delta-1\right)} \,,\\
\alpha_{E2} &= \frac{b_1^2 e^2 \left(6 \mu_\Delta^2+3 \mu_\Delta+1\right)}{12 \pi  m^5 \left(\mu_\Delta-1\right) \mu_\Delta^2 \left(\mu_\Delta+1\right)^2} \,,\\
\beta_{M2} &= -\frac{b_1^2 e^2 \left(6 \mu_\Delta^2-3 \mu_\Delta+1\right)}{12 \pi  m^5 \left(\mu_\Delta-1\right)^2 \mu_\Delta^2 \left(\mu_\Delta+1\right)} \,,\\
\alpha_{E1\nu} &= -\frac{b_1^2 e^2 \left(18 \mu_\Delta^4+15 \mu_\Delta^3+31 \mu_\Delta^2+9 \mu_\Delta+7\right)}{144 \pi  m^5 \left(\mu_\Delta-1\right)^2 \mu_\Delta^2 \left(\mu_\Delta+1\right)^3}\,,\\
\beta_{M1\nu} &= \frac{b_1^2 e^2 \left(18 \mu_\Delta^4-15 \mu_\Delta^3+31 \mu_\Delta^2-9 \mu_\Delta+7\right)}{144 \pi  m^5 \left(\mu_\Delta-1\right)^3 \mu_\Delta^2 \left(\mu_\Delta+1\right)^2}\,,\\
\alpha_{E3} &= -\frac{5 b_1^2 e^2 \left(6 \mu_\Delta^2+3 \mu_\Delta+1\right)}{4 \pi  m^7 \left(\mu_\Delta-1\right)^2 \mu_\Delta^2 \left(\mu_\Delta+1\right)^3} \,,\\
\beta_{M3} &= \frac{5 b_1^2 e^2 \left(6 \mu_\Delta^2-3 \mu_\Delta+1\right)}{4 \pi  m^7 \left(\mu_\Delta-1\right)^3 \mu_\Delta^2 \left(\mu_\Delta+1\right)^2} \,,\\
\alpha_{E2\nu} &= \frac{b_1^2 e^2 \left(42 \mu_\Delta^4+27 \mu_\Delta^3+67 \mu_\Delta^2+23 \mu_\Delta+13\right)}{18 \pi  m^7 \left(\mu_\Delta-1\right)^3 \mu_\Delta^2 \left(\mu_\Delta+1\right)^4}\,,\\
\beta_{M2\nu} &= -\frac{b_1^2 e^2 \left(42 \mu_\Delta^4-27 \mu_\Delta^3+67 \mu_\Delta^2-23 \mu_\Delta+13\right)}{18 \pi  m^7 \left(\mu_\Delta-1\right)^4 \mu_\Delta^2 \left(\mu_\Delta+1\right)^3}\,,\\
\alpha_{E1\nu^2} &= -\frac{b_1^2 e^2 \left(162 \mu_\Delta^6+141 \mu_\Delta^5+243 \mu_\Delta^4+188 \mu_\Delta^3+258 \mu_\Delta^2+71 \mu_\Delta+57\right)}{360 \pi  m^7 \left(\mu_\Delta-1\right)^4 \mu_\Delta^2 \left(\mu_\Delta+1\right)^5} \,,\\
\beta_{M1\nu^2} &= \frac{b_1^2 e^2 \left(162 \mu_\Delta^6-141 \mu_\Delta^5+243 \mu_\Delta^4-188 \mu_\Delta^3+258 \mu_\Delta^2-71 \mu_\Delta+57\right)}{360 \pi  m^7 \left(\mu_\Delta-1\right)^5 \mu_\Delta^2 \left(\mu_\Delta+1\right)^4}
\end{align*}
\begin{align*}
\gamma_{E1E1} &= -\frac{b_1^2 e^2 \left(\mu_\Delta^3+\mu_\Delta^2-1\right)}{18 \pi  m^4 \left(\mu_\Delta-1\right) \mu_\Delta\left(\mu_\Delta+1\right)^2} \,,\\
\gamma_{M1M1} &= \frac{b_1^2 e^2 \left(2 \mu_\Delta^4-3 \mu_\Delta^3+\mu_\Delta^2+3 \mu_\Delta-1\right)}{36 \pi  m^4 \left(\mu_\Delta-1\right)^2 \mu_\Delta^2 \left(\mu_\Delta+1\right)} \,,\\
\gamma_{E1M2} &= -\frac{b_1^2 e^2 \left(2 \mu_\Delta-1\right)}{36 \pi  m^4 \left(\mu_\Delta^2-1\right)} \,,\\
\gamma_{M1E2} &= \frac{b_1^2 e^2 \left(2 \mu_\Delta^3+1\right)}{36 \pi  m^4 \mu_\Delta^2 \left(\mu_\Delta^2-1\right)} \,,\\
\gamma_{E2E2} &= -\frac{b_1^2 e^2 \left(4 \mu_\Delta^6+3 \mu_\Delta^5-17 \mu_\Delta^4-8 \mu_\Delta^3+4 \mu_\Delta^2+57 \mu_\Delta+13\right)}{1728 \pi  m^6 \left(\mu_\Delta-1\right)^2 \mu_\Delta^2 \left(\mu_\Delta+1\right)^3} \,,\\
\gamma_{M2M2} &= \frac{b_1^2 e^2 \left(4 \mu_\Delta^6-5 \mu_\Delta^5-15 \mu_\Delta^4+12 \mu_\Delta^3-59 \mu_\Delta+15\right)}{1728 \pi  m^6 \left(\mu_\Delta-1\right)^3 \left(\mu^{\Delta
  }\right)^2 \left(\mu_\Delta+1\right)^2}\,,\\
\gamma_{E2M3} &= -\frac{b_1^2 e^2 \left(4 \mu_\Delta^5-\mu_\Delta^4-16 \mu_\Delta^3+8 \mu_\Delta^2-4 \mu
  ^{\Delta}+1\right)}{864 \pi  m^6 \mu_\Delta^2 \left(\mu_\Delta^2-1\right)^2} \,,\\
\gamma_{M2E3} &= \frac{b_1^2 e^2 \left(4 \mu_\Delta^5-\mu_\Delta^4-16 \mu_\Delta^3-4 \mu_\Delta^2-4 \mu
  ^{\Delta}-3\right)}{864 \pi  m^6 \mu_\Delta^2 \left(\mu_\Delta^2-1\right)^2} \,,\\
\gamma_{E1E1\nu} &= \frac{b_1^2 e^2 \left(12 \mu_\Delta^8+9 \mu_\Delta^7-63 \mu_\Delta^6-45 \mu_\Delta^5-53 \mu_\Delta^4-69 \mu_\Delta^3-93 \mu_\Delta^2+73 \mu_\Delta+37\right)}{576 \pi  m^6 \left(\mu_\Delta-1\right)^3 \mu_\Delta^2 \left(\mu_\Delta+1\right)^4} \,,\\
\gamma_{M1M1\nu} &= -\frac{b_1^2 e^2 \left(12 \mu_\Delta^8-15 \mu_\Delta^7-57 \mu_\Delta^6+63 \mu_\Delta^5-71 \mu_\Delta^4+51 \mu_\Delta^3-75 \mu_\Delta^2-67 \mu_\Delta+31\right)}{576 \pi  m^6 \left(\mu_\Delta-1\right)^4 \mu_\Delta^2 \left(\mu_\Delta+1\right)^3} \,,\\
\gamma_{E1M2\nu} &= \frac{b_1^2 e^2 \left(36 \mu_\Delta^7-9 \mu_\Delta^6-180 \mu_\Delta^5+33 \mu_\Delta^4-152 \mu_\Delta^3-11 \mu_\Delta^2-104 \mu_\Delta+67\right)}{1440 \pi  m^6 \mu_\Delta^2 \left(\mu_\Delta^2-1\right)^3} \,,\\
\gamma_{M1E2\nu} &= -\frac{b_1^2 e^2 \left(36 \mu_\Delta^7-9 \mu_\Delta^6-180 \mu_\Delta^5+21 \mu_\Delta^4-152 \mu_\Delta^3-43 \mu_\Delta^2-104 \mu_\Delta-49\right)}{1440 \pi  m^6 \mu_\Delta^2 \left(\mu_\Delta^2-1\right)^3}\,.
\end{align*}

\section{\texorpdfstring{$\epsilon^3$}{eps3}-Loop values}\label{sec:delta3-loop} 
Due to the length of the expressions, we only provide here the
expressions for the first-order polarizabilities.
In addition to the definition from Appendix \ref{sec:q3-loop}, we now have 
$m_\Delta$ as an additional mass scale and it is convenient to
introduce 
$\mu_\Delta = m_\Delta/m$, $\Xi_3 = \ln (\mu_\Delta)$ as well as $\Xi_4 = \Xi \left(m^2,m_\Delta,M \right)/(\mu^2-4)$.

\subsection{Proton values}\label{sub:proton-e3} 

\subsubsection{Spin-independent first order polarizabilities}\label{subsub:proton-spin-independent-1-e3} 

\begin{align*} 
&\alpha_{\text{E1}}^{(p)} 
  =  -\frac{e^2 h_A^2 \Xi_4}{7776 F^2 m \pi ^3 \mu_\Delta^6 \left(-\mu^2+\mu_\Delta^2-2 \mu_\Delta+1\right)^2 \left(-\mu^2+\mu_\Delta^2+2 \mu_\Delta+1\right)} \left[\vphantom{\frac{e}{m}}-81 \mu_\Delta^{16}\right.\vphantom{\frac{e}{m}}\\
&\;\;\;\;\left.\left.+228 \mu_\Delta^{15}+\left(486 \mu^2-56\right)\mu_\Delta^{14}-2 \left(522 \mu^2+325\right) \mu_\Delta^{13}+\left(-1215 \mu^4-136 \mu^2\right.\right.\right.\vphantom{\frac{e}{m}}\\
&\;\;\;\;\left.\left.+445\right) \mu_\Delta^{12}+6 \left(300 \mu^4+301 \mu^2+148\right) \mu_\Delta^{11}+30 \left(54 \mu^6+34 \mu^4+32 \mu^2-11\right)\mu_\Delta^{10}\right. \vphantom{\frac{e}{m}}\\
&\;\;\;\;\left.\left.-6 \left(220 \mu^6+204 \mu^4+343 \mu^2+126\right) \mu_\Delta^9-\left(1215 \mu^8+1586 \mu^6+3036\mu^4+666 \mu^2\right.\right.\right. \vphantom{\frac{e}{m}}\\
&\;\;\;\;\left.\left.+407\right) \mu_\Delta^8+2 \left(90 \mu^8-325 \mu^6+573 \mu^4+3 \mu^2+581\right) \mu_\Delta^7+2\left(243 \mu^{10}+412 \mu^8\right.\right.\vphantom{\frac{e}{m}}\\
&\;\;\;\;\left.\left.+692 \mu^6-315 \mu^4+295 \mu^2-25\right) \mu_\Delta^6+6 \left(42 \mu^{10}+164 \mu^8-121 \mu^6+6\mu^4+171 \mu^2\right.\right. \vphantom{\frac{e}{m}}\\
&\;\;\;\;\left.\left.-262\right) \mu_\Delta^5-3 \left(\mu^2-1\right)^2 \left(27 \mu^8+34 \mu^6-32 \mu^4-462 \mu^2-449\right)\mu_\Delta^4\right. \vphantom{\frac{e}{m}}\\
&\;\;\;\;\left.\left.-6 \left(\mu^2-1\right)^3 \left(16 \mu^6+90 \mu^4+83 \mu^2+91\right) \mu_\Delta^3-28 \left(\mu^2-1\right)^4 \left(5 \mu^4+17 \mu^2+32\right) \mu_\Delta^2\right.\right. \vphantom{\frac{e}{m}}\\
&\;\;\;\;\left.-14 \left(\mu^2-1\right)^5 \left(\mu^2+11\right) \mu_\Delta+14\left(\mu^2-1\right)^6 \left(\mu^2+2\right)\vphantom{\frac{e}{m}}\right]\\
&\;+\frac{e^2 h_A^2 \Xi_1}{7776 F^2 m \pi ^3 \mu_\Delta^6} \left[\vphantom{\frac{e}{m}}81 \mu_\Delta^{12}-66 \mu_\Delta^{11}-4 \left(81 \mu^2+19\right) \mu_\Delta^{10}+6 \left(17 \mu^2+56\right)\mu_\Delta^9+\left(486 \mu^4\right.\right. \vphantom{\frac{e}{m}}\\
&\;\;\;\;\left.\left.+482 \mu^2+440\right) \mu_\Delta^8+6 \left(15 \mu^4+29 \mu^2+13\right) \mu_\Delta^7-2 \left(162 \mu^6+326 \mu^4+248 \mu^2\right.\right. \vphantom{\frac{e}{m}}\\
&\;\;\;\;\left.\left.+131\right) \mu_\Delta^6-6 \left(37 \mu^6+138 \mu^4+22 \mu^2+52\right)\mu_\Delta^5+\left(81 \mu^8+148 \mu^6-84 \mu^4-324 \mu^2\right.\right. \vphantom{\frac{e}{m}}\\
&\;\;\;\;\left.\left.+67\right) \mu_\Delta^4+6 \left(16 \mu^8+46 \mu^6-35 \mu^4+104 \mu^2-77\right) \mu_\Delta^3+28 \left(4 \mu^8+4 \mu^6-3 \mu^4\right.\right. \vphantom{\frac{e}{m}}\\
&\;\;\;\;\left.\left.+22 \mu^2-17\right) \mu_\Delta^2+42\left(\mu^8+2 \mu^6+12 \mu^4-14 \mu^2+5\right) \mu_\Delta-14 \left(\mu^{10}-2 \mu^8-2 \mu^6\right.\right. \vphantom{\frac{e}{m}}\\
&\;\;\;\;\left.\left.-8 \mu^4+7 \mu^2-2\right)\vphantom{\frac{e}{m}}\right]\\
&+\frac{e^2 h_A^2 \Xi_3}{7776 F^2 m \pi ^3 \mu_\Delta^6}\left[\vphantom{\frac{e}{m}}-81 \mu_\Delta^{12}+66 \mu_\Delta^{11}+4 \left(81 \mu^2+19\right)\mu_\Delta^{10}-6 \left(17 \mu^2+56\right) \mu_\Delta^9\right. \vphantom{\frac{e}{m}}\\
&\;\;\;\;\left.-2 \left(243 \mu^4+241 \mu^2+220\right) \mu_\Delta^8-6 \left(15 \mu^4+29 \mu^2+13\right) \mu_\Delta^7+\left(324 \mu^6+652 \mu^4\right.\right. \vphantom{\frac{e}{m}}\\
&\;\;\;\;\left.\left.+496 \mu^2+262\right) \mu_\Delta^6+6 \left(37 \mu^6+138 \mu^4+22 \mu^2-170\right) \mu_\Delta^5+\left(-81 \mu^8-148 \mu^6+84 \mu^4\right.\right. \vphantom{\frac{e}{m}}\\
&\;\;\;\;\left.\left.+324 \mu^2+31\right)\mu_\Delta^4-6 \left(16 \mu^8+46 \mu^6-35 \mu^4-104 \mu^2+77\right) \mu_\Delta^3-28 \left(\mu^2-1\right)^2\left(4 \mu^4\right.\right. \vphantom{\frac{e}{m}}\\
&\;\;\;\;\left.\left.+12 \mu^2+17\right) \mu_\Delta^2-42 \left(\mu^2-1\right)^3 \left(\mu^2+5\right) \mu_\Delta+14 \left(\mu^2-1\right)^4 \left(\mu^2+2\right)\vphantom{\frac{e}{m}}\right] \\
&\;+\frac{e^2 h_A^2}{46656 F^2 m \pi ^3 \mu_\Delta^6 \left(-\mu^2+\mu_\Delta^2-2 \mu_\Delta+1\right)}\left[\vphantom{\frac{e}{m}}486 \mu_\Delta^{12}-1368 \mu_\Delta^{11}-3 \left(648 \mu^2-355\right) \mu_\Delta^{10}\right.\vphantom{\frac{e}{m}}\\
&\;\;\;\;\left.\left.+168 \left(21 \mu^2+11\right) \mu_\Delta^9+3 \left(972\mu^4+253 \mu^2-425\right) \mu_\Delta^8-12 \left(198 \mu^4+325 \mu^2\right.\right.\right. \vphantom{\frac{e}{m}}\\
&\;\;\;\;\left.\left.+432\right) \mu_\Delta^7-\left(1944 \mu^6+4209 \mu^4+4278 \mu^2-6401\right) \mu_\Delta^6-4 \left(90 \mu^6+144 \mu^4-1773 \mu^2\right.\right. \vphantom{\frac{e}{m}}\\
&\;\;\;\;\left.\left.+598\right) \mu_\Delta^5+\left(486 \mu^8+1797 \mu^6+3573 \mu^4-4251 \mu^2+1501\right) \mu_\Delta^4+2 \left(288 \mu^8+1272 \mu^6\right.\right. \vphantom{\frac{e}{m}}\\
&\;\;\;\;\left.\left.-999 \mu^4-818\mu^2+82\right) \mu_\Delta^3+14 \left(48 \mu^8+57 \mu^6-510 \mu^4+427 \mu^2-85\right) \mu_\Delta^2+14 \left(6\mu^8\right.\right. \vphantom{\frac{e}{m}}\\
&\;\;\;\;\left.\left.-135 \mu^6+85 \mu^4+4 \mu^2-14\right) \mu_\Delta-14 \left(6 \mu^{10}-15 \mu^8+73 \mu^6-108 \mu^4+54 \mu^2-10\right)\vphantom{\frac{e}{m}}\right] 
\,, \end{align*}

\begin{align*} 
&\beta_{\text{M1}}^{(p)}
  =  -\frac{e^2 h_A^2 \Xi_4}{7776 F^2 m \pi ^3 \mu_\Delta^6 \left(-\mu^2+\mu_\Delta^2-2 \mu_\Delta+1\right)}\left[\vphantom{\frac{e}{m}}-243 \mu_\Delta^{12}+456\mu_\Delta^{11}+2 \left(486 \mu^2 \right.\right. \\
&\;\;\;\;\left.\left.+71\right) \mu_\Delta^{10}-8 \left(147 \mu^2+136\right) \mu_\Delta^9-2 \left(729 \mu^4+385 \mu^2-477\right) \mu_\Delta^8+2 \left(396 \mu^4+715 \mu^2 \right.\right. \\
&\;\;\;\;\left.\left.+388\right) \mu_\Delta^7+2 \left(486 \mu^6+479 \mu^4+44 \mu^2-950\right) \mu_\Delta^6+2 \left(60 \mu^6-48 \mu^4-613 \mu^2\right.\right. \vphantom{\frac{e}{m}}\\
&\;\;\;\;\left.\left.+477\right) \mu_\Delta^5+\left(-243 \mu^8-160 \mu^6-228 \mu^4+620 \mu^2+95\right) \mu_\Delta^4-4 \left(48 \mu^8+58 \mu^6\right.\right. \vphantom{\frac{e}{m}}\\
&\;\;\;\;\left.\left.-145 \mu^4-97 \mu^2+136\right) \mu_\Delta^3-4 \left(\mu^2-1\right)^2 \left(46 \mu^4+35 \mu^2-102\right) \mu_\Delta^2\right. \vphantom{\frac{e}{m}}\\
&\;\;\;\;\left.\left.-14\left(\mu^2-1\right)^4 \mu_\Delta+14 \left(\mu^2-1\right)^5\vphantom{\frac{e}{m}}\right]\right.\\
&\;-\frac{e^2 h_A^2 \Xi_1}{7776 F^2 m \pi ^3 \mu_\Delta^6} \left[\vphantom{\frac{e}{m}}-243 \mu_\Delta^{12}-30 \mu_\Delta^{11}+\left(972 \mu^2+568\right) \mu_\Delta^{10}+6 \left(47 \mu^2-63\right) \mu_\Delta^9 \right. \\
&\;\;\;\;\left.\left.-2 \left(729 \mu^4+619 \mu^2+256\right)\mu_\Delta^8-6 \left(111 \mu^4+59 \mu^2-203\right) \mu_\Delta^7+2 \left(486 \mu^6+308 \mu^4\right.\right.\right. \vphantom{\frac{e}{m}}\\
&\;\;\;\;\left.\left.+62 \mu^2+47\right)\mu_\Delta^6+6 \left(101 \mu^6+111 \mu^4-58 \mu^2-142\right) \mu_\Delta^5+\left(-243 \mu^8+224 \mu^6\right.\right. \vphantom{\frac{e}{m}}\\
&\;\;\;\;\left.\left.+336 \mu^4+197\right) \mu_\Delta^4-12 \left(16 \mu^8-9 \mu^6-41 \mu^4-52 \mu^2+18\right) \mu_\Delta^3-4 \left(46 \mu^8-71 \mu^6\right.\right. \vphantom{\frac{e}{m}}\\
&\;\;\;\;\left.\left.-84 \mu^4-197 \mu^2+88\right) \mu_\Delta^2-42 \left(\mu^8-4 \mu^6-18 \mu^4+4 \mu^2-1\right) \mu_\Delta+14\left(\mu^{10}-5 \mu^8\right.\right. \vphantom{\frac{e}{m}}\\
&\;\;\;\;\left.\left.+10 \mu^6+22 \mu^4-5 \mu^2+1\right)\vphantom{\frac{e}{m}}\right]\\
&\;+\frac{e^2 h_A^2 \Xi_3}{7776 F^2m \pi ^3 \mu_\Delta^6}\left[\vphantom{\frac{e}{m}}-243 \mu_\Delta^{12}-30 \mu_\Delta^{11}+\left(972 \mu^2+568\right) \mu_\Delta^{10}+6 \left(47 \mu^2-63\right) \mu_\Delta^9 \right. \\
&\;\;\;\;\left.\left.-2 \left(729 \mu^4+619 \mu^2+256\right) \mu_\Delta^8-6 \left(111 \mu^4+59 \mu^2-203\right) \mu_\Delta^7+2 \left(486 \mu^6+308 \mu^4\right.\right.\right. \vphantom{\frac{e}{m}}\\
&\;\;\;\;\left.\left.+62 \mu^2+47\right) \mu_\Delta^6+6 \left(101 \mu^6+111 \mu^4-58 \mu^2+80\right) \mu_\Delta^5+\left(-243 \mu^8+224 \mu^6+336 \mu^4\right.\right. \vphantom{\frac{e}{m}}\\
&\;\;\;\;\left.\left.-233\right) \mu_\Delta^4-12 \left(16 \mu^8-9 \mu^6-41 \mu^4+52 \mu^2-18\right) \mu_\Delta^3-4 \left(\mu^2-1\right)^2 \left(46 \mu^4+21 \mu^2\right.\right. \vphantom{\frac{e}{m}}\\
&\;\;\;\;\left.\left.-88\right) \mu_\Delta^2-42 \left(\mu^2-1\right)^4 \mu_\Delta+14 \left(\mu^2-1\right)^5\vphantom{\frac{e}{m}}\right] \\
&\;+\frac{e^2  h_A^2}{46656 F^2 m \pi^3 \mu_\Delta^6} \left[\vphantom{\frac{e}{m}}1458 \mu_\Delta^{10}+180 \mu_\Delta^9-3 \left(1458 \mu^2+893\right) \mu_\Delta^8-18\left(84 \mu^2-131\right) \mu_\Delta^7\right. \\
&\;\;\;\;\left.\left.+6 \left(729 \mu^4+670 \mu^2+309\right) \mu_\Delta^6+18 \left(138 \mu^4+217\mu^2-162\right) \mu_\Delta^5+\left(-1458 \mu^6\right.\right.\right. \vphantom{\frac{e}{m}}\\
&\;\;\;\;\left.\left.-405 \mu^4+1656 \mu^2+1445\right) \mu_\Delta^4-12 \left(96 \mu^6+15 \mu^4+90 \mu^2-119\right) \mu_\Delta^3-2 \left(510 \mu^6\right.\right. \vphantom{\frac{e}{m}}\\
&\;\;\;\;\left.\left.-450 \mu^4-370 \mu^2+503\right) \mu_\Delta^2-42\left(6 \mu^6+39 \mu^4-22 \mu^2+4\right) \mu_\Delta+14 \left(6 \mu^8-27 \mu^6\right.\right. \vphantom{\frac{e}{m}}\\
&\;\;\;\;\left.\left.-65 \mu^4+16 \mu^2-2\right)\vphantom{\frac{e}{m}}\right]
\,. \end{align*}

\subsubsection{Spin-dependent first order polarizabilities}\label{subsub:proton-spin-dependent-1-e3} 

\begin{align*} 
&\gamma_{\text{E1E1}}^{(p)}  =  \;
\frac{e^2 h_A^2 \Xi_4}{15552 F^2 m^2 \pi ^3 \mu_\Delta^6 \left(\mu^2-\mu_\Delta^2+2 \mu_\Delta-1\right)^3 \left(\mu^2-\mu_\Delta^2-2 \mu_\Delta-1\right)^2} \left(-81 \mu_\Delta^{20}\right.\vphantom{\frac{e}{m}}\\
&\;\;\;\;\left.\left.+282 \mu_\Delta^{19}+\left(648 \mu^2-361\right)\mu_\Delta^{18}-2 \left(723 \mu^2-94\right) \mu_\Delta^{17}-4 \left(567 \mu^4-501 \mu^2-886\right) \mu_\Delta^{16}\right.\right.\vphantom{\frac{e}{m}}\\
&\;\;\;\;\left.\left.+\left(2226 \mu^4-3596 \mu^2-6942\right) \mu_\Delta^{15}+\left(4536 \mu^6-5007 \mu^4-13196 \mu^2-10489\right) \mu_\Delta^{14}\right.\right.\vphantom{\frac{e}{m}}\\
&\;\;\;\;\left.\left.+6 \left(203 \mu^6+2281 \mu^4+3724 \mu^2+3951\right) \mu_\Delta^{13}+\left(-5670\mu^8+8181 \mu^6+19029 \mu^4\right.\right.\right.\vphantom{\frac{e}{m}}\\
&\;\;\;\;\left.\left.+25959 \mu^2+16267\right) \mu_\Delta^{12}-2 \left(4305 \mu^8+10535 \mu^6+11493 \mu^4+14259 \mu^2+19673\right) \mu_\Delta^{11}\right.\vphantom{\frac{e}{m}}\\
&\;\;\;\;\left.\left.+\left(4536 \mu^{10}-10677 \mu^8-12326 \mu^6-18558 \mu^4-23154 \mu^2-14543\right) \mu_\Delta^{10}+2 \left(6279 \mu^{10}\right.\right.\right.\vphantom{\frac{e}{m}}\\
&\;\;\;\;\left.\left.+6380 \mu^8+2604 \mu^6+804 \mu^4+47 \mu^2+19122\right) \mu_\Delta^9+\left(-2268 \mu^{12}+11121 \mu^{10}\right.\right.\vphantom{\frac{e}{m}}\\
&\;\;\;\;\left.\left.-1169 \mu^8+230 \mu^6-78 \mu^4+12487 \mu^2+6389\right) \mu_\Delta^8-6 \left(1519 \mu^{12}-212 \mu^{10}-511\mu^8\right.\right.\vphantom{\frac{e}{m}}\\
&\;\;\;\;\left.\left.-776 \mu^6-403 \mu^4-3676 \mu^2+3835\right) \mu_\Delta^7+\left(648 \mu^{14}-7989 \mu^{12}+10752 \mu^{10}\right.\right.\vphantom{\frac{e}{m}}\\
&\;\;\;\;\left.\left.-1203 \mu^8-1636\mu^6+4107 \mu^4-6000 \mu^2+1321\right) \mu_\Delta^6+2 \left(\mu^2-1\right)^2 \left(1707 \mu^{10}\right.\right.\vphantom{\frac{e}{m}}\\
&\;\;\;\;\left.\left.+1057 \mu^8+425 \mu^6-1038 \mu^4-760 \mu^2+3901\right) \mu_\Delta^5-\left(\mu^2-1\right)^3 \left(81 \mu^{10}-3084 \mu^8\right.\right.\vphantom{\frac{e}{m}}\\
&\;\;\;\;\left.\left.+38 \mu^6+150 \mu^4-2265 \mu^2-3194\right) \mu_\Delta^4-2 \left(\mu^2-1\right)^4 \left(264 \mu^8+319 \mu^6+13 \mu^4\right.\right.\vphantom{\frac{e}{m}}\\
&\;\;\;\;\left.\left.-869 \mu^2+252\right) \mu_\Delta^3-2 \left(\mu^2-1\right)^5 \left(303 \mu^6+34 \mu^4-137 \mu^2+556\right) \mu_\Delta^2\right.\vphantom{\frac{e}{m}}\\
&\;\;\;\;\left.\left.-84 \left(\mu^2-1\right)^6\left(\mu^2+5\right) \mu_\Delta+7 \left(\mu^2-1\right)^7 \left(\mu^4-2 \mu^2-5\right)\right)\right.\\
&-\frac{e^2 h_A^2 \Xi_1 }{15552 F^2 m^2 \pi ^3 \mu_\Delta^6}\left(-81 \mu_\Delta^{12}+120 \mu_\Delta^{11}+\left(324 \mu^2-283\right) \mu_\Delta^{10}+12 \left(14 \mu^2+29\right) \mu_\Delta^9\right.\vphantom{\frac{e}{m}}\\
&\;\;\;\;\left.\left.+\left(-486 \mu^4+1136 \mu^2+2954\right) \mu_\Delta^8+\left(-1224 \mu^4+158 \mu^2+874\right) \mu_\Delta^7+\left(324 \mu^6\right.\right.\right.\vphantom{\frac{e}{m}}\\
&\;\;\;\;\left.\left.-1987 \mu^4-1888 \mu^2-4039\right) \mu_\Delta^6+6 \left(244 \mu^6-166 \mu^4-117 \mu^2-354\right) \mu_\Delta^5+\left(-81 \mu^8\right.\right.\vphantom{\frac{e}{m}}\\
&\;\;\;\;\left.\left.+1705 \mu^6-825 \mu^4-165 \mu^2+1030\right) \mu_\Delta^4-6 \left(88 \mu^8-84 \mu^6+26 \mu^4-24 \mu^2-75\right) \mu_\Delta^3\right.\vphantom{\frac{e}{m}}\\
&\;\;\;\;\left.\left.+\left(-578 \mu^8+496 \mu^6+300 \mu^4+340 \mu^2-482\right) \mu_\Delta^2-14 \left(\mu^8+2 \mu^6-30 \mu^4+46 \mu^2\right.\right.\right.\vphantom{\frac{e}{m}}\\
&\;\;\;\;\left.\left.-25\right) \mu_\Delta+7\left(\mu^{10}-5 \mu^8+4 \mu^6+16 \mu^4+13 \mu^2-5\right)\right) \\
& -\frac{e^2 h_A^2 \Xi_3 }{15552 F^2 m^2 \pi ^3 \mu_\Delta^6}\left(81 \mu_\Delta^{12}-120 \mu_\Delta^{11}+\left(283-324 \mu^2\right) \mu_\Delta^{10}-12 \left(14 \mu^2+29\right) \mu_\Delta^9\right.\vphantom{\frac{e}{m}}\\
&\;\;\;\;\left.\left.+2 \left(243 \mu^4-568 \mu^2-1477\right) \mu_\Delta^8+2\left(612 \mu^4-79 \mu^2-437\right) \mu_\Delta^7+\left(-324 \mu^6\right.\right.\right.\vphantom{\frac{e}{m}}\\
&\;\;\;\;\left.\left.+1987 \mu^4+1888 \mu^2+4039\right) \mu_\Delta^6+\left(-1464 \mu^6+996 \mu^4+702 \mu^2+3048\right) \mu_\Delta^5+\left(81 \mu^8\right.\right.\vphantom{\frac{e}{m}}\\
&\;\;\;\;\left.\left.-1705 \mu^6+825 \mu^4+165 \mu^2+820\right) \mu_\Delta^4+6 \left(88 \mu^8-84 \mu^6+26 \mu^4+84 \mu^2+75\right) \mu_\Delta^3\right.\vphantom{\frac{e}{m}}\\
&\;\;\;\;\left.\left.+\left(578 \mu^8-496 \mu^6-300 \mu^4+700 \mu^2-482\right) \mu_\Delta^2+14 \left(\mu^2-1\right)^2 \left(\mu^4+4 \mu^2+25\right) \mu_\Delta\right.\right.\vphantom{\frac{e}{m}}\\
&\;\;\;\;\left.\left.-7 \left(\mu^2-1\right)^3 \left(\mu^4-2 \mu^2-5\right)\right) \right.\\
&-\frac{e^2 h_A^2}{93312 F^2 m^2 \pi ^3 \mu_\Delta^6 \left(\mu^2-\mu_\Delta^2+2 \mu_\Delta-1\right)^2 \left(\mu^2-\mu_\Delta^2-2 \mu_\Delta-1\right)}\left(486 \mu_\Delta^{16}-1692 \mu_\Delta^{15}\right.\vphantom{\frac{e}{m}}\\
&\;\;\;\;\left.\left.-3 \left(972 \mu^2-965\right) \mu_\Delta^{14}+6 \left(882 \mu^2-611\right) \mu_\Delta^{13}+3 \left(2430 \mu^4-3293 \mu^2-5708\right) \mu_\Delta^{12}\right.\right.\vphantom{\frac{e}{m}}\\
&\;\;\;\;\left.\left.-12 \left(90 \mu^4-1637 \mu^2-3453\right) \mu_\Delta^{11}+\left(-9720 \mu^6+13950 \mu^4+36153 \mu^2+26911\right)\mu_\Delta^{10}\right.\right.\vphantom{\frac{e}{m}}\\
&\;\;\;\;\left.\left.-2 \left(7380 \mu^6+16422 \mu^4+27279 \mu^2+49276\right) \mu_\Delta^9+\left(7290 \mu^8-14952 \mu^6-23619 \mu^4\right.\right.\right.\vphantom{\frac{e}{m}}\\
&\;\;\;\;\left.\left.-30092 \mu^2+225\right) \mu_\Delta^8+2 \left(11610 \mu^8+8664 \mu^6+6543 \mu^4+8679 \mu^2+46169\right) \mu_\Delta^7\right.\vphantom{\frac{e}{m}}\\
&\;\;\;\;\left.\left.-\left(2916 \mu^{10}-16563 \mu^8+3972 \mu^6+10382 \mu^4+13357 \mu^2+25208\right) \mu_\Delta^6+\left(-14148 \mu^{10}\right.\right.\right.\vphantom{\frac{e}{m}}\\
&\;\;\;\;\left.\left.+3630 \mu^8+9990 \mu^6-34308 \mu^4+68166 \mu^2-28466\right) \mu_\Delta^5+\left(486 \mu^{12}-12087 \mu^{10}\right.\right.\vphantom{\frac{e}{m}}\\
&\;\;\;\;\left.\left.+20124 \mu^8-22834 \mu^6+28369 \mu^4-21493 \mu^2+6763\right) \mu_\Delta^4+2 \left(1584 \mu^{12}-2046 \mu^{10}\right.\right.\vphantom{\frac{e}{m}}\\
&\;\;\;\;\left.\left.-7245 \mu^8+10135 \mu^6-4430\mu^4-1415 \mu^2+393\right) \mu_\Delta^3+\left(3552 \mu^{12}-11877 \mu^{10}\right.\right.\vphantom{\frac{e}{m}}\\
&\;\;\;\;\left.\left.+16691 \mu^8-25025 \mu^6+31935 \mu^4-20510 \mu^2+5234\right) \mu_\Delta^2+168 \left(\mu^2-1\right)^2 \left(27 \mu^6\right.\right.\vphantom{\frac{e}{m}}\\
&\;\;\;\;\left.\left.-52 \mu^4+47 \mu^2-13\right) \mu_\Delta-7 \left(\mu^2-1\right)^3 \left(6 \mu^8-27 \mu^6-137 \mu^4+76 \mu^2-26\right)\right)
\,, \end{align*}

\begin{align*} 
& \gamma_{\text{M1M1}}^{(p)}  =  \;
\frac{e^2 h_A^2 \Xi_4}{15552 F^2 m^2 \pi ^3 \mu_\Delta^6\left(\mu^2-\mu_\Delta^2+2 \mu_\Delta-1\right)^2 \left(\mu^2-\mu_\Delta^2-2 \mu_\Delta-1\right)}\left(-567 \mu_\Delta^{16}\right.\vphantom{\frac{e}{m}}\\
&\;\;\;\;\left.\left.+1446 \mu_\Delta^{15}+\left(3402 \mu^2+1649\right) \mu_\Delta^{14}-6\left(1149 \mu^2+893\right) \mu_\Delta^{13}-\left(8505 \mu^4+6710 \mu^2\right.\right.\right.\vphantom{\frac{e}{m}}\\
&\;\;\;\;\left.\left.+2767\right) \mu_\Delta^{12}+4 \left(3195\mu^4+3432 \mu^2+2113\right) \mu_\Delta^{11}+2 \left(5670 \mu^6+5025 \mu^4+4973 \mu^2\right.\right.\vphantom{\frac{e}{m}}\\
&\;\;\;\;\left.\left.+2336\right) \mu_\Delta^{10}-2 \left(5550 \mu^6+4365 \mu^4+5225 \mu^2+4044\right) \mu_\Delta^9-\left(8505 \mu^8+6073 \mu^6\right.\right.\vphantom{\frac{e}{m}}\\
&\;\;\;\;\left.\left.+10931 \mu^4+8455 \mu^2+6140\right) \mu_\Delta^8+6 \left(645 \mu^8-433 \mu^6+208 \mu^4+339 \mu^2+1116\right) \mu_\Delta^7\right.\vphantom{\frac{e}{m}}\\
&\;\;\;\;\left.\left.+\left(3402\mu^{10}+277 \mu^8+3576 \mu^6+892 \mu^4+3154 \mu^2+3885\right) \mu_\Delta^6+2 \left(117 \mu^{10}+1632 \mu^8\right.\right.\right.\vphantom{\frac{e}{m}}\\
&\;\;\;\;\left.\left.-335 \mu^6+971 \mu^4+1116 \mu^2-2381\right) \mu_\Delta^5-\left(567 \mu^{12}-1128 \mu^{10}+343 \mu^8-1548 \mu^6\right.\right.\vphantom{\frac{e}{m}}\\
&\;\;\;\;\left.\left.+657 \mu^4+850 \mu^2+259\right)\mu_\Delta^4-2 \left(\mu^2-1\right)^2 \left(168 \mu^8+489 \mu^6-26 \mu^4+\mu^2-849\right) \mu_\Delta^3\right.\vphantom{\frac{e}{m}}\\
&\;\;\;\;\left.\left.-2\left(\mu^2-1\right)^3 \left(164 \mu^6+215 \mu^4+302 \mu^2-261\right) \mu_\Delta^2-84 \left(\mu^2-1\right)^4 \left(3 \mu^2+1\right) \mu_\Delta\right.\right.\vphantom{\frac{e}{m}}\\
&\;\;\;\;\left.\left.+7 \left(\mu^2-1\right)^5 \left(\mu^4-7\right)\right) \right. \\
& +\frac{e^2 h_A^2 \Xi_1}{15552 F^2 m^2 \pi ^3 \mu_\Delta^6}\left(567 \mu_\Delta^{12}-312 \mu_\Delta^{11}-\left(2268 \mu^2+2273\right) \mu_\Delta^{10}+\left(600 \mu^2-322\right) \mu_\Delta^9\right.\vphantom{\frac{e}{m}}\\
&\;\;\;\;\left.\left.+\left(3402 \mu^4+5122 \mu^2+3314\right) \mu_\Delta^8+\left(72 \mu^4+2062 \mu^2+2410\right) \mu_\Delta^7-\left(2268 \mu^6\right.\right.\right.\vphantom{\frac{e}{m}}\\
&\;\;\;\;\left.\left.+3125 \mu^4+3048 \mu^2+1481\right) \mu_\Delta^6-6 \left(116 \mu^6+353 \mu^4+361 \mu^2+304\right) \mu_\Delta^5+\left(567 \mu^8\right.\right.\vphantom{\frac{e}{m}}\\
&\;\;\;\;\left.\left.-31 \mu^6-159 \mu^4-321 \mu^2+986\right) \mu_\Delta^4+\left(336 \mu^8+364 \mu^6-366 \mu^4+324 \mu^2+612\right) \mu_\Delta^3\right.\vphantom{\frac{e}{m}}\\
&\;\;\;\;\left.\left.+\left(314 \mu^8+88 \mu^6-204 \mu^4+692 \mu^2-494\right) \mu_\Delta^2+14 \left(\mu^8+16 \mu^6+12 \mu^4-20 \mu^2\right.\right.\right. \vphantom{\frac{e}{m}}\\
&\;\;\;\;\left.\left.+1\right) \mu_\Delta-7\left(\mu^{10}-3 \mu^8-4 \mu^6-20 \mu^4+21 \mu^2-7\right)\right) \\
&+\frac{e^2 h_A^2 \Xi_3}{15552 F^2 m^2 \pi ^3 \mu_\Delta^6}\left(-567 \mu_\Delta^{12}+312 \mu_\Delta^{11}+\left(2268 \mu^2+2273\right) \mu_\Delta^{10}+\left(322-600 \mu^2\right) \mu_\Delta^9\right.\vphantom{\frac{e}{m}}\\
&\;\;\;\;\left.\left.-2 \left(1701 \mu^4+2561 \mu^2+1657\right) \mu_\Delta^8-2\left(36 \mu^4+1031 \mu^2+1205\right) \mu_\Delta^7+\left(2268 \mu^6\right.\right.\right.\vphantom{\frac{e}{m}}\\
&\;\;\;\;\left.\left.+3125 \mu^4+3048 \mu^2+1481\right) \mu_\Delta^6+6 \left(116 \mu^6+353 \mu^4+361 \mu^2+458\right) \mu_\Delta^5+\left(-567 \mu^8\right.\right.\vphantom{\frac{e}{m}}\\
&\;\;\;\;\left.\left.+31 \mu^6+159 \mu^4+321 \mu^2+1196\right) \mu_\Delta^4+\left(-336 \mu^8-364 \mu^6+366 \mu^4+324 \mu^2\right.\right.\vphantom{\frac{e}{m}}\\
&\;\;\;\;\left.\left.+612\right) \mu_\Delta^3+\left(-314 \mu^8-88 \mu^6+204 \mu^4+692 \mu^2-494\right) \mu_\Delta^2-14 \left(\mu^2-1\right)^2 \left(\mu^4\right.\right.\vphantom{\frac{e}{m}}\\
&\;\;\;\;\left.\left.+18 \mu^2-1\right) \mu_\Delta+7\left(\mu^2-1\right)^3 \left(\mu^4-7\right)\right) \\
&-\frac{e^2 h_A^2}{93312 F^2 m^2 \pi ^3 \mu_\Delta^6 \left(\mu^2-\mu_\Delta^2+2 \mu_\Delta-1\right)}  \left(3402 \mu_\Delta^{12}-8676 \mu_\Delta^{11}-3 \left(4536 \mu^2\right.\right.\vphantom{\frac{e}{m}}\\
&\;\;\;\;\left.\left.+1597\right) \mu_\Delta^{10}+18 \left(1334 \mu^2+1063\right) \mu_\Delta^9+3 \left(6804 \mu^4+5123 \mu^2+2666\right) \mu_\Delta^8\right.\vphantom{\frac{e}{m}}\\
&\;\;\;\;\left.\left.-6 \left(3330 \mu^4+3516 \mu^2+3829\right) \mu_\Delta^7-\left(13608 \mu^6+14565 \mu^4+18741 \mu^2+6716\right) \mu_\Delta^6\right.\right.\vphantom{\frac{e}{m}}\\
&\;\;\;\;\left.\left.+\left(2628 \mu^6-2898 \mu^4+9408 \mu^2+23984\right) \mu_\Delta^5+\left(3402 \mu^8+2145 \mu^6+6003 \mu^4-843\mu^2\right.\right.\right.\vphantom{\frac{e}{m}}\\
&\;\;\;\;\left.\left.-14987\right) \mu_\Delta^4+\left(2016 \mu^8+4860 \mu^6+3270 \mu^4-8018 \mu^2+298\right) \mu_\Delta^3+\left(1884 \mu^8-267 \mu^6\right.\right.\vphantom{\frac{e}{m}}\\
&\;\;\;\;\left.\left.-3495 \mu^4-1038 \mu^2+2090\right) \mu_\Delta^2-84 \left(30 \mu^6-41 \mu^4+28 \mu^2-8\right) \mu_\Delta-7 \left(6 \mu^{10}\right.\right.\vphantom{\frac{e}{m}}\\
&\;\;\;\;\left.\left.-21 \mu^8+196 \mu^6-349 \mu^4+218 \mu^2-50\right)\right)
\,, \end{align*}

\begin{align*} 
&\gamma_{\text{E1M2}}^{(p)}  =  
-\frac{e^2 h_A^2 \Xi_4}{15552 F^2 m^2 \pi ^3 \mu_\Delta^6 \left(\mu^2-\mu_\Delta^2+2 \mu_\Delta-1\right)^2 \left(\mu^2-\mu_\Delta^2-2 \mu_\Delta-1\right)}\left(81 \mu_\Delta^{16}\right.\vphantom{\frac{e}{m}}\\
&\;\;\;\;\left.\left.-210 \mu_\Delta^{15}-\left(486 \mu^2+475\right) \mu_\Delta^{14}+6 \left(183 \mu^2+178\right)\mu_\Delta^{13}+\left(1215 \mu^4+1714 \mu^2\right.\right.\right.\vphantom{\frac{e}{m}}\\
&\;\;\;\;\left.\left.+1583\right) \mu_\Delta^{12}-4 \left(585 \mu^4+642 \mu^2+428\right)\mu_\Delta^{11}-2 \left(810 \mu^6+1077 \mu^4+1765 \mu^2\right.\right.\vphantom{\frac{e}{m}}\\
&\;\;\;\;\left.\left.+2159\right) \mu_\Delta^{10}+2 \left(1290 \mu^6+654 \mu^4-511 \mu^2+901\right) \mu_\Delta^9+\left(1215 \mu^8+983 \mu^6+1733 \mu^4\right.\right.\vphantom{\frac{e}{m}}\\
&\;\;\;\;\left.\left.+5733 \mu^2+5652\right) \mu_\Delta^8+\left(-1530 \mu^8+762 \mu^6+3840 \mu^4+1570 \mu^2-804\right) \mu_\Delta^7\right.\vphantom{\frac{e}{m}}\\
&\;\;\;\;\left.\left.-\left(486 \mu^{10}+47 \mu^8-968 \mu^6+1340\mu^4-546 \mu^2+3607\right) \mu_\Delta^6+2 \left(225 \mu^{10}-237 \mu^8\right.\right.\right.\vphantom{\frac{e}{m}}\\
&\;\;\;\;\left.\left.-587 \mu^6+1003 \mu^4-352 \mu^2-248\right) \mu_\Delta^5+\left(81 \mu^{12}+48 \mu^{10}-937 \mu^8+1484 \mu^6\right.\right.\vphantom{\frac{e}{m}}\\
&\;\;\;\;\left.\left.+483 \mu^4-2438 \mu^2+1279\right) \mu_\Delta^4-2 \left(\mu^2-1\right)^2 \left(24 \mu^8+117 \mu^6+92 \mu^4+209 \mu^2\right.\right.\vphantom{\frac{e}{m}}\\
&\;\;\;\;\left.\left.-197\right) \mu_\Delta^3-2 \left(\mu^2-1\right)^3 \left(38 \mu^6+19 \mu^4+205 \mu^2-94\right) \mu_\Delta^2\right.\vphantom{\frac{e}{m}}\\
&\;\;\;\;\left.\left.+42 \left(\mu^2-1\right)^5 \left(\mu^2+1\right) \mu_\Delta+7 \left(\mu^2-1\right)^5\left(\mu^4+4 \mu^2+1\right)\right) \right.\\
&+\frac{e^2 h_A^2 \Xi_1}{15552 F^2 m^2 \pi ^3 \mu_\Delta^6} \left(81 \mu_\Delta^{12}-48\mu_\Delta^{11}-\left(324 \mu^2+571\right) \mu_\Delta^{10}+4 \left(48 \mu^2-59\right) \mu_\Delta^9\right.\vphantom{\frac{e}{m}}\\
&\;\;\;\;\left.\left.+2 \left(243 \mu^4+607 \mu^2+644\right) \mu_\Delta^8+\left(-288 \mu^4+758 \mu^2+1958\right) \mu_\Delta^7-\left(324 \mu^6+763 \mu^4\right.\right.\right.\vphantom{\frac{e}{m}}\\
&\;\;\;\;\left.\left.+300 \mu^2+501\right) \mu_\Delta^6+2 \left(96 \mu^6-255 \mu^4-611 \mu^2-1006\right)\mu_\Delta^5+\left(81 \mu^8+175 \mu^6\right.\right.\vphantom{\frac{e}{m}}\\
&\;\;\;\;\left.\left.-587 \mu^4-179 \mu^2-1000\right) \mu_\Delta^4-2 \left(24 \mu^8+20 \mu^6+75\mu^4+270 \mu^2-96\right) \mu_\Delta^3+\left(-62 \mu^8\right.\right.\vphantom{\frac{e}{m}}\\
&\;\;\;\;\left.\left.+24 \mu^6-246 \mu^4-800 \mu^2+300\right) \mu_\Delta^2+28\left(\mu^8-4 \mu^6-21 \mu^4-2 \mu^2+2\right) \mu_\Delta+7 \left(\mu^{10}\right.\right.\vphantom{\frac{e}{m}}\\
&\;\;\;\;\left.\left.+\mu^8-8 \mu^6-32 \mu^4+\mu^2+1\right)\right)\\
&-\frac{e^2 h_A^2 \Xi_3}{15552 F^2 m^2 \pi ^3 \mu_\Delta^6}\left(81 \mu_\Delta^{12}-48 \mu_\Delta^{11}-\left(324 \mu^2+571\right) \mu_\Delta^{10}+4 \left(48 \mu^2-59\right) \mu_\Delta^9\right.\vphantom{\frac{e}{m}}\\
&\;\;\;\;\left.\left.+2 \left(243 \mu^4+607 \mu^2+644\right) \mu_\Delta^8+\left(-288 \mu^4+758 \mu^2+1958\right) \mu_\Delta^7-\left(324 \mu^6+763 \mu^4\right.\right.\right.\vphantom{\frac{e}{m}}\\
&\;\;\;\;\left.\left.+300 \mu^2+501\right) \mu_\Delta^6+2 \left(96 \mu^6-255 \mu^4-611 \mu^2-544\right) \mu_\Delta^5+\left(81 \mu^8+175 \mu^6\right.\right.\vphantom{\frac{e}{m}}\\
&\;\;\;\;\left.\left.-587 \mu^4-179 \mu^2+1182\right) \mu_\Delta^4-2 \left(24 \mu^8+20 \mu^6+75 \mu^4-54 \mu^2+96\right) \mu_\Delta^3+\left(-62 \mu^8\right.\right.\vphantom{\frac{e}{m}}\\
&\;\;\;\;\left.\left.+24 \mu^6-246 \mu^4+584 \mu^2-300\right) \mu_\Delta^2+28 \left(\mu^2-1\right)^2 \left(\mu^4-2\mu^2-2\right) \mu_\Delta\right.\vphantom{\frac{e}{m}}\\
&\;\;\;\;\left.+7 \left(\mu^2-1\right)^3 \left(\mu^4+4 \mu^2+1\right)\right) \\
&-\frac{e^2  h_A^2}{93312 F^2 m^2 \pi ^3 \mu_\Delta^6\left(\mu^2-\mu_\Delta^2+2 \mu_\Delta-1\right)} \left(486 \mu_\Delta^{12}-1260 \mu_\Delta^{11}-3 \left(648 \mu^2+707\right) \mu_\Delta^{10}\right.\vphantom{\frac{e}{m}}\\
&\;\;\;\;\left.\left.+18 \left(226 \mu^2+251\right) \mu_\Delta^9+3 \left(972 \mu^4+1285 \mu^2+2038\right) \mu_\Delta^8-12\left(387 \mu^4+180 \mu^2\right.\right.\right.\vphantom{\frac{e}{m}}\\
&\;\;\;\;\left.\left.-134\right) \mu_\Delta^7-\left(1944 \mu^6+1635 \mu^4+5055 \mu^2+23338\right) \mu_\Delta^6+2 \left(1062 \mu^6-927 \mu^4\right.\right.\vphantom{\frac{e}{m}}\\
&\;\;\;\;\left.\left.-8307 \mu^2+4691\right) \mu_\Delta^5+\left(486 \mu^8+231 \mu^6-4683 \mu^4+465 \mu^2+4651\right) \mu_\Delta^4-2 \left(144 \mu^8\right.\right.\vphantom{\frac{e}{m}}\\
&\;\;\;\;\left.\left.+378 \mu^6+2154 \mu^4+11 \mu^2-1707\right) \mu_\Delta^3-3 \left(124\mu^8-229 \mu^6-629 \mu^4-980 \mu^2\right.\right.\vphantom{\frac{e}{m}}\\
&\;\;\;\;\left.\left.+944\right) \mu_\Delta^2+126 \left(2 \mu^8+27 \mu^6-29 \mu^4+16 \mu^2-2\right) \mu_\Delta+7 \left(6 \mu^{10}+3 \mu^8+160 \mu^6\right.\right.\vphantom{\frac{e}{m}}\\
&\;\;\;\;\left.\left.-249 \mu^4+102 \mu^2-22\right)\right)
\,, \end{align*}

\begin{align*} 
& \gamma_{\text{M1E2}}^{(p)}  =  
-\frac{e^2 h_A^2 \Xi_4}{15552 F^2 m^2 \pi ^3 \mu_\Delta^6 \left(\mu^2-\mu_\Delta^2+2 \mu_\Delta-1\right)^2 \left(\mu^2-\mu_\Delta^2-2 \mu_\Delta-1\right)}\left(-81\mu_\Delta^{16}\right.\vphantom{\frac{e}{m}}\\
&\;\;\;\;\left.\left.+114 \mu_\Delta^{15}+\left(486 \mu^2+233\right) \mu_\Delta^{14}-2 \left(261 \mu^2+347\right) \mu_\Delta^{13}+\left(-1215 \mu^4-932 \mu^2\right.\right.\right.\vphantom{\frac{e}{m}}\\
&\;\;\;\;\left.\left.+41\right) \mu_\Delta^{12}+4 \left(225 \mu^4+533 \mu^2+453\right) \mu_\Delta^{11}+2 \left(810 \mu^6+687 \mu^4+317 \mu^2-763\right) \mu_\Delta^{10}\right.\vphantom{\frac{e}{m}}\\
&\;\;\;\;\left.\left.-2 \left(330 \mu^6+1124 \mu^4+2069 \mu^2+442\right) \mu_\Delta^9+\left(-1215 \mu^8-829 \mu^6-1623 \mu^4+1709 \mu^2\right.\right.\right.\vphantom{\frac{e}{m}}\\
&\;\;\;\;\left.\left.+1544\right) \mu_\Delta^8+2 \left(45 \mu^8+425 \mu^6+1725 \mu^4+455 \mu^2-781\right) \mu_\Delta^7+\left(486 \mu^{10}+61 \mu^8\right.\right.\vphantom{\frac{e}{m}}\\
&\;\;\;\;\left.\left.+1324 \mu^6-1744 \mu^4-894 \mu^2+613\right) \mu_\Delta^6+2 \left(63 \mu^{10}+14 \mu^8-705 \mu^6+360 \mu^4\right.\right.\vphantom{\frac{e}{m}}\\
&\;\;\;\;\left.\left.-572 \mu^2+924\right) \mu_\Delta^5+\left(-81 \mu^{12}+138 \mu^{10}-541 \mu^8+508 \mu^6-1005 \mu^4+2404 \mu^2\right.\right.\vphantom{\frac{e}{m}}\\
&\;\;\;\;\left.\left.-1423\right) \mu_\Delta^4-2\left(\mu^2-1\right)^2 \left(24 \mu^8+103 \mu^6+39 \mu^4+63 \mu^2+296\right) \mu_\Delta^3\right.\vphantom{\frac{e}{m}}\\
&\;\;\;\;\left.\left.-2 \left(\mu^2-1\right)^3 \left(26 \mu^6-15 \mu^4-153 \mu^2+310\right) \mu_\Delta^2+42 \left(\mu^2-1\right)^4 \left(\mu^4+4 \mu^2-1\right) \mu_\Delta\right.\right.\vphantom{\frac{e}{m}}\\
&\;\;\;\;\left.\left.+7 \left(\mu^2-1\right)^5 \left(\mu^4+2 \mu^2+3\right)\right)\right. \\
&+\frac{e^2 h_A^2 \Xi_1}{15552 F^2 m^2\pi ^3 \mu_\Delta^6} \left(-81 \mu_\Delta^{12}-48 \mu_\Delta^{11}+\left(324 \mu^2+137\right) \mu_\Delta^{10}+6 \left(32 \mu^2-43\right) \mu_\Delta^9\right.\vphantom{\frac{e}{m}}\\
&\;\;\;\;\left.\left.-2 \left(243 \mu^4+170 \mu^2+230\right) \mu_\Delta^8+\left(-288 \mu^4+242 \mu^2+570\right) \mu_\Delta^7+\left(324 \mu^6+245 \mu^4\right.\right.\right.\vphantom{\frac{e}{m}}\\
&\;\;\;\;\left.\left.+796 \mu^2+267\right) \mu_\Delta^6+6 \left(32 \mu^6+8 \mu^4-91 \mu^2+52\right) \mu_\Delta^5-\left(81 \mu^8+11 \mu^6+315 \mu^4\right.\right.\vphantom{\frac{e}{m}}\\
&\;\;\;\;\left.\left.+285 \mu^2-568\right) \mu_\Delta^4-6 \left(8 \mu^8+10 \mu^6-40 \mu^4-48\mu^2+59\right) \mu_\Delta^3+\left(-38 \mu^8+40 \mu^6\right.\right.\vphantom{\frac{e}{m}}\\
&\;\;\;\;\left.\left.+66 \mu^4+520 \mu^2-452\right) \mu_\Delta^2+28 \left(\mu^8+3 \mu^6+12 \mu^4-11 \mu^2+3\right) \mu_\Delta+7 \left(\mu^{10}-\mu^8\right.\right.\vphantom{\frac{e}{m}}\\
&\;\;\;\;\left.\left.+4 \mu^4-7 \mu^2+3\right)\right)\\
&+\frac{e^2 h_A^2 \Xi_3}{15552 F^2 m^2 \pi ^3 \mu_\Delta^6}\left(81 \mu_\Delta^{12}+48 \mu_\Delta^{11}-\left(324 \mu^2+137\right) \mu_\Delta^{10}-6 \left(32 \mu^2-43\right) \mu_\Delta^9\right.\vphantom{\frac{e}{m}}\\
&\;\;\;\;\left.\left.+\left(486 \mu^4+340 \mu^2+460\right)\mu_\Delta^8+\left(288 \mu^4-242 \mu^2-570\right) \mu_\Delta^7-\left(324 \mu^6+245 \mu^4\right.\right.\right.\vphantom{\frac{e}{m}}\\
&\;\;\;\;\left.\left.+796 \mu^2+267\right)\mu_\Delta^6-6 \left(32 \mu^6+8 \mu^4-91 \mu^2-102\right) \mu_\Delta^5+\left(81 \mu^8+11 \mu^6+315 \mu^4\right.\right.\vphantom{\frac{e}{m}}\\
&\;\;\;\;\left.\left.+285\mu^2+1282\right) \mu_\Delta^4+6 \left(8 \mu^8+10 \mu^6-40 \mu^4+60 \mu^2-59\right) \mu_\Delta^3+\left(38 \mu^8-40 \mu^6\right.\right.\vphantom{\frac{e}{m}}\\
&\;\;\;\;\left.\left.-66 \mu^4+520 \mu^2-452\right) \mu_\Delta^2-28 \left(\mu^2-1\right)^2 \left(\mu^4+5 \mu^2-3\right) \mu_\Delta\right.\vphantom{\frac{e}{m}}\\
&\;\;\;\;\left.\left.-7\left(\mu^2-1\right)^3 \left(\mu^4+2 \mu^2+3\right)\right) \right.\\
&-\frac{e^2  h_A^2}{93312 F^2 m^2 \pi ^3 \mu_\Delta^6 \left(\mu^2-\mu_\Delta^2+2 \mu_\Delta-1\right)}\left(-486 \mu_\Delta^{12}+684 \mu_\Delta^{11}+3 \left(648 \mu^2+223\right) \mu_\Delta^{10}\right.\vphantom{\frac{e}{m}}\\
&\;\;\;\;\left.\left.-6 \left(294 \mu^2+523\right) \mu_\Delta^9-3 \left(972 \mu^4+689 \mu^2-484\right) \mu_\Delta^8+6\left(198 \mu^4+816 \mu^2+217\right) \mu_\Delta^7\right.\right.\vphantom{\frac{e}{m}}\\
&\;\;\;\;\left.\left.+\left(1944 \mu^6+1983 \mu^4+3315 \mu^2+1874\right) \mu_\Delta^6+2 \left(90 \mu^6-711 \mu^4-1944 \mu^2-3338\right) \mu_\Delta^5\right.\right.\vphantom{\frac{e}{m}}\\
&\;\;\;\;\left.\left.+\left(-486 \mu^8-399 \mu^6-2499 \mu^4+4659 \mu^2+9271\right) \mu_\Delta^4+\left(-288 \mu^8-588 \mu^6+5916 \mu^4\right.\right.\right.\vphantom{\frac{e}{m}}\\
&\;\;\;\;\left.\left.+970 \mu^2-4090\right) \mu_\Delta^3+\left(-228\mu^8+711 \mu^6-4995 \mu^4+4248 \mu^2-1444\right) \mu_\Delta^2+42 \left(6 \mu^8\right.\right.\vphantom{\frac{e}{m}}\\
&\;\;\;\;\left.\left.-87 \mu^6+115 \mu^4-56 \mu^2+10\right) \mu_\Delta+7 \left(6 \mu^{10}-9 \mu^8-146 \mu^6+313 \mu^4-218 \mu^2+54\right)\right) 
\,. \end{align*}

\subsection{Neutron values}\label{sub:neutron-e3} 

\subsubsection{Spin-independent first order polarizabilities}\label{subsub:neutron-spin-independent-1-e3} 

\begin{align*} 
&\alpha_{\text{E1}}^{(n)}  =  
-\frac{e^2 h_A^2 \Xi_4}{3888 F^2 m \pi ^3 \mu_\Delta^6 \left(-\mu^2+\mu_\Delta^2-2\mu_\Delta+1\right)^2 \left(-\mu^2+\mu_\Delta^2+2 \mu_\Delta+1\right)} \left[\vphantom{\frac{e}{m}}-8 \mu_\Delta^{14}\right.\vphantom{\frac{e}{m}}\\
&\;\;\;\;\left.\left.+70 \mu_\Delta^{13}+\left(50 \mu^2-239\right) \mu_\Delta^{12}+\left(105-348 \mu^2\right)\mu_\Delta^{11}+\left(-132 \mu^4+936 \mu^2+429\right) \mu_\Delta^{10}\right.\right.\vphantom{\frac{e}{m}}\\
&\;\;\;\;\left.\left.+3 \left(230 \mu^4-135 \mu^2-93\right)\mu_\Delta^9+2 \left(95 \mu^6-681 \mu^4-150 \mu^2-157\right) \mu_\Delta^8+\left(-680 \mu^6\right.\right.\right.\vphantom{\frac{e}{m}}\\
&\;\;\;\;\left.\left.+597 \mu^4+90 \mu^2+133\right) \mu_\Delta^7+\left(-160 \mu^8+868 \mu^6-651 \mu^4+98 \mu^2+217\right) \mu_\Delta^6\right.\vphantom{\frac{e}{m}}\\
&\;\;\;\;\left.\left.+3 \left(110 \mu^8-137 \mu^6+81 \mu^4+5 \mu^2-59\right) \mu_\Delta^5+3 \left(\mu^2-1\right)^2 \left(26 \mu^6-17 \mu^4+104 \mu^2\right.\right.\right.\vphantom{\frac{e}{m}}\\
&\;\;\;\;\left.\left.+13\right)\mu_\Delta^4-6 \left(\mu^2-1\right)^3 \left(10 \mu^4+9 \mu^2+21\right) \mu_\Delta^3-4 \left(\mu^2-1\right)^4\left(5 \mu^4+17 \mu^2+32\right) \mu_\Delta^2\right.\vphantom{\frac{e}{m}}\\
&\;\;\;\;\left.\left.-2 \left(\mu^2-1\right)^5 \left(\mu^2+11\right) \mu_\Delta+2 \left(\mu^2-1\right)^6 \left(\mu^2+2\right)\vphantom{\frac{e}{m}}\right]\right. \\
&+\frac{e^2 h_A^2 \Xi_1}{3888 F^2 m \pi ^3 \mu_\Delta^6} \left[\vphantom{\frac{e}{m}}-2 \mu^{10}+4 \mu^8+4 \mu^6+16 \mu^4-14 \mu^2+8 \mu_\Delta^{10}-54 \mu_\Delta^9+\left(131\right.\right.\vphantom{\frac{e}{m}}\\
&\;\;\;\;\left.\left.-34 \mu^2\right) \mu_\Delta^8+3 \left(52 \mu^2+47\right) \mu_\Delta^7+\left(56 \mu^4-238 \mu^2-31\right) \mu_\Delta^6-3 \left(48 \mu^4+41 \mu^2\right.\right.\vphantom{\frac{e}{m}}\\
&\;\;\;\;\left.\left.+33\right) \mu_\Delta^5-\left(44 \mu^6-87 \mu^4+126 \mu^2+35\right) \mu_\Delta^4+6 \left(6 \mu^6-5 \mu^4+4 \mu^2-3\right) \mu_\Delta^3+4 \left(4 \mu^8\right.\right.\vphantom{\frac{e}{m}}\\
&\;\;\;\;\left.\left.+4 \mu^6-3 \mu^4+22 \mu^2-17\right) \mu_\Delta^2+6 \left(\mu^8+2 \mu^6+12 \mu^4-14 \mu^2+5\right)\mu_\Delta+4\vphantom{\frac{e}{m}}\right]\\
&+ \frac{e^2 h_A^2 \Xi_3}{3888 F^2m \pi ^3 \mu_\Delta^6} \left[\vphantom{\frac{e}{m}}-8 \mu_\Delta^{10}+54 \mu_\Delta^9+\left(34 \mu^2-131\right) \mu_\Delta^8-3 \left(52 \mu^2+47\right) \mu_\Delta^7+\left(-56 \mu^4\right.\right.\vphantom{\frac{e}{m}}\\
&\;\;\;\;\left.\left.+238 \mu^2+31\right) \mu_\Delta^6+3\left(48 \mu^4+41 \mu^2-71\right) \mu_\Delta^5+\left(44 \mu^6-87 \mu^4+126 \mu^2-53\right) \mu_\Delta^4\right.\vphantom{\frac{e}{m}}\\
&\;\;\;\;\left.\left.-6\left(6 \mu^6-5 \mu^4-4 \mu^2+3\right) \mu_\Delta^3-4 \left(\mu^2-1\right)^2 \left(4 \mu^4+12 \mu^2+17\right) \mu_\Delta^2-6 \left(\mu^2-1\right)^3 \left(\mu^2\right.\right.\right.\vphantom{\frac{e}{m}}\\
&\;\;\;\;\left.\left.+5\right) \mu_\Delta+2 \left(\mu^2-1\right)^4 \left(\mu^2+2\right)\vphantom{\frac{e}{m}}\right]\\
&+\frac{e^2 h_A^2}{23328 F^2 m \pi ^3 \mu_\Delta^6\left(-\mu^2+\mu_\Delta^2-2 \mu_\Delta+1\right)} \left[\vphantom{\frac{e}{m}}48 \mu_\Delta^{10}-420 \mu_\Delta^9-6 \left(34 \mu^2-251\right) \mu_\Delta^8\right.\vphantom{\frac{e}{m}}\\
&\;\;\;\;\left.\left.+12 \left(104 \mu^2-123\right) \mu_\Delta^7+\left(336\mu^4-2838 \mu^2+413\right) \mu_\Delta^6+\left(-1224 \mu^4+1764 \mu^2\right.\right.\right.\vphantom{\frac{e}{m}}\\
&\;\;\;\;\left.\left.+62\right) \mu_\Delta^5+\left(-264 \mu^6+1188 \mu^4-399 \mu^2+1\right) \mu_\Delta^4+2 \left(192 \mu^6-9 \mu^4-230 \mu^2\right.\right.\vphantom{\frac{e}{m}}\\
&\;\;\;\;\left.\left.+22\right) \mu_\Delta^3+2\left(48 \mu^8+57 \mu^6-510 \mu^4+427 \mu^2-85\right) \mu_\Delta^2+2 \left(6 \mu^8-135 \mu^6+85 \mu^4\right.\right.\vphantom{\frac{e}{m}}\\
&\;\;\;\;\left.\left.+4 \mu^2-14\right) \mu_\Delta+2 \left(-6 \mu^{10}+15 \mu^8-73 \mu^6+108 \mu^4-54 \mu^2+10\right)\vphantom{\frac{e}{m}}\right]
\,, \end{align*}

\begin{align*} 
& \beta_{\text{M1}}^{(n)}  =   
-\frac{e^2 h_A^2 \Xi_4}{3888 F^2 m \pi ^3 \mu_\Delta^6 \left(-\mu^2+\mu_\Delta^2-2 \mu_\Delta+1\right)} \left[\vphantom{\frac{e}{m}}-8 \mu_\Delta^{10}+70 \mu_\Delta^9+\left(34 \mu^2-183\right) \mu_\Delta^8\right.\vphantom{\frac{e}{m}}\\
&\;\;\;\;\left.+\left(245-208 \mu^2\right) \mu_\Delta^7+\left(-56 \mu^4+358 \mu^2-241\right) \mu_\Delta^6+\left(204 \mu^4-305 \mu^2+45\right) \mu_\Delta^5\right.\vphantom{\frac{e}{m}}\\
&\;\;\;\;\left.+\left(44 \mu^6-177 \mu^4+38 \mu^2+107\right) \mu_\Delta^4+\left(-64 \mu^6+52 \mu^4+100 \mu^2-88\right) \mu_\Delta^3\right.\vphantom{\frac{e}{m}}\\
&\;\;\;\;\left.-4 \left(\mu^2-1\right)^2 \left(4 \mu^4+5 \mu^2-12\right) \mu_\Delta^2-2 \left(\mu^2-1\right)^4 \mu_\Delta+2 \left(\mu^2-1\right)^5\vphantom{\frac{e}{m}}\right]\\
&-\frac{e^2 h_A^2 \Xi_1}{3888 F^2 m \pi ^3 \mu_\Delta^6} \left[\vphantom{\frac{e}{m}}-8 \mu_\Delta^{10}+54 \mu_\Delta^9+\left(34 \mu^2-59\right) \mu_\Delta^8+\left(3-156 \mu^2\right) \mu_\Delta^7+\left(-56 \mu^4\right.\right.\vphantom{\frac{e}{m}}\\
&\;\;\;\;\left.\left.+118 \mu^2+7\right) \mu_\Delta^6+3\left(48 \mu^4-29 \mu^2-63\right) \mu_\Delta^5+\left(44 \mu^6-69 \mu^4+18 \mu^2-37\right) \mu_\Delta^4\right.\vphantom{\frac{e}{m}}\\
&\;\;\;\;\left.+12 \mu^2\left(-3 \mu^4+5 \mu^2+2\right) \mu_\Delta^3+\left(-16 \mu^8+20 \mu^6+48 \mu^4+92 \mu^2-40\right) \mu_\Delta^2-6 \left(\mu^8\right.\right.\vphantom{\frac{e}{m}}\\
&\;\;\;\;\left.\left.-4 \mu^6-18 \mu^4+4 \mu^2-1\right) \mu_\Delta+2 \left(\mu^{10}-5 \mu^8+10 \mu^6+22 \mu^4-5 \mu^2+1\right)\vphantom{\frac{e}{m}}\right] \\
&+\frac{e^2 h_A^2 \Xi_3}{3888 F^2 m \pi ^3 \mu_\Delta^6}\left[\vphantom{\frac{e}{m}}-8 \mu_\Delta^{10}+54\mu_\Delta^9+\left(34 \mu^2-59\right) \mu_\Delta^8+\left(3-156 \mu^2\right) \mu_\Delta^7+\left(-56\mu^4\right.\right.\vphantom{\frac{e}{m}}\\
&\;\;\;\;\left.\left.+118 \mu^2+7\right) \mu_\Delta^6+3 \left(48 \mu^4-29 \mu^2+41\right) \mu_\Delta^5+\left(44 \mu^6-69 \mu^4+18 \mu^2+19\right) \mu_\Delta^4\right.\vphantom{\frac{e}{m}}\\
&\;\;\;\;\left.-12 \mu^2 \left(3 \mu^4-5 \mu^2+2\right) \mu_\Delta^3-4 \left(\mu^2-1\right)^2\left(4 \mu^4+3 \mu^2-10\right) \mu_\Delta^2\right.\vphantom{\frac{e}{m}}\\
&\;\;\;\;\left.-6 \left(\mu^2-1\right)^4 \mu_\Delta+2 \left(\mu^2-1\right)^5\vphantom{\frac{e}{m}}\right] \\
&+ \frac{e^2  h_A^2}{23328 F^2 m \pi ^3 \mu_\Delta^6} \left[\vphantom{\frac{e}{m}}48 \mu_\Delta^8-324 \mu_\Delta^7-6 \left(26 \mu^2-63\right) \mu_\Delta^6+36 \left(17\mu^2+1\right) \mu_\Delta^5+\left(180 \mu^4\right.\right.\vphantom{\frac{e}{m}}\\
&\;\;\;\;\left.\left.-360 \mu^2+371\right) \mu_\Delta^4-12 \left(21 \mu^4+6 \mu^2-17\right) \mu_\Delta^3-2 \left(42 \mu^6-18 \mu^4-94 \mu^2+77\right) \mu_\Delta^2\right.\vphantom{\frac{e}{m}}\\
&\;\;\;\;\left.-6 \left(6 \mu^6+39 \mu^4-22\mu^2+4\right) \mu_\Delta+2 \left(6 \mu^8-27 \mu^6-65 \mu^4+16 \mu^2-2\right)\vphantom{\frac{e}{m}}\right]
\,. \end{align*}

\subsubsection{Spin-dependent first order polarizabilities}\label{subsub:neutron-spin-dependent-1-e3} 

\begin{align*} 
&\gamma_{\text{E1E1}}^{(n)}  =  \;
\frac{e^2 h_A^2 \Xi_4}{7776 F^2 m^2 \pi ^3 \mu_\Delta^6 \left(\mu^2-\mu_\Delta^2+2 \mu_\Delta-1\right)^3 \left(\mu^2-\mu_\Delta^2-2 \mu_\Delta-1\right)^2} \left[\vphantom{\frac{e}{m}}-34 \mu_\Delta^{18}\right.\vphantom{\frac{e}{m}}\\
&\;\;\;\;\left.\left.+122 \mu_\Delta^{17}+\left(273 \mu^2-38\right) \mu_\Delta^{16}-7 \left(122 \mu^2+51\right) \mu_\Delta^{15}+\left(-960 \mu^4+316 \mu^2-49\right) \mu_\Delta^{14}\right.\right.\vphantom{\frac{e}{m}}\\
&\;\;\;\;\left.\left.+3 \left(854 \mu^4+457 \mu^2+347\right) \mu_\Delta^{13}+\left(1932 \mu^6-1107 \mu^4+708 \mu^2+733\right) \mu_\Delta^{12}-2 \left(2135 \mu^6\right.\right.\right.\vphantom{\frac{e}{m}}\\
&\;\;\;\;\left.\left.+744 \mu^4+1347 \mu^2+1265\right) \mu_\Delta^{11}-2 \left(1218 \mu^8-1067 \mu^6+1116 \mu^4+507 \mu^2+625\right) \mu_\Delta^{10}\right.\vphantom{\frac{e}{m}}\\
&\;\;\;\;\left.\left.+\left(4270 \mu^8-630 \mu^6+2142 \mu^4+898 \mu^2+3456\right) \mu_\Delta^9+\left(1974 \mu^{10}-2465 \mu^8+2858 \mu^6\right.\right.\right.\vphantom{\frac{e}{m}}\\
&\;\;\;\;\left.\left.-1008 \mu^4+946 \mu^2+935\right) \mu_\Delta^8-3 \left(854 \mu^{10}-825 \mu^8+228 \mu^6-482 \mu^4-754 \mu^2\right.\right.\vphantom{\frac{e}{m}}\\
&\;\;\;\;\left.\left.+915\right) \mu_\Delta^7+\left(-1008 \mu^{12}+1728 \mu^{10}-1557 \mu^8+1256 \mu^6+540 \mu^4-804 \mu^2-155\right) \mu_\Delta^6\right.\vphantom{\frac{e}{m}}\\
&\;\;\;\;\left.\left.+\left(\mu^2-1\right)^2 \left(854 \mu^8-125 \mu^6-579 \mu^4+97 \mu^2+1265\right) \mu_\Delta^5+\left(\mu^2-1\right)^3 \left(300 \mu^8\right.\right.\right.\vphantom{\frac{e}{m}}\\
&\;\;\;\;\left.\left.+199 \mu^6-51 \mu^4+411 \mu^2+323\right) \mu_\Delta^4-2 \left(\mu^2-1\right)^4 \left(61 \mu^6+7 \mu^4-167 \mu^2+96\right) \mu_\Delta^3\right.\vphantom{\frac{e}{m}}\\
&\;\;\;\;\left.\left.-2\left(\mu^2-1\right)^5 \left(21 \mu^6+34 \mu^4-35 \mu^2+88\right) \mu_\Delta^2-12 \left(\mu^2-1\right)^6 \left(\mu^2+5\right)\mu_\Delta\right.\right.\vphantom{\frac{e}{m}}\\
&\;\;\;\;\left.\left.+\left(\mu^2-1\right)^7 \left(\mu^4-2 \mu^2-5\right)\vphantom{\frac{e}{m}}\right]\right.\\
&+\frac{e^2 h_A^2 \Xi_1}{7776 F^2 m^2 \pi ^3\mu_\Delta^6}\left[\vphantom{\frac{e}{m}}-\mu^{10}+5 \mu^8-4 \mu^6-16 \mu^4-13 \mu^2+34 \mu_\Delta^{10}-54 \mu_\Delta^9-\left(137 \mu^2\right.\right.\vphantom{\frac{e}{m}}\\
&\;\;\;\;\left.\left.+2\right)\mu_\Delta^8+\left(160 \mu^2+41\right) \mu_\Delta^7+\left(208 \mu^4+46 \mu^2+451\right) \mu_\Delta^6+\left(-156 \mu^4+39 \mu^2\right.\right.\vphantom{\frac{e}{m}}\\
&\;\;\;\;\left.\left.+159\right) \mu_\Delta^5-\left(142 \mu^6+81 \mu^4-60 \mu^2+151\right) \mu_\Delta^4+6 \left(8 \mu^6-14 \mu^4+12 \mu^2-25\right) \mu_\Delta^3\right.\vphantom{\frac{e}{m}}\\
&\;\;\;\;\left.\left.+2 \left(19 \mu^8+16 \mu^6-66 \mu^4-14 \mu^2+43\right)\mu_\Delta^2+2 \left(\mu^8+2 \mu^6-30 \mu^4+46 \mu^2\right.\right.\right.\vphantom{\frac{e}{m}}\\
&\;\;\;\;\left.\left.-25\right) \mu_\Delta+5\vphantom{\frac{e}{m}}\right] \\
&-\frac{e^2 h_A^2 \Xi_3}{7776 F^2 m^2 \pi ^3 \mu_\Delta^6}\left[\vphantom{\frac{e}{m}}34 \mu_\Delta^{10}-54 \mu_\Delta^9-\left(137 \mu^2+2\right) \mu_\Delta^8+\left(160 \mu^2+41\right) \mu_\Delta^7+\left(208 \mu^4\right.\right.\vphantom{\frac{e}{m}}\\
&\;\;\;\;\left.\left.+46 \mu^2+451\right) \mu_\Delta^6+\left(-156 \mu^4+39 \mu^2+279\right) \mu_\Delta^5+\left(-142 \mu^6-81 \mu^4+60 \mu^2\right.\right.\vphantom{\frac{e}{m}}\\
&\;\;\;\;\left.\left.+121\right) \mu_\Delta^4+6 \left(8 \mu^6-14 \mu^4+8 \mu^2+25\right) \mu_\Delta^3+2 \left(19 \mu^8+16 \mu^6-66 \mu^4+74 \mu^2\right.\right.\vphantom{\frac{e}{m}}\\
&\;\;\;\;\left.\left.-43\right)\mu_\Delta^2+2 \left(\mu^2-1\right)^2 \left(\mu^4+4 \mu^2+25\right) \mu_\Delta-\left(\mu^2-1\right)^3 \left(\mu^4-2 \mu^2-5\right)\vphantom{\frac{e}{m}}\right]\\
&+\frac{e^2 h_A^2}{46656 F^2 m^2 \pi ^3 \mu_\Delta^6 \left(\mu^2-\mu_\Delta^2+2 \mu_\Delta-1\right)^2 \left(\mu^2-\mu_\Delta^2-2 \mu_\Delta-1\right)}  \left[\vphantom{\frac{e}{m}}-204 \mu_\Delta^{14}+732 \mu_\Delta^{13}\right.\vphantom{\frac{e}{m}}\\
&\;\;\;\;\left.\left.+6 \left(205 \mu^2-89\right) \mu_\Delta^{12}-12 \left(305 \mu^2+132\right) \mu_\Delta^{11}+\left(-3096 \mu^4+2367 \mu^2+230\right) \mu_\Delta^{10}\right.\right.\vphantom{\frac{e}{m}}\\
&\;\;\;\;\left.\left.+\left(7320 \mu^4+3186 \mu^2+5678\right) \mu_\Delta^9+\left(4170 \mu^6-4173 \mu^4+2261 \mu^2-1530\right) \mu_\Delta^8\right.\right.\vphantom{\frac{e}{m}}\\
&\;\;\;\;\left.\left.-2 \left(3660 \mu^6-297 \mu^4+4827 \mu^2+4253\right) \mu_\Delta^7+\left(-3180 \mu^8+3702 \mu^6-5173 \mu^4\right.\right.\right.\vphantom{\frac{e}{m}}\\
&\;\;\;\;\left.\left.+4783 \mu^2+4820\right) \mu_\Delta^6+\left(3660 \mu^8-5058 \mu^6+5706 \mu^4-8298 \mu^2+2966\right) \mu_\Delta^5\right.\vphantom{\frac{e}{m}}\\
&\;\;\;\;\left.\left.+\left(1314 \mu^{10}-1728 \mu^8+2605 \mu^6-6124 \mu^4+6079 \mu^2-2050\right) \mu_\Delta^4+\left(-732 \mu^{10}\right.\right.\right.\vphantom{\frac{e}{m}}\\
&\;\;\;\;\left.\left.+3510 \mu^8-4274 \mu^6+3220 \mu^4-1262 \mu^2+402\right) \mu_\Delta^3+\left(-240 \mu^{12}+411 \mu^{10}+115 \mu^8\right.\right.\vphantom{\frac{e}{m}}\\
&\;\;\;\;\left.\left.+1127 \mu^6-3369 \mu^4+2714 \mu^2-758\right) \mu_\Delta^2-24 \left(\mu^2-1\right)^2 \left(27 \mu^6-52 \mu^4+47 \mu^2\right.\right.\vphantom{\frac{e}{m}}\\
&\;\;\;\;\left.\left.-13\right) \mu_\Delta+\left(\mu^2-1\right)^3 \left(6 \mu^8-27 \mu^6-137 \mu^4+76 \mu^2-26\right)\vphantom{\frac{e}{m}}\right]
\,, \end{align*}

\begin{align*} 
& \gamma_{\text{M1M1}}^{(n)}  =  \;
\frac{e^2 h_A^2 \Xi_4}{7776 F^2 m^2 \pi ^3 \mu_\Delta^6 \left(\mu^2-\mu_\Delta^2+2 \mu_\Delta-1\right)^2 \left(\mu^2-\mu_\Delta^2-2 \mu_\Delta-1\right)}\left[\vphantom{\frac{e}{m}}-34 \mu_\Delta^{14}\right.\vphantom{\frac{e}{m}}\\
&\;\;\;\;\left.\left.+102 \mu_\Delta^{13}+\left(205 \mu^2-262\right) \mu_\Delta^{12}+\left(139-510 \mu^2\right) \mu_\Delta^{11}+\left(-516 \mu^4+1114 \mu^2+613\right) \mu_\Delta^{10}\right.\right.\vphantom{\frac{e}{m}}\\
&\;\;\;\;\left.\left.+\left(1020 \mu^4-493 \mu^2-543\right) \mu_\Delta^9+\left(695 \mu^6-1841\mu^4-727 \mu^2-659\right) \mu_\Delta^8-3 \left(340 \mu^6\right.\right.\right.\vphantom{\frac{e}{m}}\\
&\;\;\;\;\left.\left.-227 \mu^4+62 \mu^2-193\right) \mu_\Delta^7+\left(-530\mu^8+1464 \mu^6-497 \mu^4+562 \mu^2+567\right) \mu_\Delta^6\right.\vphantom{\frac{e}{m}}\\
&\;\;\;\;\left.\left.+5 \left(102 \mu^8-95 \mu^6+149 \mu^4+15 \mu^2-107\right)\mu_\Delta^5+\left(219 \mu^{10}-556 \mu^8+726 \mu^6-360 \mu^4\right.\right.\right.\vphantom{\frac{e}{m}}\\
&\;\;\;\;\left.\left.+149 \mu^2-178\right) \mu_\Delta^4-2 \left(\mu^2-1\right)^2 \left(51 \mu^6+10 \mu^4+43 \mu^2-135\right) \mu_\Delta^3-2 \left(\mu^2\right.\right.\vphantom{\frac{e}{m}}\\
&\;\;\;\;\left.\left.-1\right)^3 \left(20 \mu^6+17 \mu^4+50 \mu^2-27\right) \mu_\Delta^2-12 \left(\mu^2-1\right)^4 \left(3 \mu^2+1\right) \mu_\Delta\right.\vphantom{\frac{e}{m}}\\
&\;\;\;\;\left.\left.+\left(\mu^2-1\right)^5 \left(\mu^4-7\right)\vphantom{\frac{e}{m}}\right]\right. \\
&+\frac{e^2 h_A^2 \Xi_1}{7776 F^2 m^2 \pi ^3 \mu_\Delta^6} \left[\vphantom{\frac{e}{m}}-\mu^{10}+3 \mu^8+4 \mu^6+20 \mu^4-21 \mu^2+34\mu_\Delta^{10}-34 \mu_\Delta^9+\left(194\right.\right.\vphantom{\frac{e}{m}}\\
&\;\;\;\;\left.\left.-137 \mu^2\right) \mu_\Delta^8+\left(100 \mu^2+181\right)\mu_\Delta^7+\left(208 \mu^4-390 \mu^2-149\right) \mu_\Delta^6-3 \left(32 \mu^4+81 \mu^2\right.\right.\vphantom{\frac{e}{m}}\\
&\;\;\;\;\left.\left.+59\right) \mu_\Delta^5+\left(-142 \mu^6+201 \mu^4-96 \mu^2+137\right) \mu_\Delta^4+2 \left(14 \mu^6+15 \mu^4-6 \mu^2+78\right)\mu_\Delta^3\right.\vphantom{\frac{e}{m}}\\
&\;\;\;\;\left.\left.+2 \left(19 \mu^8-4 \mu^6+6 \mu^4+46 \mu^2-25\right) \mu_\Delta^2+2 \left(\mu^8+16 \mu^6+12 \mu^4-20 \mu^2+1\right) \mu_\Delta+7\vphantom{\frac{e}{m}}\right]\right.\\
&+\frac{e^2 h_A^2 \Xi_3}{7776 F^2 m^2 \pi ^3 \mu_\Delta^6}\left[\vphantom{\frac{e}{m}}-34 \mu_\Delta^{10}+34 \mu_\Delta^9+\left(137 \mu^2-194\right) \mu_\Delta^8-\left(100 \mu^2+181\right) \mu_\Delta^7\right.\vphantom{\frac{e}{m}}\\
&\;\;\;\;\left.\left.+\left(-208 \mu^4+390 \mu^2+149\right) \mu_\Delta^6+3 \left(32 \mu^4+81 \mu^2+99\right) \mu_\Delta^5+\left(142 \mu^6-201 \mu^4\right.\right.\right.\vphantom{\frac{e}{m}}\\
&\;\;\;\;\left.\left.+96 \mu^2+167\right) \mu_\Delta^4-2 \left(14 \mu^6+15 \mu^4+6 \mu^2-78\right) \mu_\Delta^3-2 \left(19 \mu^8-4 \mu^6+6 \mu^4-46 \mu^2\right.\right.\vphantom{\frac{e}{m}}\\
&\;\;\;\;\left.\left.+25\right) \mu_\Delta^2-2 \left(\mu^2-1\right)^2 \left(\mu^4+18 \mu^2-1\right) \mu_\Delta+\left(\mu^2-1\right)^3 \left(\mu^4-7\right)\vphantom{\frac{e}{m}}\right] \\
&-\frac{e^2  h_A^2}{46656 F^2 m^2 \pi ^3 \mu_\Delta^6 \left(\mu^2-\mu_\Delta^2+2 \mu_\Delta-1\right)}\left[\vphantom{\frac{e}{m}}-6 \mu^{10}+21 \mu^8-196 \mu^6+349 \mu^4-218 \mu^2\right.\vphantom{\frac{e}{m}}\\
&\;\;\;\;\left.\left.+204 \mu_\Delta^{10}-612 \mu_\Delta^9+\left(1878-822 \mu^2\right) \mu_\Delta^8+12 \left(153 \mu^2-191\right) \mu_\Delta^7+\left(1248 \mu^4\right.\right.\right.\vphantom{\frac{e}{m}}\\
&\;\;\;\;\left.\left.-3855 \mu^2-242\right) \mu_\Delta^6+\left(-1836 \mu^4+1650 \mu^2+2390\right) \mu_\Delta^5-\left(852 \mu^6-2097\mu^4+285 \mu^2\right.\right.\vphantom{\frac{e}{m}}\\
&\;\;\;\;\left.\left.+2024\right) \mu_\Delta^4+2 \left(306 \mu^6+501 \mu^4-727 \mu^2+11\right) \mu_\Delta^3+\left(228\mu^8-141 \mu^6-129 \mu^4\right.\right.\vphantom{\frac{e}{m}}\\
&\;\;\;\;\left.\left.-498 \mu^2+422\right) \mu_\Delta^2-12 \left(30 \mu^6-41 \mu^4+28 \mu^2-8\right) \mu_\Delta+50\vphantom{\frac{e}{m}}\right] 
\,, \end{align*}

\begin{align*} 
&\gamma_{\text{E1M2}}^{(n)}  =  
-\frac{e^2 h_A^2 \Xi_4}{7776 F^2 m^2\pi ^3 \mu_\Delta^6 \left(\mu^2-\mu_\Delta^2+2 \mu_\Delta-1\right)^2 \left(\mu^2-\mu_\Delta^2-2 \mu_\Delta-1\right)} \left[\vphantom{\frac{e}{m}}2 \mu_\Delta^{14}\right.\vphantom{\frac{e}{m}}\\
&\;\;\;\;\left.\left.+48 \mu_\Delta^{13}-\left(11 \mu^2+148\right) \mu_\Delta^{12}+\left(247-246 \mu^2\right) \mu_\Delta^{11}+\left(24 \mu^4+574 \mu^2-229\right)\mu_\Delta^{10}\right.\right.\vphantom{\frac{e}{m}}\\
&\;\;\;\;\left.\left.+\left(510 \mu^4-881 \mu^2-325\right) \mu_\Delta^9+\left(-25 \mu^6-817 \mu^4+741 \mu^2+669\right)\mu_\Delta^8+\left(-540 \mu^6\right.\right.\right.\vphantom{\frac{e}{m}}\\
&\;\;\;\;\left.\left.+1185 \mu^4-50 \mu^2+63\right) \mu_\Delta^7+\left(10 \mu^8+488 \mu^6-785 \mu^4+234\mu^2-397\right) \mu_\Delta^6+\left(300 \mu^8\right.\right.\vphantom{\frac{e}{m}}\\
&\;\;\;\;\left.\left.-739 \mu^6+653 \mu^4-173 \mu^2-97\right) \mu_\Delta^5+\left(3 \mu^{10}-82 \mu^8+254 \mu^6-156 \mu^4-143 \mu^2\right.\right.\vphantom{\frac{e}{m}}\\
&\;\;\;\;\left.\left.+124\right) \mu_\Delta^4-2 \left(\mu^2-1\right)^2 \left(39 \mu^6-28 \mu^4+59 \mu^2-35\right) \mu_\Delta^3-2 \left(\mu^2-1\right)^3 \left(2 \mu^6+13 \mu^4\right.\right.\vphantom{\frac{e}{m}}\\
&\;\;\;\;\left.\left.+19 \mu^2-10\right) \mu_\Delta^2+6\left(\mu^2-1\right)^5 \left(\mu^2+1\right) \mu_\Delta+\left(\mu^2-1\right)^5 \left(\mu^4+4 \mu^2+1\right)\vphantom{\frac{e}{m}}\right]\\
&+\frac{e^2 h_A^2 \Xi_1}{7776 F^2 m^2 \pi ^3 \mu_\Delta^6} \left[\vphantom{\frac{e}{m}}\mu^{10}+\mu^8-8 \mu^6-32 \mu^4-2 \left(32 \mu^4-51 \mu^2+30\right) \mu_\Delta^3 \mu^2+\mu^2+2 \mu_\Delta^{10}\right.\vphantom{\frac{e}{m}}\\
&\;\;\;\;\left.\left.+52 \mu_\Delta^9-\left(7 \mu^2+44\right) \mu_\Delta^8-5\left(32 \mu^2-31\right) \mu_\Delta^7+\left(8 \mu^4+66 \mu^2-21\right) \mu_\Delta^6+\left(168 \mu^4\right.\right.\right.\vphantom{\frac{e}{m}}\\
&\;\;\;\;\left.\left.-241 \mu^2-227\right) \mu_\Delta^5+\left(-2 \mu^6+\mu^4-68 \mu^2-139\right) \mu_\Delta^4-2 \left(\mu^8+12 \mu^6-3 \mu^4+64 \mu^2\right.\right.\vphantom{\frac{e}{m}}\\
&\;\;\;\;\left.\left.-18\right) \mu_\Delta^2+4 \left(\mu^8-4 \mu^6-21 \mu^4-2 \mu^2+2\right) \mu_\Delta+1\vphantom{\frac{e}{m}}\right]\\
&-\frac{e^2 h_A^2 \Xi_3}{7776 F^2 m^2 \pi ^3 \mu_\Delta^6}\left[\vphantom{\frac{e}{m}}2 \mu_\Delta^{10}+52 \mu_\Delta^9-\left(7 \mu^2+44\right) \mu_\Delta^8-5 \left(32 \mu^2-31\right) \mu_\Delta^7+\left(8 \mu^4+66 \mu^2\right.\right.\vphantom{\frac{e}{m}}\\
&\;\;\;\;\left.\left.-21\right) \mu_\Delta^6+\left(168 \mu^4-241 \mu^2-107\right) \mu_\Delta^5+\left(-2 \mu^6+\mu^4-68 \mu^2+165\right) \mu_\Delta^4-2 \mu^2 \left(32 \mu^4\right.\right.\vphantom{\frac{e}{m}}\\
&\;\;\;\;\left.\left.-51 \mu^2+42\right) \mu_\Delta^3-2 \left(\mu^8+12 \mu^6-3 \mu^4-28 \mu^2+18\right) \mu_\Delta^2+4 \left(\mu^2-1\right)^2 \left(\mu^4-2 \mu^2\right.\right.\vphantom{\frac{e}{m}}\\
&\;\;\;\;\left.\left.-2\right) \mu_\Delta+\left(\mu^2-1\right)^3\left(\mu^4+4 \mu^2+1\right)\vphantom{\frac{e}{m}}\right] \\
&-\frac{e^2 h_A^2}{46656 F^2 m^2 \pi ^3 \mu_\Delta^6 \left(\mu^2-\mu_\Delta^2+2 \mu_\Delta-1\right)}\left[\vphantom{\frac{e}{m}}6 \mu^{10}+3 \mu^8+160 \mu^6-249 \mu^4+102 \mu^2\right.\vphantom{\frac{e}{m}}\\
&\;\;\;\;\left.\left.+12 \mu_\Delta^{10}+288 \mu_\Delta^9-6 \left(7 \mu^2+145\right) \mu_\Delta^8+\left(2454-900 \mu^2\right) \mu_\Delta^7+\left(48 \mu^4+1659 \mu^2\right.\right.\right.\vphantom{\frac{e}{m}}\\
&\;\;\;\;\left.\left.-3388\right)\mu_\Delta^6+4 \left(243 \mu^4-684 \mu^2+328\right) \mu_\Delta^5+\left(-12 \mu^6-705 \mu^4+627 \mu^2+172\right)\mu_\Delta^4\right.\vphantom{\frac{e}{m}}\\
&\;\;\;\;\left.\left.-2 \left(198 \mu^6+102 \mu^4+125 \mu^2-285\right) \mu_\Delta^3-3 \left(4 \mu^8+29 \mu^6-179 \mu^4-92 \mu^2+128\right) \mu_\Delta^2\right.\right.\vphantom{\frac{e}{m}}\\
&\;\;\;\;\left.\left.+18 \left(2 \mu^8+27 \mu^6-29 \mu^4+16 \mu^2-2\right) \mu_\Delta-22\vphantom{\frac{e}{m}}\right]\right.
\,, \end{align*}

\begin{align*} 
& \gamma_{\text{M1E2}}^{(n)}  =  
-\frac{e^2 h_A^2 \Xi_4}{7776F^2 m^2 \pi ^3 \mu_\Delta^6 \left(\mu^2-\mu_\Delta^2+2 \mu_\Delta-1\right)^2 \left(\mu^2-\mu_\Delta^2-2 \mu_\Delta-1\right)} \left[\vphantom{\frac{e}{m}}2 \mu_\Delta^{14}\right.\vphantom{\frac{e}{m}}\\
&\;\;\;\;\left.\left.+44 \mu_\Delta^{13}-\left(11 \mu^2+184\right) \mu_\Delta^{12}+\left(219-226 \mu^2\right) \mu_\Delta^{11}+\left(24 \mu^4+730 \mu^2+67\right) \mu_\Delta^{10}\right.\right.\vphantom{\frac{e}{m}}\\
&\;\;\;\;\left.\left.+\left(470 \mu^4-811 \mu^2-281\right) \mu_\Delta^9+\left(-25 \mu^6-1083 \mu^4+269 \mu^2+161\right) \mu_\Delta^8+\left(-500 \mu^6\right.\right.\right.\vphantom{\frac{e}{m}}\\
&\;\;\;\;\left.\left.+1119 \mu^4+46 \mu^2-131\right) \mu_\Delta^7+\left(10 \mu^8+712 \mu^6-673 \mu^4-54 \mu^2-17\right) \mu_\Delta^6+\left(280 \mu^8\right.\right.\vphantom{\frac{e}{m}}\\
&\;\;\;\;\left.\left.-681 \mu^6+333 \mu^4-199 \mu^2+291\right) \mu_\Delta^5+\left(3 \mu^{10}-178 \mu^8+274 \mu^6-150 \mu^4+169 \mu^2\right.\right.\vphantom{\frac{e}{m}}\\
&\;\;\;\;\left.\left.-118\right) \mu_\Delta^4-2 \left(\mu^2-1\right)^2 \left(37 \mu^6-3 \mu^4-27 \mu^2+68\right) \mu_\Delta^3-2 \left(\mu^2-1\right)^3 \left(2 \mu^6+3 \mu^4\right.\right.\vphantom{\frac{e}{m}}\\
&\;\;\;\;\left.\left.-27 \mu^2+46\right) \mu_\Delta^2+6\left(\mu^2-1\right)^4 \left(\mu^4+4 \mu^2-1\right) \mu_\Delta+\left(\mu^2-1\right)^5 \left(\mu^4+2 \mu^2+3\right)\vphantom{\frac{e}{m}}\right]\\
&+\frac{e^2 h_A^2 \Xi_1}{7776 F^2 m^2 \pi ^3 \mu_\Delta^6} \left[\vphantom{\frac{e}{m}}\mu^{10}-\mu^8+4 \mu^4-7 \mu^2+2 \mu_\Delta^{10}+48 \mu_\Delta^9-\left(7 \mu^2+88\right) \mu_\Delta^8+\left(39\right.\right.\vphantom{\frac{e}{m}}\\
&\;\;\;\;\left.\left.-148 \mu^2\right) \mu_\Delta^7+\left(8 \mu^4+166 \mu^2+51\right) \mu_\Delta^6+3 \left(52 \mu^4-41 \mu^2+15\right) \mu_\Delta^5-\left(2 \mu^6+69 \mu^4\right.\right.\vphantom{\frac{e}{m}}\\
&\;\;\;\;\left.\left.+24 \mu^2-85\right) \mu_\Delta^4-6 \left(10 \mu^6-12 \mu^4+4 \mu^2+1\right) \mu_\Delta^3-2 \left(\mu^8+4 \mu^6-15 \mu^4-44 \mu^2\right.\right.\vphantom{\frac{e}{m}}\\
&\;\;\;\;\left.\left.+34\right) \mu_\Delta^2+4 \left(\mu^8+3 \mu^6+12 \mu^4-11 \mu^2+3\right) \mu_\Delta+3\vphantom{\frac{e}{m}}\right]\\
&-\frac{e^2 h_A^2 \Xi_3}{7776 F^2 m^2 \pi ^3 \mu_\Delta^6}\left[\vphantom{\frac{e}{m}}2 \mu_\Delta^{10}+48 \mu_\Delta^9-\left(7 \mu^2+88\right) \mu_\Delta^8+\left(39-148 \mu^2\right) \mu_\Delta^7+\left(8 \mu^4+166 \mu^2\right.\right.\vphantom{\frac{e}{m}}\\
&\;\;\;\;\left.\left.+51\right) \mu_\Delta^6+3 \left(52 \mu^4-41 \mu^2-25\right) \mu_\Delta^5-\left(2 \mu^6+69 \mu^4+24 \mu^2+187\right) \mu_\Delta^4+\left(-60 \mu^6\right.\right.\vphantom{\frac{e}{m}}\\
&\;\;\;\;\left.\left.+72 \mu^4+6\right) \mu_\Delta^3-2 \left(\mu^8+4 \mu^6-15 \mu^4+44 \mu^2-34\right) \mu_\Delta^2+4 \left(\mu^2-1\right)^2 \left(\mu^4+5 \mu^2\right.\right.\vphantom{\frac{e}{m}}\\
&\;\;\;\;\left.\left.-3\right) \mu_\Delta+\left(\mu^2-1\right)^3 \left(\mu^4+2 \mu^2+3\right)\vphantom{\frac{e}{m}}\right] \\
& -\frac{e^2  h_A^2}{46656F^2 m^2 \pi ^3 \mu_\Delta^6 \left(\mu^2-\mu_\Delta^2+2 \mu_\Delta-1\right)}\left[\vphantom{\frac{e}{m}}6 \mu^{10}-9 \mu^8-146 \mu^6+313 \mu^4-218 \mu^2\right.\vphantom{\frac{e}{m}}\\
&\;\;\;\;\left.\left.+12 \mu_\Delta^{10}+264 \mu_\Delta^9-6 \left(7 \mu^2+181\right) \mu_\Delta^8-18 \left(46 \mu^2-65\right) \mu_\Delta^7+\left(48 \mu^4+2163 \mu^2\right.\right.\right.\vphantom{\frac{e}{m}}\\
&\;\;\;\;\left.\left.+8\right) \mu_\Delta^6+4 \left(225 \mu^4-483 \mu^2-256\right) \mu_\Delta^5+\left(-12 \mu^6-1077 \mu^4+1275 \mu^2+1444\right)\mu_\Delta^4\right.\vphantom{\frac{e}{m}}\\
&\;\;\;\;\left.\left.-2 \left(186 \mu^6-642 \mu^4+61 \mu^2+299\right) \mu_\Delta^3+\left(-12 \mu^8+9 \mu^6-621 \mu^4+576\mu^2-196\right) \mu_\Delta^2\right.\right.\vphantom{\frac{e}{m}}\\
&\;\;\;\;\left.\left.+6 \left(6 \mu^8-87 \mu^6+115 \mu^4-56 \mu^2+10\right) \mu_\Delta+54\vphantom{\frac{e}{m}}\right] \right.
\,. \end{align*}

\section{\texorpdfstring{$q^4$}{q4}-Values}\label{sec:q4-loop} 

Here, we  use the notation $$\tilde{e}_{117} \equiv e_{117}+2e_{90}+e_{94} + g_A^2\frac{2c_6+3c_7}{128m\pi^2F^2} \text{ and } \tilde{e}_{118} \equiv e_{118}+2e_{89}+e_{93} - g_A^2\frac{13c_6+15c_7}{512m\pi^2F^2} \punkt$$ 

\subsection{Proton values}\label{sub:proton} 

\subsubsection{Spin-independent first order polarizabilities}\label{subsub:proton-spin-independent-1} 

\begin{align*}
&\alpha_{\text{E1}}^{\text{p}}
 =  
-\frac{e^2 g_A^2}{64 \pi ^3 F^2 m}\left[\vphantom{\frac{e}{m}}
c_6 \Xi_2 \left(5 \mu ^4-17 \mu ^2+4\right)+3 c_7 \Xi_2 \left(3 \mu ^4-11 \mu ^2+4\right) + 5 c_6 \mu^2 \right. \\
& \left.\;\;\;\;+ 9 c_7 \mu^2\vphantom{\frac{e}{m}}\right]
-\frac{e^2 (\tilde{e}_{117} + 2\tilde{e}_{118} + e_{92} + 2e_{91})}{\pi} 
-\frac{e^2 (4 c_1+c_2-2 c_3)}{192 \pi ^3 F^2} \\
& \;\;
+\frac{e^2\Xi_1}{192 \pi ^3 F^2 m} \left[\vphantom{\frac{e}{m}}
 3 g_A^2c_6 \mu^2 \left(5 \mu ^2-7\right)+  9 g_A^2c_7 \mu ^2 \left(3 \mu ^2-5\right)-2c_2m\vphantom{\frac{e}{m}}\right] \,,\\
 \\
&\beta_{\text{M1}}^{\text{p}}
 =  \frac{e^2 g_A^2 }{64 \pi ^3 F^2\left(\mu ^2-4\right) m}  \left[\vphantom{\frac{e}{m}}
 c_6 \Xi_2\left(-15 \mu ^6+113 \mu ^4-230 \mu ^2+96\right)
 +c_7 \Xi_2\left(-21 \mu ^6 \right.\right.\\
 &\;\;\;\;\left.\left. +159 \mu ^4-328 \mu ^2+144\right)+c_6 \left(15 \mu ^4-55 \mu ^2+4\right)+c_7 \left(21 \mu ^4-78 \mu^2+8\right)\vphantom{\frac{e}{m}}\right] \\
 &\;\;+\frac{2 e^2(\tilde{e}_{118} + e_{91})}{\pi }
+\frac{e^2 (4 c_1-c_2-2 c_3)}{192 \pi ^3 F^2} \\
&\;\;+\frac{e^2}{192 \pi ^3 F^2 m} \Xi_1\left[\vphantom{\frac{e}{m}}
 3c_6g_A^2\left(15 \mu ^4-23 \mu ^2+2\right)
+3c_7g_A^2\left(21 \mu ^4-33 \mu ^2+4\right)
-2c_2 m\vphantom{\frac{e}{m}}\right]\,.
\end{align*}

\subsubsection{Spin-independent second order polarizabilities}\label{subsub:proton-spin-independent-2} 

\begin{align*}
&\alpha_{\text{E2}}^{(p)}  =  \;
\frac{e^2 g_A^2}{64 \pi ^3 F^2 \left(\mu ^2-4\right) m^3}\left[\vphantom{\frac{e}{m}} c_6 \Xi_2 \left(-45 \mu ^6+340 \mu ^4-696 \mu ^2+276\right) + c_7 \Xi_2\left(-81 \mu ^6 \right.\right. \vphantom{\frac{e}{m}} \\
&\;\;\;\; \left.\left.+628 \mu ^4-1356 \mu ^2+636\right) - c_7\left(81 \mu ^4-358 \mu^2+212\right) - c_6\left(45 \mu ^4-190\mu ^2+92\right)\vphantom{\frac{e}{m}} \right]\\
&\;+ \frac{e^2 \Xi_1}{64 \pi ^3 F^2 m^3} \left[\vphantom{\frac{e}{m}}2c_2 m + g_A^2c_6 \left(45 \mu ^4-70\mu ^2+6\right)
+ g_A^2c_7 \left(81 \mu ^4-142 \mu ^2+18\right)\vphantom{\frac{e}{m}}\right] \\
&\;+ \frac{e^2\left(-32 c_1+15 \mu ^2 c_2-44c_3\right)}{960 \pi ^3 F^2 \mu ^2 m^2} 
+ \frac{3 e^2\tilde{e}_{117}}{2 \pi m^2}\,,\\
\\
&\beta_{\text{M2}}^{(p)}  =  \;
\frac{e^2 g_A^2 }{64 \pi ^3F^2 \mu ^2 \left(\mu ^2-4\right)^2 m^3} \left[\vphantom{\frac{e}{m}}  c_6 \Xi_2\left(-105\mu ^{10}+1208 \mu ^8-4756 \mu ^6+7100 \mu ^4\right.\right. \vphantom{\frac{e}{m}} \\
&\;\;\;\; \left.\left.  -2920 \mu ^2+64\right) +c_7 \Xi_2\left(-153 \mu ^{10}+1772 \mu ^8-7056 \mu ^6+10780 \mu ^4 - 4720 \mu^2+160\right) \right. \vphantom{\frac{e}{m}} \\
&\;\;\;\; \left.\left. +c_6\left(-105 \mu ^6+848 \mu ^4-1882\mu ^2+848\right) + c_7\left(-153 \mu ^6+1250 \mu ^4-2848 \mu ^2+1400\right)\vphantom{\frac{e}{m}} \right]\right. \\
&\; +\frac{e^2 \Xi_1}{64 \pi ^3 F^2m^3} \left[\vphantom{\frac{e}{m}}2c_2 m + g_A^2c_6 \left(105 \mu ^4-158 \mu ^2+26\right)
 +g_A^2c_7 \left(153 \mu ^4-242 \mu ^2+46\right)  \vphantom{\frac{e}{m}}\right]\\
&\; +\frac{e^2 \left(32 c_1+\left(15\mu ^2+8\right) c_2+44 c_3\right)}{960 \pi ^3 F^2 \mu^2 m^2} + \frac{3e^2 \tilde{e}_{117}}{2 \pi m^2} \,,\\
\\
&\alpha_{\text{E1}\nu}^{(p)}  =  \; 
\frac{e^2 g_A^2}{768 \pi ^3 F^2 \mu^2 \left(\mu^2-4\right)^2 m^3} \left[\vphantom{\frac{e}{m}} c_6\Xi_2  \left(-567 \mu^{10}+6616 \mu ^8-26660 \mu ^6  +  41620 \mu ^4\right.\right. \vphantom{\frac{e}{m}} \\
&\;\;\;\; \left.\left.-19080 \mu ^2+576\right) +  c_7\Xi_2 \left(1312- 5 \mu ^2\left(171 \mu ^8-2004 \mu ^6  +8136 \mu ^4-12900\mu^2\right.\right.\right. \vphantom{\frac{e}{m}} \\
&\;\;\;\; \left.\left.+6176\right)\right) +c_6 \left(-567\mu ^6+4624 \mu ^4-10434 \mu ^2+4784\right)  + 3c_7 \left(-285 \mu ^6+2338 \mu ^4 \right. \vphantom{\frac{e}{m}} \\
&\;\;\;\; \left.\left.-5348 \mu^2 + 2568\right)\vphantom{\frac{e}{m}} \right] \\
&\;+ \frac{e^2 \Xi_1}{768 \pi ^3 F^2 m^3} \left[\vphantom{\frac{e}{m}}
g_A^2 c_6 \left(567 \mu ^4-946 \mu ^2+190\right)
+ 15 g_A^2 c_7 \left(57 \mu ^4-98 \mu ^2+22\right)
-  2c_2 m\vphantom{\frac{e}{m}}\right]\\
&\;+ \frac{e^2 \left(96c_1+\left(16-15 \mu ^2\right) c_2+132c_3\right)}{11520 \pi ^3 F^2 \mu ^2 m^2} 
- \frac{e^2\tilde{e}_{117}}{8 \pi m^2} \,,\\
\\
&\beta_{\text{M1}\nu}^{(p)}  = \;
 -\frac{e^2 g_A^2}{768 \pi ^3 F^2 \mu ^2 \left(\mu ^2-4\right)^3 m^3}  \left[\vphantom{\frac{e}{m}}c_6 \Xi_2 \left(627 \mu^{12}-9764 \mu ^{10}+57904 \mu ^8 - 159740 \mu ^6\right.\right. \vphantom{\frac{e}{m}} \\
&\;\;\;\; \left.\left.+196496 \mu ^4-80992 \mu^2+4352\right) +  c_7 \Xi_2 \left(927 \mu ^{12} -14484 \mu ^{10}+86300 \mu ^8-239764 \mu ^6\right.\right. \vphantom{\frac{e}{m}} \\
&\;\;\;\; \left.\left.+298416 \mu^4-125504 \mu ^2+6400\right)-c_6 \left(-627 \mu ^{10}+7562 \mu ^8-31376 \mu ^6+49112 \mu ^4\right.\right. \vphantom{\frac{e}{m}} \\
&\;\;\;\; \left.\left.-19392 \mu^2+256\right) -c_7 \left(-927 \mu ^{10}+11226 \mu ^8-46880 \mu ^6+74336 \mu ^4\right.\right. \vphantom{\frac{e}{m}} \\
&\;\;\;\; \left.\left.-30464 \mu^2+512\right)\vphantom{\frac{e}{m}}\right]\\
&+ \frac{e^2 \Xi_1}{768 \pi^3F^2 m^3} \left[\vphantom{\frac{e}{m}}g_A^2  c_6\left(627 \mu ^4-986 \mu ^2+210\right)
+ g_A^2 c_7 \left(927 \mu ^4-1506 \mu ^2+326\right)
- 2c_2 m \vphantom{\frac{e}{m}}\right]\\
&-\frac{e^2 \left(32c_1+\left(5 \mu ^2+8\right) c_2+44 c_3\right)}{3840\pi ^3 F^2 \mu ^2 m^2} -\frac{e^2 \tilde{e}_{117}}{8 \pi m^2} \,.
\end{align*}

\subsubsection{Spin-dependent first order polarizabilities}\label{subsub:proton-spin-dependent-1} 

\begin{align*}
&\gamma_{\text{E1E1}}^{(p)}
 =  \; 
\frac{e^2 g_A^2}{384 \pi^3 F^2 \left(\mu^2-4\right)^2 m^2} \left[\vphantom{\frac{e}{m}} -c_6 \Xi_2 \left(54 \mu^8-635\mu^6+2582 \mu^4-4056 \mu^2\right.\right. \vphantom{\frac{e}{m}} \\
&\;\;\;\; \left.\left.+1704\right)  - c_7 \Xi_2 \left(72 \mu^8-851 \mu^6+3488 \mu^4-5556 \mu^2+2400\right) +2c_6 \left(-27 \mu^6+223 \mu^4 \right.\right. \vphantom{\frac{e}{m}} \\
&\;\;\;\; \left.\left.-517 \mu^2+216\right) +c_7 \left(-72 \mu^6+599 \mu^4-1408 \mu^2+608\right) \vphantom{\frac{e}{m}} \right] \\
&\; +\frac{e^2 g_A^2 \Xi_1}{384 \pi^3 F^2m^2} \left[\vphantom{\frac{e}{m}}c_6 \left(54 \mu^4-95 \mu^2+12\right)
+ c_7 \left(72 \mu^4-131 \mu^2+18\right)\vphantom{\frac{e}{m}}\right]\,,\\
\\
&\gamma_{\text{M1M1}}^{(p)}
 =  \; 
 \frac{e^2 g_A^2}{384 \pi^3 F^2 \left(\mu^2-4\right)^2 m^2 \mu^2} \left[ \vphantom{\frac{e}{m}} c_6 \Xi_2 \left(-144 \mu^{10}+1651 \mu^8-6468 \mu^6+9584 \mu^4 \right.\right. \vphantom{\frac{e}{m}} \\
&\;\;\;\; \left.\left.-3968 \mu^2+128\right) +c_7 \Xi_2 \left(-180 \mu^{10}+2053\mu^8-7972 \mu^6+11604 \mu^4 -4592 \mu^2+128\right) \right. \vphantom{\frac{e}{m}} \\
&\;\;\;\; \left.+c_6 \left(-144 \mu^8+1147 \mu^6-2456 \mu^4+976\mu^2\right) +c_7 \left(-180 \mu^8+1423 \mu^6\right.\right. \vphantom{\frac{e}{m}} \\
&\;\;\;\; \left.\left.-2996 \mu^4+1120\mu^2\right) \vphantom{\frac{e}{m}} \right]\\
&\; +\frac{e^2 g_A^2 \Xi_1}{384 \pi^3 F^2 m^2} \left[\vphantom{\frac{e}{m}}c_6 \left(144 \mu^4-211 \mu^2+38\right)
+ c_7 \left(180\mu^4-253 \mu^2+42\right)\vphantom{\frac{e}{m}}\right]\,,\\
\\
&\gamma_{\text{E1M2}}^{(p)}
 =  \; 
 \frac{e^2 g_A^2}{384 \pi^3 F^2 \left(\mu^2-4\right)^2 m^2} \left[ \vphantom{\frac{e}{m}} -3c_6 \Xi_2 \left(12 \mu^8-139\mu^6+552 \mu^4-832 \mu^2+312\right) \right. \vphantom{\frac{e}{m}} \\
&\;\;\;\; \left.-c_7 \Xi_2 \left(36 \mu^8-411 \mu^6+1592 \mu^4-2276 \mu^2+672\right)+c_6 \left(-36 \mu^6+291 \mu^4  -650 \mu^2\right.\right. \vphantom{\frac{e}{m}} \\
&\;\;\;\; \left.\left.+224\right) +c_7 \left(-36 \mu^6+285 \mu^4-608 \mu^2+128\right) \vphantom{\frac{e}{m}} \right] \\
&\; +\frac{e^2 g_A^2 \Xi_1}{128 \pi^3 F^2m^2} \left[\vphantom{\frac{e}{m}}c_6 \left(12 \mu^4-19 \mu^2+2\right)
+ c_7 \left(36 \mu^4-51 \mu^2+2\right)\vphantom{\frac{e}{m}}\right]\,,\\
\\
&\gamma_{\text{M1E2}}^{(p)}
 =  \; 
 \frac{e^2 g_A^2}{384 \pi^3 F^2 \left(\mu^2-4\right) m^2 \mu^2} \left[\vphantom{\frac{e}{m}}  c_6 \Xi_2 \left(-6 \mu^8+49\mu^6-114 \mu^4+48\mu^2\right)+c_7 \Xi_2 \left(5 \mu^6 \right.\right. \vphantom{\frac{e}{m}} \\
&\;\;\;\; \left.\left.-24 \mu^4-12 \mu^2+16\right)  -2c_6 \left(3 \mu^6-14 \mu^4+4\mu^2\right) +c_7 \left(5 \mu^4+4\mu^2\right) \vphantom{\frac{e}{m}} \right]\\
&\; +\frac{e^2 g_A^2 \Xi_1}{384 \pi^3 F^2 m^2} \left[\vphantom{\frac{e}{m}}c_6 \mu^2 \left(6 \mu^2-13\right)
- c_7 \left(5\mu^2+6\right)\vphantom{\frac{e}{m}}\right]\,.
\end{align*}

\subsubsection{Spin-dependent second order polarizabilities}\label{subsub:proton-spin-dependent-2} 

\begin{align*} 
&\gamma_{\text{E2E2}}^{(p)}  =  
\; \frac{e^2 g_A^2}{276480 \pi^3 F^2 \mu^2 \left(\mu^2-4\right)^3 m^4} \left[\vphantom{\frac{e}{m}}
- 6c_6 \Xi_2 \left(2815 \mu^{12}-44860 \mu^{10}+274613 \mu^8 \right.\right. \vphantom{\frac{e}{m}}\\
&\;\;\;\;\left.\left. -793312 \mu^6+1048500 \mu^4-478000 \mu^2+17920\right)
- 6c_7 \Xi_2 \left(4530 \mu^{12}  -72615 \mu^{10}\right.\right. \vphantom{\frac{e}{m}}\\
&\;\;\;\;\left.\left.+448178 \mu^8-1310752 \mu^6+1768680 \mu^4-843040 \mu^2+38400\right)  
- c_6 \mu^2 \left(16890 \mu^8\right.\right. \vphantom{\frac{e}{m}}\\
&\;\;\;\;\left.\left.-210045 \mu^6+914298 \mu^4-1564712 \mu^2+742400\right)
- 4c_7 \mu^2 \left(6795 \mu^8-85140 \mu^6\right.\right. \vphantom{\frac{e}{m}}\\
&\;\;\;\;\left.\left.+374947 \mu^4-655268 \mu^2+328480\right)\vphantom{\frac{e}{m}}\right] \vphantom{\frac{e}{m}}\\
&\;\;+\frac{e^2 g_A^2 \Xi_1}{46080 \pi^3 F^2 m^4} \left[ \vphantom{\frac{e}{m}}c_6 \left(2815 \mu^4-5450 \mu^2+1263\right)
 +c_7 \left(4530 \mu^4 -9195 \mu^2+2348\right) \vphantom{\frac{e}{m}}\right] \,,\\
 \\
&\gamma_{\text{M2M2}}^{(p)}  =  
 \; \frac{e^2 g_A^2}{276480 \pi^3 F^2 \mu^2 \left(\mu^2-4\right)^3 m^4} \left[\vphantom{\frac{e}{m}}
6 c_7 \Xi_2 \left(-12270 \mu^{12}+189729 \mu^{10}-1114038 \mu^8 \right.\right. \vphantom{\frac{e}{m}}\\
&\;\;\;\;\left.\left. +3028128\mu^6-3641480\mu^4+1458080\mu^2-78720\right)
- 6 c_6 \Xi_2 \left(9265 \mu^{12}  -143484 \mu^{10}\right.\right. \vphantom{\frac{e}{m}}\\
&\;\;\;\;\left.\left.+844403 \mu^8-2303488 \mu^6+2788060 \mu^4-1132080 \mu^2+61440\right) + c_6 \left(-55590\mu^{10}\right.\right. \vphantom{\frac{e}{m}}\\
&\;\;\;\;\left.\left.+666339 \mu^8-2738494 \mu^6+4234376 \mu^4-1680256 \mu^2 +23040\right)
+ 4c_7 \left(-18405 \mu^{10}\right.\right. \vphantom{\frac{e}{m}}\\
&\;\;\;\;\left.\left.+220176 \mu^8-901871 \mu^6+1385404 \mu^4-538544 \mu^2+7680\right)\vphantom{\frac{e}{m}}\right] \vphantom{\frac{e}{m}}\\
&\;\;+\frac{e^2 g_A^2 \Xi_1}{46080 \pi^3 F^2 m^4} \left[\vphantom{\frac{e}{m}} c_6 \left(9265 \mu^4-13774 \mu^2+3017\right)
 +3 c_7 \left(4090 \mu^4 -5983 \mu^2+1284\right)\vphantom{\frac{e}{m}} \right]\,,\\
 \\
&\gamma_{\text{E2M3}}^{(p)}  =  
 \; \frac{e^2 g_A^2}{138240 \pi^3 F^2 \mu^2 \left(\mu^2-4\right)^3 m^4} \left[\vphantom{\frac{e}{m}}
6c_7 \Xi_2 \mu^2 \left(-930 \mu^{10}+14319 \mu^8-83402 \mu^6 \right.\right. \vphantom{\frac{e}{m}}\\
&\;\;\;\;\left.\left. +222784 \mu^4-255000 \mu^2+77920\right)
- 6c_6 \Xi_2 \left(895 \mu^{12}-13924 \mu^{10}+82397 \mu^8 \right.\right. \vphantom{\frac{e}{m}}\\
&\;\;\;\;\left.\left. -226144 \mu^6+274020 \mu^4-102640 \mu^2+2560\right) 
- c_6 \mu^2 \left(5370\mu^8-64749 \mu^6  +268578 \mu^4\right.\right. \vphantom{\frac{e}{m}}\\
&\;\;\;\;\left.\left.-420872 \mu^2+148736\right)
- 4c_7 \mu^2 \left(1395 \mu^8-16596 \mu^6+67237 \mu^4 -99788 \mu^2+25504\right)\vphantom{\frac{e}{m}}\right] \\
&\;\;+\frac{e^2 g_A^2 \Xi_1}{23040 \pi^3 F^2 m^4} \left[\vphantom{\frac{e}{m}} c_6 \left(895 \mu^4-1394 \mu^2+231\right)
 + c_7 \left(930 \mu^4-1299 \mu^2+116\right) \vphantom{\frac{e}{m}}\right] \,,\\
 \\
&\gamma_{\text{M2E3}}^{(p)}  =  
 \; \frac{e^2 g_A^2}{138240 \pi^3 F^2 \mu^2 \left(\mu^2-4\right)^2 m^4} \left[\vphantom{\frac{e}{m}}
- 6 c_6 \Xi_2 \mu^2 \left(145 \mu^8-1808 \mu^6+7971 \mu^4 \right.\right. \vphantom{\frac{e}{m}}\\
&\;\;\;\;\left.\left.-14020\mu^2+7020\right)
- 6 c_7 \Xi_2 \left(30 \mu^{10}-513 \mu^8+2946 \mu^6-6120 \mu^4+1640 \mu^2 +1920\right)\right.\vphantom{\frac{e}{m}}\\
&\;\;\;\;\left.
+ c_6 \mu^2 \left(-870 \mu^6+7803 \mu^4-20218 \mu^2+11008\right)
+ 4c_7 \left(-45 \mu^8  +612 \mu^6-2177 \mu^4\right.\right. \vphantom{\frac{e}{m}}\\
&\;\;\;\;\left.\left.+1112 \mu^2+480\right)\vphantom{\frac{e}{m}}\right] \\
&\;\;+\frac{e^2 g_A^2 \Xi_1}{23040 \pi^3 F^2 m^4} \left[\vphantom{\frac{e}{m}} c_6 \left(145 \mu^4-358 \mu^2+41\right)
 + 3c_7 \left(10 \mu^4-71 \mu^2-28\right) \vphantom{\frac{e}{m}}\right] \,,\\
 \\
&\gamma_{\text{E1E1}\nu}^{(p)}  =  
 \; \frac{e^2 g_A^2}{92160 \pi^3 F^2 \mu^2 \left(\mu^2-4\right)^4 m^4} \left[\vphantom{\frac{e}{m}}
6 c_6 \Xi_2 \left(-12795 \mu^{14}+252768 \mu^{12}-2022553 \mu^{10}\right.\right. \vphantom{\frac{e}{m}}\\
&\;\;\;\;\left.\left.+8314020 \mu^8-18266004 \mu^6+19956640 \mu^4-8462400 \mu^2+586240\right)\right. \vphantom{\frac{e}{m}}\\
&\;\;\;\;\left.+ 6c_7 \Xi_2 \left(-17370 \mu^{14}+343603 \mu^{12}-2754238 \mu^{10}+11349720 \mu^8 -25027944 \mu^6\right.\right. \vphantom{\frac{e}{m}}\\
&\;\;\;\;\left.\left.+27512320 \mu^4-11801600 \mu^2+837120\right) 
- c_6 \left(76770\mu^{12}  -1247913 \mu^{10}\right.\right. \vphantom{\frac{e}{m}}\\
&\;\;\;\;\left.\left.+7772910 \mu^8-22691488 \mu^6+29783392 \mu^4-12724992 \mu^2  +276480\right)\right. \vphantom{\frac{e}{m}}\\
&\;\;\;\;\left.
- 4c_7 \left(26055 \mu^{12}-424212 \mu^{10}+2648475 \mu^8-7760752 \mu^6  +10258144 \mu^4\right.\right. \vphantom{\frac{e}{m}}\\
&\;\;\;\;\left.\left.-4459456 \mu^2+111616\right)\vphantom{\frac{e}{m}}\right]  \vphantom{\frac{e}{m}}\\
&\;\;+\frac{e^2 g_A^2 \Xi_1}{15360 \pi^3 F^2 m^4} \left[\vphantom{\frac{e}{m}} c_6 \left(12795 \mu^4-22458 \mu^2+6139\right)
 + c_7 \left(17370 \mu^4 -30943 \mu^2+8644\right) \vphantom{\frac{e}{m}}\right] \,,\\
 \\
&\gamma_{\text{M1M1}\nu}^{(p)}  =  
 \; \frac{e^2 g_A^2}{30720 \pi^3 F^2 \mu^4 \left(\mu^2-4\right)^4 m^4} \left[\vphantom{\frac{e}{m}}
2 c_6 \Xi_2 \left(-18645 \mu^{16}+365512 \mu^{14}-2895039 \mu^{12} \right.\right. \vphantom{\frac{e}{m}}\\
&\;\;\;\;\left.\left. +11734236 \mu^{10}-25255068 \mu^8+26691840 \mu^6-10715712 \mu^4+690176 \mu^2 +12288\right)\right. \vphantom{\frac{e}{m}}\\
&\;\;\;\;\left.
+ 2 c_7 \Xi_2 \left(-24390 \mu^{16}+477637 \mu^{14}-3777690 \mu^{12}+15279624 \mu^{10}  -32775144 \mu^8\right.\right. \vphantom{\frac{e}{m}}\\
&\;\;\;\;\left.\left.+34427520 \mu^6-13633152 \mu^4+843776 \mu^2+12288\right) 
- c_6 \mu^2\left(37290 \mu^{12}-600509 \mu^{10}\right.\right. \vphantom{\frac{e}{m}}\\
&\;\;\;\;\left.\left.+3691774 \mu^8-10567696 \mu^6+13374368 \mu^4 -5340160 \mu^2+145408\right)\right. \vphantom{\frac{e}{m}}\\
&\;\;\;\;\left.
- 4c_7 \mu^2\left(12195 \mu^{12}-196136 \mu^{10}+1203489 \mu^8  -3433800 \mu^6+4317072 \mu^4\right.\right. \vphantom{\frac{e}{m}}\\
&\;\;\;\;\left.\left.-1691520 \mu^2+40448\right)\vphantom{\frac{e}{m}}\right]  \vphantom{\frac{e}{m}}\\
&\;\;+\frac{e^2 g_A^2 \Xi_1}{15360 \pi^3 F^2 m^4} \left[\vphantom{\frac{e}{m}} c_6 \left(18645 \mu^4-29902\mu^2+7533\right)
 + c_7 \left(24390 \mu^4 -38617 \mu^2+9444\right) \vphantom{\frac{e}{m}}\right] \,,\\
 \\
&\gamma_{\text{E1M2}\nu}^{(p)}  =  
 \; \frac{e^2 g_A^2}{230400 \pi^3 F^2 \mu^2 \left(\mu^2-4\right)^4 m^4} \left[\vphantom{\frac{e}{m}}
6 c_6 \Xi_2 \left(-28605 \mu^{14}+558752 \mu^{12}-4403315 \mu^{10} \right.\right. \vphantom{\frac{e}{m}}\\
&\;\;\;\;\left.\left. +17712684 \mu^8-37646844 \mu^6+38881120 \mu^4-14719680 \mu^2+787200\right) \right. \vphantom{\frac{e}{m}}\\
&\;\;\;\;\left.
+ 6 c_7 \Xi_2 \left(-34470 \mu^{14}+670917 \mu^{12}-5260850 \mu^{10}+21003984 \mu^8  -44091984 \mu^6\right.\right. \vphantom{\frac{e}{m}}\\
&\;\;\;\;\left.\left.+44459680 \mu^4-15800320 \mu^2+593920\right)
- c_6 \left(171630\mu^{12}  -2751807 \mu^{10}\right.\right. \vphantom{\frac{e}{m}}\\
&\;\;\;\;\left.\left.+16799538 \mu^8-47489120 \mu^6+58760928 \mu^4-21684352 \mu^2 +307200\right)\right. \vphantom{\frac{e}{m}}\\
&\;\;\;\;\left.
- 4c_7 \left(51705 \mu^{12}-825408 \mu^{10}+5005722 \mu^8-13991740 \mu^6 +16921472 \mu^4\right.\right. \vphantom{\frac{e}{m}}\\
&\;\;\;\;\left.\left.-5799168 \mu^2+28160\right)\vphantom{\frac{e}{m}}\right]  \vphantom{\frac{e}{m}}\\
&\;\;+\frac{e^2 g_A^2 \Xi_1}{38400 \pi^3 F^2 m^4} \left[\vphantom{\frac{e}{m}} c_6 \left(28605 \mu^4-43862 \mu^2+9569\right)
 + c_7 \left(34470 \mu^4 -50457 \mu^2+9404\right)\vphantom{\frac{e}{m}} \right]\,,\\
 \\
&\gamma_{\text{M1E2}\nu}^{(p)}  =  
 \; \frac{e^2 g_A^2}{230400 \pi^3 F^2 \mu^4 \left(\mu^2-4\right)^3 m^4} \left[\vphantom{\frac{e}{m}}
6 c_7 \Xi_2 \left(-17370 \mu^{14}+275051 \mu^{12}-1668778 \mu^{10} \right.\right. \vphantom{\frac{e}{m}}\\
&\;\;\;\;\left.\left. +4756616 \mu^8-6145600 \mu^6+2653280 \mu^4-74880 \mu^2+5120\right)  
- 6 c_6 \Xi_2 \mu^2\left(14355 \mu^{12}\right.\right. \vphantom{\frac{e}{m}}\\
&\;\;\;\;\left.\left.-226956 \mu^{10}+1374373 \mu^8-3909056 \mu^6+5044180 \mu^4 -2201360 \mu^2+77440\right) \right. \vphantom{\frac{e}{m}}\\
&\;\;\;\;\left.
- c_6 \mu^4 \left(86130 \mu^8-1060281 \mu^6+4545762 \mu^4  -7588888 \mu^2+3381184\right)\right. \vphantom{\frac{e}{m}}\\
&\;\;\;\;\left.
- 4c_7 \mu^2\left(26055 \mu^{10}-321384 \mu^8+1381428 \mu^6  -2311252 \mu^4+1010576 \mu^2+16640\right)\vphantom{\frac{e}{m}}\right] \vphantom{\frac{e}{m}}\\
&\;\;+\frac{e^2 g_A^2 \Xi_1}{38400 \pi^3 F^2 m^4} \left[\vphantom{\frac{e}{m}} c_6 \left(14355 \mu^4-25986 \mu^2+5719\right)
 + c_7 \left(17370 \mu^4 -31871 \mu^2+6684\right) \vphantom{\frac{e}{m}}\right] \,.
\end{align*}

\subsection{Neutron values}\label{sub:neutron} 

\subsubsection{Spin-independent first order polarizabilities}\label{subsub:neutron-spin-independent-1} 

\begin{align*}  
&\alpha_{\text{E1}}^{(n)}  =  
-\frac{e^2 g_A^2}{32 \pi ^3 F^2 m} \left(
 2c_6\Xi_2 \left(\mu ^2-2\right) + 3c_7\Xi_2 \left(\mu ^2-2\right)\right)
 -\frac{e^2(\tilde{e}_{117}+ 2\tilde{e}_{118}-e_{92}-2e_{91})}{\pi} \\
 &\;\; + \frac{e^2 \Xi_1}{96 \pi ^3 F^2 m} \left(6 c_6 g_A^2 \mu^2 +9 c_7 g_A^2 \mu^2 -c_2m\right)
-\frac{e^2 (4c_1+c_2-2c_3)}{192 \pi ^3 F^2} \,,\\
\\
&\beta_{\text{M1}}^{(n)}  =   
-\frac{e^2 g_A^2}{64 \pi ^3 F^2 \left(\mu ^2-4\right) m}\left(
2c_6\Xi_2 \left(\mu ^4-7 \mu ^2+8\right) + 2c_7\Xi_2 \left(3 \mu ^4-20 \mu ^2+24\right) \right. \\
&\;\;\;\;\left. - c_6 \left(4-3 \mu ^2\right) - 2 c_7 \left(4-3 \mu ^2\right)\right)
+\frac{2 e^2 (\tilde{e}_{118} - e_{91})}{\pi} +\frac{e^2}{192 \pi ^3 F^2} (4c_1-c_2-2c_3)\\
&\;\;+\frac{e^2 \Xi_1}{96 \pi ^3 F^2 m}\left(
 3g_A^2c_6 \left(\mu ^2-1\right) + 3g_A^2c_7 \left(3 \mu ^2-2\right) - c_2 m\right)\,.
\end{align*}

\subsubsection{Spin-independent second order polarizabilities}\label{subsub:neutron-spin-independent-2} 

\begin{align*}  
&\alpha_{\text{E2}}^{(n)}  =  
 -\frac{e^2 g_A^2}{128 \pi ^3 F^2 \left(\mu ^2-4\right)m^3} \left[\vphantom{\frac{e}{m}}8c_6 \Xi_2 \left(7\mu ^4-45 \mu ^2+60\right)+4c_7 \Xi_2 \left(20 \mu ^4-129 \mu ^2 \right.\right. \vphantom{\frac{e}{m}} \\
 &\;\;\;\; \left.\left. +174\right) -c_6 \left(196-65 \mu ^2\right) -2 c_7 \left(152-49 \mu ^2\right)\vphantom{\frac{e}{m}}\right]  +\frac{e^2 \left(-32 c_1+15 \mu ^2c_2-44 c_3\right)}{960 \pi ^3 F^2 \mu ^2 m^2}  \vphantom{\frac{e}{m}} \\
 &\;\; +\frac{e^2\Xi_1}{32 \pi ^3 F^2 m^3} \left[\vphantom{\frac{e}{m}}2g_A^2 c_6\left(7 \mu ^2-3\right)+g_A^2 c_7 \left(20 \mu ^2-9\right)+c_2 m\vphantom{\frac{e}{m}}\right] +\frac{3 e^2(\tilde{e}_{117}-e_{92})}{2 \pi m^2} \vphantom{\frac{e}{m}} \,,\\
 \\
&\beta_{\text{M2}}^{(n)}  =  
 -\frac{e^2 g_A^2}{128 \pi ^3 F^2 \mu ^2 \left(\mu ^2-4\right)^2 m^3} \left[\vphantom{\frac{e}{m}}24c_6 \Xi_2 \left(2 \mu ^8-21\mu ^6+70 \mu ^4-66 \mu ^2+8\right) \right. \vphantom{\frac{e}{m}} \\
 &\;\;\;\; \left.+ 4c_7 \Xi_2 \left(22 \mu ^8-231 \mu ^6+770 \mu ^4-740 \mu^2+80\right) + c_6  \left(57 \mu ^6-412 \mu^4+544\mu^2\right) \right. \vphantom{\frac{e}{m}} \\
 &\;\; \left. +2c_7 \left(53 \mu ^6-382 \mu ^4+536\mu^2\right)\vphantom{\frac{e}{m}}\right] +\frac{e^2 \left(32 c_1+\left(15 \mu ^2+8\right) c_2+44c_3\right)}{960 \pi ^3 F^2 \mu ^2 m^2} +\frac{3 e^2(\tilde{e}_{117} -e_{92})}{2 \pi m^2} \vphantom{\frac{e}{m}} \\
 &\;\; +\frac{e^2 \Xi_1}{32 \pi ^3 F^2 m^3} \left[\vphantom{\frac{e}{m}}6 g_A^2 c_6 \left(2 \mu ^2-1\right) +11 g_A^2 c_7\left(2 \mu ^2-1\right) +c_2 m \vphantom{\frac{e}{m}}\right] \,,
\end{align*}

\begin{align*}
&\alpha_{\text{E1}\nu}^{(n)}  =  
 -\frac{e^2 g_A^2}{1536 \pi ^3 F^2 \mu ^2 \left(\mu ^2-4\right)^2 m^3} \left[\vphantom{\frac{e}{m}}8c_6 \Xi_2  \left(18 \mu ^8-197\mu ^6+710 \mu ^4-826 \mu ^2+184\right) \right. \vphantom{\frac{e}{m}} \\
 &\;\;\;\; \left. + 4c_7 \Xi_2 \left(90 \mu ^8-969 \mu ^6+3390 \mu ^4-3740 \mu^2+656\right)  + c_6 \left(135 \mu ^4-1012 \mu^2+1408\right) \right. \vphantom{\frac{e}{m}} \\
 &\;\;\;\; \left. + 6c_7 \left(57 \mu ^4-414 \mu ^2+584\right)\vphantom{\frac{e}{m}}\right] + \frac{e^2 \Xi_1}{384 \pi ^3 F^2 m^3}\left[\vphantom{\frac{e}{m}} 2 g_A^2 c_6 \left(18 \mu ^2-17\right)+ 3 g_A^2 c_7 \left(30 \mu ^2 \right.\right. \vphantom{\frac{e}{m}} \\
 &\;\;\;\; \left.\left. -23\right)
- c_2 m\vphantom{\frac{e}{m}}\right]+\frac{e^2 \left(96 c_1+\left(16-15 \mu ^2\right) c_2+132c_3\right)}{11520 \pi ^3 F^2 \mu ^2 m^2} -\frac{e^2(\tilde{e}_{117}-e_{92})}{8 \pi m^2}\,,\\
\\
&\beta_{\text{M1}\nu}^{(n)}  = \;
 \frac{e^2 g_A^2}{1536 \pi ^3 F^2 \mu ^2\left(\mu ^2-4\right)^3 m^3} \left[\vphantom{\frac{e}{m}}8c_6 \Xi_2 \left(-29 \mu^{10}+419 \mu ^8-2212 \mu ^6+5016 \mu ^4 \right.\right. \vphantom{\frac{e}{m}} \\
 &\;\;\;\; \left.\left. - 3976 \mu ^2+128\right) + 4c_7 \Xi_2 \left(-108 \mu ^{10}+1567\mu ^8-8330 \mu ^6+19104 \mu ^4-15520 \mu ^2 \right.\right. \vphantom{\frac{e}{m}} \\
 &\;\;\;\; \left.\left. +896\right) - c_6 \left(223 \mu ^8-2420 \mu ^6+8768 \mu ^4-9792 \mu ^2+512\right) - 2c_7 \left(207 \mu^8-2264 \mu ^6 \right.\right. \vphantom{\frac{e}{m}} \\
 &\;\;\;\; \left.\left. +8240 \mu ^4-9152 \mu ^2+512\right)\vphantom{\frac{e}{m}}\right]+ \frac{e^2 \Xi_1}{384 \pi ^3 F^2 m^3} \left[2 g_A^2 c_6 \left(29 \mu ^2-13\right)
+ g_A^2c_7 \left(108 \mu ^2 \right.\right. \vphantom{\frac{e}{m}} \\
&\;\;\;\; \left.\left. -55\right)
- c_2 m\right]- \frac{e^2 \left(32 c_1+\left(5 \mu ^2+8\right)c_2+44 c_3\right)}{3840 \pi ^3 F^2 \mu ^2 m^2} 
- \frac{e^2 (\tilde{e}_{117}-e_{92})}{8 \pi m^2}\,.
\end{align*}

\subsubsection{Spin-dependent first order polarizabilities}\label{subsub:neutron-spin-dependent-1} 

\begin{align*}
&\gamma_{\text{E1E1}}^{(n)}
 =  \; 
 \frac{e^2 g_A^2}{192 \pi^3 F^2 \left(\mu^2-4\right)^2 m^2} \left[\vphantom{\frac{e}{m}}
 c_6 \Xi_2 \left(-4 \mu^6+43\mu^4-150 \mu^2+180\right) +c_7 \Xi_2 \left(-13 \mu^6 \right.\right. \vphantom{\frac{e}{m}} \\
 &\;\;\;\; \left.\left.+139 \mu^4-474 \mu^2+516\right)
-c_6 \left(4 \mu^4-29\mu^2+64\right) - c_7 \left(13 \mu^4-89 \mu^2+160\right)\vphantom{\frac{e}{m}} \right]  \vphantom{\frac{e}{m}} \\
&\;\;\;\; +\frac{e^2 g_A^2 \Xi_1}{192 \pi^3 F^2 m^2} \left[\vphantom{\frac{e}{m}}c_6 \left(4 \mu^2-3\right)
+ c_7 \left(13 \mu^2-9\right)\vphantom{\frac{e}{m}}\right]\,,\\
\\
&\gamma_{\text{M1M1}}^{(n)}
 =  \; 
 \frac{e^2 g_A^2}{192 \pi^3 F^2 \left(\mu^2-4\right)^2 m^2 \mu^2} \left[\vphantom{\frac{e}{m}}
 c_6 \Xi_2 \left(-9 \mu^8+92\mu^6-290 \mu^4+216\mu^2\right)\right. \vphantom{\frac{e}{m}} \\
 &\;\;\;\; \left. -c_7 \Xi_2 \left(23 \mu^8-239 \mu^6+786 \mu^4-712 \mu^2+64\right)
-c_6 \left(9 \mu^6-62 \mu^4+56\mu^2\right) \right. \vphantom{\frac{e}{m}} \\
&\;\;\;\; \left.- c_7 \left(23 \mu^6-166 \mu^4+200\mu^2\right) \vphantom{\frac{e}{m}}\right] +\frac{e^2 g_A^2 \Xi_1}{192 \pi^3 F^2 m^2} \left[\vphantom{\frac{e}{m}}c_6 \left(9 \mu^2-2\right)
+ c_7 \left(23 \mu^2-9\right)\vphantom{\frac{e}{m}}\right]\,,\\ 
\\
&\gamma_{\text{E1M2}}^{(n)}
 =  \; 
 \frac{e^2 g_A^2}{192 \pi^3 F^2 \left(\mu^2-4\right)^2 m^2} \left[\vphantom{\frac{e}{m}}
c_6 \Xi_2 \left(3 \mu^6-32\mu^4+110 \mu^2-108\right)+ c_7 \Xi_2 \left(-3 \mu^6 \right.\right. \vphantom{\frac{e}{m}} \\
&\;\;\;\; \left.\left. +31 \mu^4-94 \mu^2+84\right)
+ c_6 \left(3 \mu^4-23\mu^2+32\right)
- c_7 \left(3 \mu^4-13 \mu^2+16\right) \vphantom{\frac{e}{m}}\right] \vphantom{\frac{e}{m}} \\
&\;\;  +\frac{e^2 g_A^2 \Xi_1}{192 \pi^3 F^2 m^2} \left[\vphantom{\frac{e}{m}}c_6 \left(2-3 \mu^2\right)
+ c_7 \left(3 \mu^2-1\right)\vphantom{\frac{e}{m}}\right]\,,\\ 
\\
&\gamma_{\text{M1E2}}^{(n)}
 =  \; 
 \frac{e^2 g_A^2}{192 \pi^3 F^2 m^2 \mu^2} \left[\vphantom{\frac{e}{m}}
c_6 \Xi_2 \left(2 \mu^4-7 \mu^2+2\right)
- c_7 \Xi_2 \left(\mu^4+\mu^2-2\right) + (2c_6 -c_7)\mu^2 \vphantom{\frac{e}{m}}\right] \vphantom{\frac{e}{m}} \\
&\;\; +\frac{e^2 g_A^2 \Xi_1}{192 \pi^3 F^2 m^2} \left[\vphantom{\frac{e}{m}}c_6 \left(3-2 \mu^2\right)
+ c_7 \left(\mu^2+3\right)\vphantom{\frac{e}{m}}\right] \,.
\end{align*}

\subsubsection{Spin-dependent second order polarizabilities}\label{subsub:neutron-spin-dependent-2} 

\begin{align*} 
&\gamma_{\text{E2E2}}^{(n)}  =  
\; \frac{e^2 g_A^2}{138240 \pi^3 F^2 \mu^2 \left(\mu^2-4\right)^3 m^4} \left[\vphantom{\frac{e}{m}}
3 c_6 \Xi_2 \left(-895 \mu^{10}+13239 \mu^8-72576 \mu^6 \right.\right. \vphantom{\frac{e}{m}} \\
&\;\;\;\; \left.\left. +175180 \mu^4-164240 \mu^2+20480\right)
+ 6 c_7 \Xi_2 \left(-795 \mu^{10}+11782 \mu^8-64748 \mu^6 \right.\right. \vphantom{\frac{e}{m}} \\
&\;\;\;\; \left.\left. +156750 \mu^4-147080 \mu^2+19200\right)
+ c_6 \mu^2 \left(-2685 \mu^6+30197 \mu^4-111268 \mu^2 \right.\right. \vphantom{\frac{e}{m}} \\
&\;\;\;\; \left.\left. +138080\right)
+ c_7 \mu^2 \left(-4770 \mu^6+53717\mu^4-197908 \mu^2+241760\right)\vphantom{\frac{e}{m}}\right]  \\
&\;\; +\frac{e^2 g_A^2 \Xi_1}{46080 \pi^3 F^2 m^4} \left[\vphantom{\frac{e}{m}}c_6 \left(895 \mu^2-709\right)
 + 2 c_7 \left(795 \mu^2-652\right)\vphantom{\frac{e}{m}} \right]\,,\\
 \\
&\gamma_{\text{M2M2}}^{(n)}  =  
\; \frac{e^2 g_A^2}{138240 \pi^3 F^2 \mu^2 \left(\mu^2-4\right)^3 m^4} \left[\vphantom{\frac{e}{m}}
3 c_6 \Xi_2 \left(-1377 \mu^{10}+19929 \mu^8-105504 \mu^6 \right.\right. \vphantom{\frac{e}{m}} \\
&\;\;\;\; \left.\left. +238100 \mu^4-181520 \mu^2+17280\right)
+ 6 c_7 \Xi_2 \left(-1341 \mu^{10}+19482 \mu^8  -103812 \mu^6\right.\right. \vphantom{\frac{e}{m}} \\
&\;\;\;\; \left.\left. +237490 \mu^4-189400 \mu^2+22080\right) 
- c_6 \left(4131 \mu^8-45451 \mu^6  +159284 \mu^4 \right.\right. \vphantom{\frac{e}{m}} \\
&\;\;\;\; \left.\left. -145024 \mu^2+3840\right)
- c_7 \left(8046 \mu^8-89011 \mu^6+315884 \mu^4  -306304 \mu^2+15360\right)\vphantom{\frac{e}{m}}\right] \\
&\;\; +\frac{e^2 g_A^2 \Xi_1}{15360 \pi^3 F^2 m^4} \left[\vphantom{\frac{e}{m}}c_6 \left(459 \mu^2-217\right)
 + 2 c_7 \left(447 \mu^2-236\right) \vphantom{\frac{e}{m}}\right]\,,\\
 \\
&\gamma_{\text{E2M3}}^{(n)}  =  
\; \frac{e^2 g_A^2}{69120 \pi^3 F^2 \mu^2 \left(\mu^2-4\right)^3 m^4} \left[\vphantom{\frac{e}{m}}
3 c_6 \Xi_2 \left(113 \mu^{10}-1689 \mu^8+9408 \mu^6-23060 \mu^4 \right.\right. \vphantom{\frac{e}{m}} \\
&\;\;\;\; \left.\left. +20080 \mu^2-2560\right)
- 6 c_7 \Xi_2 \mu^2 \left(51 \mu^8-718 \mu^6+3596 \mu^4-7230 \mu^2  +5000\right)\right.\vphantom{\frac{e}{m}} \\
&\;\;\;\; \left.+ c_6 \mu^2 \left(339 \mu^6-4003 \mu^4+15452 \mu^2-15136\right)
+ c_7 \mu^2 \left(-306 \mu^6  +2957 \mu^4-7828 \mu^2 \right.\right. \vphantom{\frac{e}{m}} \\
&\;\;\;\; \left.\left. +7904\right)\vphantom{\frac{e}{m}}\right] +\frac{e^2 g_A^2 \Xi_1}{23040 \pi^3 F^2 m^4} \left[ \vphantom{\frac{e}{m}}c_6 \left(107-113 \mu^2\right)
 + 2 c_7 \left(51 \mu^2-4\right) \vphantom{\frac{e}{m}}\right]\,,\\
 \\
&\gamma_{\text{M2E3}}^{(n)}  =  
 \; \frac{e^2 g_A^2}{69120 \pi^3 F^2 \mu^2 \left(\mu^2-4\right)^2 m^4} \left[\vphantom{\frac{e}{m}}
3 c_6 \Xi_2 \left(111 \mu^8-1227 \mu^6+4500 \mu^4-5980 \mu^2 \right.\right. \vphantom{\frac{e}{m}} \\
&\;\;\;\; \left.\left. +1920\right)
+ 6 c_7 \Xi_2 \left(3 \mu^8-66 \mu^6+420 \mu^4-1070 \mu^2+960\right)
+ c_6 \left(333 \mu^6  -2393 \mu^4 \right.\right. \vphantom{\frac{e}{m}} \\
&\;\;\;\; \left.\left. +4568 \mu^2-960\right)
+ c_7 \left(18 \mu^6-53 \mu^4+248 \mu^2-960\right)\vphantom{\frac{e}{m}}\right]  \vphantom{\frac{e}{m}} \\
&\;\;\;\; +\frac{e^2 g_A^2 \Xi_1}{7680 \pi^3 F^2 m^4} \left[\vphantom{\frac{e}{m}}c_6 \left(39-37 \mu^2\right)
 - 2 c_7 \left(\mu^2-12\right) \vphantom{\frac{e}{m}}\right]\,,\\
 \\
&\gamma_{\text{E1E1}\nu}^{(n)}  =  
\; \frac{e^2 g_A^2}{46080 \pi^3 F^2 \mu^2 \left(\mu^2-4\right)^4 m^4} \left[\vphantom{\frac{e}{m}}
- 3 c_6 \Xi_2 \left(1339 \mu^{12}-24999 \mu^{10}+184860 \mu^8 \right.\right. \vphantom{\frac{e}{m}} \\
&\;\;\;\; \left.\left. -674892 \mu^6+1211040 \mu^4-883520 \mu^2+122880\right)
- 6 c_7 \Xi_2 \left(1787 \mu^{12}  -33422 \mu^{10} \right.\right. \vphantom{\frac{e}{m}} \\
&\;\;\;\; \left.\left. +247860 \mu^8-909966 \mu^6+1654880\mu^4-1245280 \mu^2  +184320\right)
+ c_6 \left(-4017\mu^{10} \right.\right. \vphantom{\frac{e}{m}} \\
&\;\;\;\; \left.\left. +61305 \mu^8-346112 \mu^6+833840 \mu^4-727424 \mu^2  +84992\right)
+ c_7 \left(-10722\mu^{10} \right.\right. \vphantom{\frac{e}{m}} \\
&\;\;\;\; \left.\left. +163845 \mu^8-929672 \mu^6+2283920 \mu^4  -2071424 \mu^2+223232\right)\vphantom{\frac{e}{m}}\right] \\ &\;\;+\frac{e^2 g_A^2 \Xi_1}{15360 \pi^3 F^2 m^4} \left[\vphantom{\frac{e}{m}} 13 c_6 \left(103 \mu^2-69\right)
 + 2 c_7 \left(1787 \mu^2-1256\right) \vphantom{\frac{e}{m}}\right]\,,\\
 \\
&\gamma_{\text{M1M1}\nu}^{(n)}  =  
 \; \frac{e^2 g_A^2}{15360 \pi^3 F^2 \mu^4 \left(\mu^2-4\right)^4 m^4} \left[\vphantom{\frac{e}{m}}
- c_6 \Xi_2 \mu^2 \left(1413 \mu^{12}-26345 \mu^{10}+194436 \mu^8 \right.\right. \vphantom{\frac{e}{m}} \\
&\;\;\;\; \left.\left. -708756 \mu^6+1285760 \mu^4-953280 \mu^2+122880\right)
- 2 c_7 \Xi_2 \left(1997 \mu^{14}  -37266 \mu^{12} \right.\right. \vphantom{\frac{e}{m}} \\
&\;\;\;\; \left.\left. +275292 \mu^{10}-1003890 \mu^8+1814400 \mu^6-1334496 \mu^4  +168448 \mu^2+6144\right) \right. \vphantom{\frac{e}{m}} \\
&\;\;\;\; \left. + c_6 \mu^2 \left(-1413 \mu^{10}+21277 \mu^8-117992 \mu^6  +292400 \mu^4-256896 \mu^2+8192\right)\right. \vphantom{\frac{e}{m}} \\
&\;\;\;\; \left. + c_7 \mu^2 \left(-3994 \mu^{10}+60273 \mu^8  -334632 \mu^6+818160 \mu^4-707712 \mu^2+31744\right)\vphantom{\frac{e}{m}}\right] \\
&\;\;+\frac{e^2 g_A^2 \Xi_1}{15360 \pi^3 F^2 m^4} \left[ \vphantom{\frac{e}{m}}c_6 \left(1413 \mu^2-911\right)
 + 2 c_7 \left(1997 \mu^2-1320\right) \vphantom{\frac{e}{m}}\right]\,,\\
 \\
&\gamma_{\text{E1M2}\nu}^{(n)}  =  
\; \frac{e^2 g_A^2}{115200 \pi^3 F^2 \mu^2 \left(\mu^2-4\right)^4 m^4} \left[\vphantom{\frac{e}{m}}
3 c_6 \Xi_2 \left(-1621 \mu^{12}+29085 \mu^{10}-202572 \mu^8 \right.\right. \vphantom{\frac{e}{m}} \\
&\;\;\;\; \left.\left. +667452 \mu^6-949760 \mu^4+264640 \mu^2+180480\right)
- 6 c_7 \Xi_2 \left(3183 \mu^{12}  -58600 \mu^{10}\right.\right. \vphantom{\frac{e}{m}} \\
&\;\;\;\; \left.\left. +424836 \mu^8-1504806 \mu^6+2553440\mu^4-1608800 \mu^2  +76160\right)
- c_6 \left(4863\mu^{10} \right.\right. \vphantom{\frac{e}{m}} \\
&\;\;\;\; \left.\left. -71337 \mu^8+376280 \mu^6-744112 \mu^4+285888 \mu^2  +97280\right)
+ c_7 \left(-19098\mu^{10}\right.\right. \vphantom{\frac{e}{m}} \\
&\;\;\;\; \left.\left. +287277 \mu^8-1591400 \mu^6+3681712 \mu^4  -2808768 \mu^2+56320\right)\vphantom{\frac{e}{m}}\right] \\ &\;\; +\frac{e^2 g_A^2 \Xi_1}{38400 \pi^3 F^2 m^4} \left[\vphantom{\frac{e}{m}} c_6 \left(1621 \mu^2+93\right)
 + 2 c_7 \left(3183 \mu^2-1306\right) \vphantom{\frac{e}{m}}\right]\,,\\
 \\
&\gamma_{\text{M1E2}\nu}^{(n)}  =  
 \; \frac{e^2 g_A^2}{115200 \pi^3 F^2 \mu^4 \left(\mu^2-4\right)^3 m^4} \left[\vphantom{\frac{e}{m}}
-3 c_6 \Xi_2 \left(443 \mu^{12}-6359 \mu^{10}+33208 \mu^8 \right.\right. \vphantom{\frac{e}{m}} \\
&\;\;\;\; \left.\left. -74660 \mu^6+69680 \mu^4-3840 \mu^2+5120\right)
-6 c_7 \Xi_2 \left(2089 \mu^{12}-30692 \mu^{10} +166204 \mu^8 \right.\right. \vphantom{\frac{e}{m}} \\
&\;\;\;\; \left.\left. -391370 \mu^6+350200 \mu^4-42240 \mu^2+2560\right)
+c_6 \mu^2 \left(-1329\mu^8+13323 \mu^6-40892 \mu^4\right.\right. \vphantom{\frac{e}{m}} \\
&\;\;\;\; \left.\left.+51616 \mu^2+33280\right)
+c_7 \mu^2 \left(-12534 \mu^8+  137763\mu^6-482012 \mu^4+542176 \mu^2 \right.\right. \vphantom{\frac{e}{m}} \\
&\;\;\;\; \left.\left. +33280\right) \vphantom{\frac{e}{m}}\right] +\frac{e^2 g_A^2 \Xi_1}{38400 \pi^3 F^2 m^4} \left[\vphantom{\frac{e}{m}}c_6 \left(443 \mu^2-157\right)
 + 2 c_7 \left(2089 \mu^2-1446\right) \vphantom{\frac{e}{m}}\right]\,.
\end{align*}

\section{Renormalization of the nucleon magnetic moments}\label{sec:counter_terms}
Below, we provide the expressions for the LECs $c_6$, $c_7$ in terms of the
renormalized quantities $\bar c_6$ and $\bar c_7$, see subsection~\ref{sub:renormalization}.
\begin{align}
 c_6 = \bar c_6+\delta c_6^{(3)}+\delta c_6^{(4)}\,, \ c_7 = \bar c_7+\delta c_7^{(3)}+\delta c_7^{(4)}\,,
\end{align}
with
\begin{align}
\delta c_6^{(3)} &= -\frac{ g_A^2}{F^2 (4 m^2-M^2)} ((4 (\epsilon -4) m^2-3 (\epsilon -2) M^2) A_0(m) \nonumber \\
&\hspace{0.25cm} +(4 (5-2 \epsilon ) m^2+3 (\epsilon -2) M^2) A_0(M) \nonumber \\
&\hspace{0.25cm} +(16 m^4+2 (4 \epsilon -13) M^2 m^2-3(\epsilon -2) M^4) B_0(m,M,m^2)) \nonumber \\
&-\frac{h_A^2}{3 (\epsilon -2) (2 \epsilon -3)^3 F^2 m^2 m_\Delta^4}(-10 (\epsilon -2) (\epsilon -1)^2 (2 \epsilon -3) m^6-20 (\epsilon -2)^2(\epsilon -1) (2 \epsilon \nonumber \\
&\hspace{0.25cm}-1) m_\Delta m^5+(20 (\epsilon -1)^2 (2 \epsilon -5) (2 \epsilon -3) M^2+(\epsilon -2)^2 (\epsilon  (4 \epsilon  (7 \epsilon-18)-13)\nonumber \\
&\hspace{0.25cm}+47) m_\Delta^2) m^4+2 (\epsilon -2) (2 \epsilon -3) m_\Delta (10 (\epsilon -1) (3 \epsilon -4) M^2+(\epsilon  (\epsilon  (14 \epsilon-43)+10)\nonumber \\
&\hspace{0.25cm}+2) m_\Delta^2) m^3+2 (\epsilon -2) (5 (\epsilon -1)^2 (2 \epsilon -3) M^4+(7-\epsilon  (8 \epsilon  ((\epsilon -5) \epsilon +3)+25))m_\Delta^2 M^2\nonumber \\
&\hspace{0.25cm}+(\epsilon  (\epsilon  (8 \epsilon ^2-50 \epsilon +71)-51)+35) m_\Delta^4) m^2+2 (\epsilon -2) (M^2-m_\Delta^2) m_\Delta (((\epsilon \nonumber \\
&\hspace{0.25cm}-2) \epsilon (12 (\epsilon -3) \epsilon +47)+20) m_\Delta^2-20 (\epsilon -1)^2 M^2) m\nonumber \\
&\hspace{0.25cm}-3 (\epsilon -1) (2 \epsilon ^2-7 \epsilon +6)^2 m_\Delta^2 (m_\Delta^2-M^2)^2)A_0(M)\nonumber \\
&+\frac{ h_A^2}{3 (\epsilon -2) (2 \epsilon -3)^3 F^2 m^2 m_\Delta^4} (10 (\epsilon -2) (\epsilon -1)^2 (2 \epsilon -3) m^6+20 (\epsilon -2)^2 (\epsilon -1) (2 \epsilon\nonumber \\
&\hspace{0.25cm} -1) m_\Delta m^5+(-20 (\epsilon -2) (\epsilon-1)^2 (2 \epsilon -3) M^2-(\epsilon  (\epsilon  (\epsilon  (4 \epsilon  (7 \epsilon -36)+207)+111)\nonumber \\
&\hspace{0.25cm}-460)+248) m_\Delta^2) m^4+2 m_\Delta ((\epsilon (\epsilon  (\epsilon  (4 \epsilon  (13 \epsilon -94)+1035)-1306)+754)\nonumber \\
&\hspace{0.25cm}-172) m_\Delta^2-10 (\epsilon -2) (\epsilon -1) (\epsilon  (2 \epsilon -7)+4)M^2) m^3+2 (\epsilon -2) (5 (\epsilon -1)^2 (2 \epsilon -3) M^4\nonumber \\
&\hspace{0.25cm}-5 (\epsilon  (\epsilon  (4 (\epsilon -5) \epsilon +27)-13)+4) m_\Delta^2 M^2+(\epsilon (2 \epsilon  (5 \epsilon  (2 \epsilon -11)+91)-141)\nonumber \\
&\hspace{0.25cm}+62) m_\Delta^4) m^2+2 (\epsilon -2) (M-m_\Delta) m_\Delta (M+m_\Delta) (((\epsilon -2) \epsilon  (12(\epsilon -3) \epsilon +47)\nonumber \\
&\hspace{0.25cm}+20) m_\Delta^2-20 (\epsilon -1)^2 M^2) m-3 (\epsilon -1) (2 \epsilon ^2-7 \epsilon +6)^2 m_\Delta^2 (m_\Delta^2-M^2)^2)A_0(m_\Delta)\nonumber \\
&+\frac{ h_A^2 ((m+m_\Delta)^2-M^2)}{3 (2 \epsilon -3)^3 F^2 m^2 m_\Delta^4}(-10 (\epsilon -1)^2 (2 \epsilon -3) m^6+20 (\epsilon -1) m_\Delta m^5\nonumber \\
&\hspace{0.25cm}+(2 \epsilon -3) (20 (\epsilon -1)^2M^2+(\epsilon  (\epsilon  (14 \epsilon -23)-39)+38) m_\Delta^2) m^4+2 m_\Delta ((\epsilon  (4 (22\nonumber \\
&\hspace{0.25cm}-5 \epsilon ) \epsilon -159)+98) m_\Delta^2-10 (\epsilon -1)(2 \epsilon -1) M^2) m^3+2 (2 \epsilon -3) (-5 (\epsilon -1)^2 M^4\nonumber \\
&\hspace{0.25cm}+5 (\epsilon  (\epsilon  (2 \epsilon -7)+7)-4) m_\Delta^2 M^2+(7-2 (\epsilon -1) \epsilon (5 \epsilon -12)) m_\Delta^4) m^2\nonumber \\
&\hspace{0.25cm}+2 (M^2-m_\Delta^2) m_\Delta(20 (\epsilon -1)^2 M^2+(\epsilon  (4 (10-3 \epsilon ) \epsilon -59)+34) m_\Delta^2) m\nonumber \\
&\hspace{0.25cm}+3(3-2 \epsilon )^2 (\epsilon -2) (\epsilon -1) m_\Delta^2 (M^2-m_\Delta^2)^2) B_0(M,m_\Delta,m^2)  \,,  \\ \nonumber \\
\delta c_7^{(3)} &= -\frac{4  (1-\epsilon) g_A^2 m^2}{F^2(4m^2-M^2)}(2 A_0(m) - A_0(M))+ \frac{2  g_A^2 m^2 (4 m^2-2 (1-\epsilon) M^2)}{F^2(4m^2-M^2)}B_0(m,M,m^2) \nonumber \\
&- \frac{2  h_A^2}{3 F^2(\epsilon -2) (2 \epsilon -3)^3 m m_\Delta^4} ((\epsilon -2) (\epsilon -1)^2 (2 \epsilon -3) m^5 +2 (\epsilon -2)^2 (\epsilon -1)(2 \epsilon  \nonumber \\
&\hspace{0.25cm}-1) m_\Delta m^4+(-2 (\epsilon-1)^2 (2 \epsilon -5) (2 \epsilon -3) M^2-(\epsilon -2)^2 (\epsilon  (4 (\epsilon -3) \epsilon +5) \nonumber \\
&\hspace{0.25cm}+2) m_\Delta^2) m^3-(\epsilon -2) (2 \epsilon -3)m_\Delta (2 (\epsilon -1) (3 \epsilon -4) M^2+(\epsilon  (2 \epsilon  (2 \epsilon -7)+11) \nonumber \\
&\hspace{0.25cm}-5) m_\Delta^2) m^2+(\epsilon -2) (M^2-m_\Delta^2)((\epsilon(2 \epsilon -7) (2 (\epsilon -2) \epsilon +3)+7) m_\Delta^2 \nonumber \\
&\hspace{0.25cm}-(\epsilon -1)^2 (2 \epsilon -3) M^2) m+4 (\epsilon -2) (\epsilon -1)^2 m_\Delta(M^2-m_\Delta^2)^2) A_0(M) \nonumber \\
&-\frac{2  h_A^2}{3 F^2(\epsilon -2) (2\epsilon -3)^3 m m_\Delta^4} (-4 (\epsilon -2)(\epsilon -1)^2 m_\Delta^5+(\epsilon -2) (\epsilon  (2 \epsilon -7) (2 (\epsilon -2) \epsilon  \nonumber \\
&\hspace{0.25cm}+3)+7) m m_\Delta^4+(((\epsilon -1) \epsilon  (8 \epsilon ((\epsilon -6) \epsilon +12)-59)-2) m^2+8 (\epsilon -2) (\epsilon  \nonumber \\
&\hspace{0.25cm}-1)^2 M^2) m_\Delta^3+m ((-\epsilon  (2 \epsilon -5) (\epsilon  (\epsilon  (2\epsilon -7)+8)+2)-14) m^2 \nonumber \\
&\hspace{0.25cm}-(\epsilon -2) (\epsilon  (\epsilon  (4 (\epsilon -5) \epsilon +27)-13)+4) M^2) m_\Delta^2 \nonumber \\
&\hspace{0.25cm}+2 (\epsilon -2) (\epsilon-1) (m^2-M^2) ((\epsilon -2) (2 \epsilon -1) m^2+2 (\epsilon -1) M^2) m_\Delta \nonumber \\
&\hspace{0.25cm}+(\epsilon -2) (\epsilon -1)^2 (2 \epsilon -3) m (m-M)^2 (m+M)^2)A_0(m_\Delta) \nonumber \\
&-\frac{2  h_A^2((m+m_\Delta)^2-M^2)}{3F^2(2 \epsilon -3)^3 m m_\Delta^4} (-(\epsilon -1)^2(2 \epsilon -3) m^5+2 (\epsilon -1) m_\Delta m^4+(2 \epsilon -3) (2 (\epsilon  \nonumber \\
&\hspace{0.25cm}-1)^2 M^2+((2 \epsilon -5) \epsilon ^2+2) m_\Delta^2) m^3+m_\Delta ((-4 \epsilon^2+6 \epsilon -2) M^2+(\epsilon  (-4 (\epsilon -5) \epsilon  \nonumber \\
&\hspace{0.25cm}-39)+25) m_\Delta^2) m^2-(2 \epsilon -3) (M^2-m_\Delta^2) ((\epsilon -1)^2 M^2+(\epsilon  (-2(\epsilon -4) \epsilon -9) \nonumber \\
&\hspace{0.25cm}+5) m_\Delta^2) m+4 (\epsilon -1)^2 m_\Delta (M^2-m_\Delta^2)^2) B_0(m_\Delta,M,m^2)  \,,  \\ \nonumber \\
\delta c_6^{(4)} &= -\frac{ g_A^2 c_{6} M^2 B_0(m,M,m^2)}{2 F^2 (4 m^2-M^2)} (8 m^2 (2 \epsilon -2)+M^2 (5-6 \epsilon)) \nonumber \\
&+\frac{ A_0(M)}{2 F^2 m (2\epsilon -4) (4 m^2-M^2)} (c_{6} m (2 \epsilon -4) (8 m^2(g_A^2 (2 \epsilon -2)-1)+M^2 (2 \nonumber \\
&\hspace{0.25cm} -3 g_A^2 (2 \epsilon -2)))+4 (4 m^2-M^2) (2 c_{4} m^2 (2 \epsilon -4)+3 c_{2} M^2)) \nonumber \\
&-\frac{ g_A^2 c_{6} A_0(m)}{2 F^2 (4 m^2-M^2)} (4 m^2 (2 \epsilon -1)+M^2 (5-6 \epsilon ))+\frac{4 M^2}{m} (c_{1} c_{6}+4e_{106} m^2)  \,,  \\ \nonumber \\
\delta c_7^{(4)} &= -\frac{ g_A^2 M^2 B_0(m,M,m^2)}{2 F^2 (4 m^2-M^2)} (c_{6} (2 M^2-4 m^2 (2 \epsilon +1))-3 c_{7} (M^2 (2 \epsilon -3)+8 m^2)) \nonumber \\
&-\frac{ A_0(M)}{2 F^2 (4 m^2-M^2)}(c_{6} (4 m^2 (g_A^2 (2 \epsilon +1)-1)+(1-3 g_A^2) M^2)\nonumber \\
&\hspace{0.25cm} +3 g_A^2 c_{7} (M^2 (2 \epsilon -4)+8 m^2)+4 c_{4} m (4m^2-M^2))\nonumber \\
& +\frac{ g_A^2 A_0(m)}{2 F^2 (4 m^2-M^2)} (c_{6} (8 m^2 (2 \epsilon -1)-2 M^2)+3 c_{7} (4 m^2 (2 \epsilon -1)+M^2 (2 \epsilon-3)))\nonumber \\
&+\frac{4 M^2}{m} (c_{1} c_{7}+2 m^2 (2 e_{105}-e_{106}))  \,,
\end{align}
with the space-time dimension $d = 4- 2\epsilon$.
The expressions for the loop integrals are provided in
Appendix~\ref{sec:loop_integrals}.

\section{Loop integrals}\label{sec:loop_integrals}
The loop integral functions are defined as
\begin{align}
A_0(m) &=\frac{1}{i}\int\frac{d^d l}{(2\pi)^d}\frac{\mu^{4-d}}{l^2-m_0^2+i\varepsilon}\,,\nonumber\\
B_0(m_1,m_2,p^2) 
&=\frac{1}{i}\int\frac{d^d l}{(2\pi)^d}\frac{\mu^{4-d}}{(l^2-m_1^2+i\varepsilon)((l-p)^2-m_2^2+i\varepsilon)}\,.
\label{eq:loop_integrals}
\end{align}
The renormalization scale $\mu$ in all integrals is set to $\mu=m$.

\bibliography{4.0}
\bibliographystyle{apsrev}

\end{document}